\documentclass[a4paper,fleqn,10pt]{article}
\pdfoutput=1
\usepackage{amsmath}
\usepackage{amssymb}
\usepackage{array}
\usepackage{calc}
\usepackage{longtable}
\usepackage{multirow}
\usepackage{slashed}
\usepackage{pstricks}
\usepackage{graphicx}
\usepackage{xspace}
\usepackage{units}
\usepackage{tikz}
\usepackage{textcomp}

\numberwithin{equation}{section}
\usepackage{cite}
\usepackage[pdfborder={0 0 0}]{hyperref}
\usepackage[format=hang,labelfont=bf,hypcap=true]{caption}
\usepackage{subcaption}
\usepackage{sectsty}
\usepackage{enumitem}
\allsectionsfont{\sffamily}
\subsubsectionfont{\mdseries\itshape\large}
\makeatletter
\DeclareRobustCommand*{\bfseries}{%
  \not@math@alphabet\bfseries\mathbf
  \fontseries\bfdefault\selectfont
  \boldmath
}
\makeatother
\let\spreprint\empty
\newcommand{\preprint}[1]{\def\spreprint{\protect#1}}
\let\sinstitute\empty
\newcommand{\institute}[1]{\def\sinstitute{\protect#1}}
\makeatletter
\renewcommand{\maketitle}{\begingroup
  \null\thispagestyle{empty}%
    \ifx\spreprint\empty
      \vskip 5ex
    \else
      \flushright\large\spreprint\vskip 10ex
    \fi
    \vskip 5ex
    \centering
      {\sffamily\bfseries\huge\@title}\vskip 10ex
    \flushleft
      \@author\vskip 2ex
      \ifx\sinstitute\empty
      \else
        {\small\sinstitute}
      \fi
    \vskip 5ex
  \endgroup
}
\makeatother
\renewenvironment{abstract}{\begin{center}
  {\large\sffamily\bfseries Abstract: }
  \begin{minipage}[t]{0.75\textwidth}
}{\end{minipage}\end{center}\vskip 10ex}

\numberwithin{equation}{section}
\allowdisplaybreaks[2]

\newcommand{\LHAPDF}{L\protect\scalebox{0.8}{HAPDF}\xspace}

\newcommand{\MSbar}{\ensuremath{\overline{\text{MS}}}\xspace}

\newcommand{\Rivet}{R\protect\scalebox{0.8}{IVET}\xspace}

\newcommand{\Recola}{R\protect\scalebox{0.8}{ECOLA}\xspace}
\newcommand{\OpenLoops}{O\protect\scalebox{0.8}{PEN}L\protect\scalebox{0.8}{OOPS}\xspace}

\newcommand{\Sherpa}{S\protect\scalebox{0.8}{HERPA}\xspace}

\newcommand{\Amegic}{A\protect\scalebox{0.8}{MEGIC}\xspace}

\long\def\symbolfootnote[#1]#2{\begingroup%
\def\thefootnote{\fnsymbol{footnote}}\footnote[#1]{#2}\endgroup}

\newcommand{\eqsref}[2]{(\ref{#1}-\ref{#2})}

\newcommand{\done}{{\rm d}}
\newcommand{\order}{\mathcal{O}}

\newcommand{\nnb}{\nonumber}
\newcommand{\bea}{\begin{eqnarray}}
\newcommand{\eea}{\end{eqnarray}}
\newcommand{\bi}{\begin{itemize}}
\newcommand{\ei}{\end{itemize}}

\newcommand{\qT}{\ensuremath{q_\mathrm{T}}}
\newcommand{\qTvec}{\ensuremath{\vec{q}_\mathrm{T}}}

\newcommand{\dphi}{\ensuremath{\Delta\phi}}
\newcommand{\bT}{\ensuremath{b_\mathrm{T}}}
\newcommand{\bTvec}{\ensuremath{\vec{b}_\mathrm{T}}}

\newcommand{\NLO}{\ensuremath{\text{NLO}}}\xspace
\newcommand{\NNLO}{\ensuremath{\text{N$^2$LO}}}\xspace
\newcommand{\NNNLO}{\ensuremath{\text{N$^3$LO}}}\xspace
\newcommand{\NNNNLO}{\ensuremath{\text{N$^4$LO}}}\xspace
\newcommand{\NLOs}{\ensuremath{\text{NLO$_\text{s}$}}}\xspace
\newcommand{\NNLOs}{\ensuremath{\text{N$^2$LO$_\text{s}$}}}\xspace
\xspace
\newcommand{\NLL}{\ensuremath{\text{NLL}}}\xspace
\newcommand{\NNLL}{\ensuremath{\text{N$^2$LL}}}\xspace
\newcommand{\NNLLD}{\ensuremath{\text{N$^2$LL$_{\mathrm{D}}$}}}\xspace
\newcommand{\NNLLR}{\ensuremath{\text{N$^2$LL$_{\mathrm{R}}$}}}\xspace
\newcommand{\NNNLL}{\ensuremath{\text{N$^3$LL}}}\xspace
\xspace
\newcommand{\NNLLp}{\ensuremath{\text{N$^2$LL$'$}}}\xspace
\xspace
\newcommand{\NLLNLO}{\ensuremath{\text{NLL+NLO}}}\xspace
\newcommand{\NNLLNNLO}{\ensuremath{\text{N$^2$LL+N$^2$LO}}}\xspace
\newcommand{\NNLLDNNLO}{\ensuremath{\text{N$^2$LL$_{\mathrm{D}}$+N$^2$LO}}}\xspace
\newcommand{\NNLLRNNLO}{\ensuremath{\text{N$^2$LL$_{\mathrm{R}}$+N$^2$LO}}}\xspace
\xspace
\xspace
\xspace
\xspace
\xspace
\newcommand{\SCETII}{\ensuremath{\text{SCET$_{\mathrm{II}}$}}}\xspace

\newcommand{\tbar}{{\ensuremath{\bar{t}}}}
\newcommand{\ttbar}{{\ensuremath{t\tbar}}}
\newcommand{\qbar}{{\ensuremath{\bar{q}}}}
\newcommand{\qqbar}{{\ensuremath{q\qbar}}}
\newcommand{\betattbar}{\ensuremath{\beta_\ttbar}}
\newcommand{\dEttbar}{\ensuremath{\Delta E_\ttbar}}
\newcommand{\Mttbar}{\ensuremath{M_\ttbar}}
\newcommand{\MTttbar}{\ensuremath{M_{\mathrm{T}}^\ttbar}}

\newcommand{\Yttbar}{\ensuremath{Y_\ttbar}}
\newcommand{\Lttbar}{\ensuremath{L_\ttbar}}
\newcommand{\dphittbar}{\ensuremath{\Delta\phi_{\ttbar}}}
\newcommand{\dPhittbar}{\ensuremath{\Delta\Phi_{\ttbar}}}

\newcommand{\nbar}{{\ensuremath{\bar{n}}}}

\newlist{myitemize}{itemize}{3}
\setlist[myitemize]{leftmargin=14em}

\newcolumntype{C}{>{\centering\arraybackslash}p{0.14\textwidth}}

\newlength{\unitcharwidth}
\hypersetup{
  pdfauthor={Wan-Li Ju, Marek Schoenherr},
  pdftitle={The qT and dphi spectra in top-antitop hadroproduction at NNLL+NNLO: the interplay of soft-collinear resummation and Coulomb singularities}
}
\preprint{IPPP/24/40\\MCnet-24-12}
\author{Wan-Li Ju${}^{(a,c)}$, Marek Sch{\"o}nherr${}^{(b)}$}
\title{The \texorpdfstring{\qT}{qT} and \texorpdfstring{\dphittbar}{dphi} spectra
       in top-antitop hadroproduction at \texorpdfstring{\NNLLNNLO}{NNLL+NNLO}:\\[5mm]
       \Large the interplay of soft-collinear resummation and Coulomb singularities}
\institute{${}^{(a)}$~INFN, Sezione di Milano, Via Celoria 16, 20133 Milano, Italy\\
${}^{(b)}$~Institute for Particle Physics Phenomenology, Durham University, Durham DH1 3LE, United Kingdom\\
${}^{(c)}$~Department of Physics, University of Alberta, Edmonton AB T6G 2J1, Canada\\
$~$\\
Emails:  \href{ }{\color{black} wju@ualberta.ca}\,, \href{ }{\color{black} marek.schoenherr@durham.ac.uk} }
\newcommand{\changed}[1]{#1}

\begin{document}
\vspace*{10mm}
\maketitle
\vspace*{20mm}
\begin{abstract} 
  In this paper, we \changed{present} the \changed{resummation-improved}
  differential transverse momentum
  and azimuthal decorrelation cross sections,
  $\done\sigma_{t\bar{t}}/\done\qT$ and
  $\done\sigma_{t\bar{t}}/\done\dphittbar$, in top-antitop pair
  production at the LHC.
  \changed{Our calculation is based on the observation that both
  cross sections are dominated by topologies where the top-quark pair
  is well separated, expressed in their relative velocity
  $\betattbar\sim\order(1)$,
  at colliding energies of $\sqrt{s}=13\,\text{TeV}$ or higher.
  Therefore, the asymptotic behaviour in the limits $\qT\to0$ and
  $\dphi_\ttbar\to0$ can mostly be captured by the soft and collinear
  resummation in the HQET$+$SCET framework.
  Nevertheless, starting at \NNLL, Coulomb singularities emerge in the
  threshold regime, $\betattbar\to0$, in both the hard sector and its
  evolution kernels, leading to unphysical results upon integration
  over the entire $\betattbar$ range.}
  To this end, two prescriptions, dubbed the D- and R-\changed{prescription},
  are introduced to \changed{regularise these Coulomb singularities.}
  \changed{They embody two fundamentally different methods to
  truncate the threshold enhanced terms, rendering their contribution finite.
  In the absence of a combined threshold and small-transverse-momentum
  resummation, we present a quantitative assessment of the ambiguity
  introduced by the choice of prescription, itself a test of the sensitivity
  of our calculation to such threshold enhancements, for both the
  $\done\sigma_{t\bar{t}}/\done\qT$ and
  $\done\sigma_{t\bar{t}}/\done\dphittbar$ spectra.  }
\end{abstract}
\newpage
\tableofcontents
\section{Introduction}
\label{sec:intro}
   
The investigation of top-antitop pair (\ttbar) production at hadron
colliders has drawn both experimental and theoretical attention in the
past decades.
This has facilitated the precise determination of the top quark mass $m_t$
as an input parameter
of the Standard Model (SM) as well as the exploration of many possible
new physics scenarios.
In the recent experiments carried out at Large Hadron Collider (LHC),
the total cross sections of the top-antitop pair hadroproduction have
been measured at a variety of colliding energies, for instance,
$\sqrt{s}=5.02\,\text{TeV}$~\cite{ATLAS:2022jbj,CMS:2017zpm,
  ATLAS:2021xhc,CMS:2021gwv},
$7\,\text{TeV}$~\cite{CMS:2013yjt,ATLAS:2014nxi,ATLAS:2012ptu,
  CMS:2016yys,CMS:2016csa,CMS:2012hcu,LHCb:2015nta,
  ATLAS:2022aof},
$8\,\text{TeV}$~\cite{ATLAS:2019guf,ATLAS:2016pal,CMS:2016poo,
  CMS:2015auz,ATLAS:2014nxi,CMS:2016yys,CMS:2016csa,LHCb:2015nta,
  ATLAS:2022aof},
~$13\,\text{TeV}$~\cite{ATLAS:2023gsl,CMS:2015yky,LHCb:2018usb,
  ATLAS:2016qjg,ATLAS:2019hau,ATLAS:2020aln,CMS:2019snc,
  CMS:2016rtp,CMS:2018fks},
and~$13.6\,\text{TeV}$~\cite{CMS:2022elr,ATLAS:2023slx}.
In addition, many properties of the final state have been measured in
single and double-differential distributions
\cite{ATLAS:2012vvt,CMS:2012hkm,ATLAS:2014ipf,ATLAS:2015dbj,
  ATLAS:2016pal,CMS:2017iqf,CMS:2016poo,CMS:2015rld,
  ATLAS:2015lsn,ATLAS:2015mip,CMS:2024ybg,CMS:2022uae,
  ATLAS:2023gsl,CMS:2016oae,ATLAS:2016pbv,CMS:2017xio,
  ATLAS:2019hau,CMS:2019esx,CMS:2020tvq,ATLAS:2020ccu,
  ATLAS:2022xfj,ATLAS:2022mlu,CMS:2021vhb,CMS:2018htd,
  CMS:2018adi,ATLAS:2019hxz}, among them the transverse momentum
\qT\ of the \ttbar\ system, its invariant mass \Mttbar\ or the separation
in the azimuthal plane \dPhittbar.
Simultaneously, precise theoretical predictions were developed and
the first \NLO\ accurate calculations, including first-order QCD
corrections, became available over 30 years ago
\cite{Nason:1987xz,Beenakker:1988bq,Beenakker:1990maa,Mangano:1991jk}.
More recently, the precision of the theoretical predictions has been
further increased by including second-order corrections at \NNLO\ accuracy
in QCD \cite{Czakon:2013goa,Czakon:2015owf,Czakon:2016ckf,
  Czakon:2017dip,Czakon:2017wor,Catani:2019iny,Catani:2020tko,
  Czakon:2020coa,Czakon:2021ohs,Catani:2019hip,Garzelli:2023rvx} 
 and first-order \NLO\ electroweak (EW)
effects \cite{Bernreuther:2010ny,Kuhn:2006vh,Bernreuther:2006vg,
  Kuhn:2013zoa,Hollik:2011ps,Pagani:2016caq,Gutschow:2018tuk,
  Denner:2016jyo,Czakon:2017wor,Bernreuther:2024ltu,Frederix:2021zsh}.
Alongside, corrections to top-quark decays and off-shell corrections
were included \cite{Gao:2012ja,Brucherseifer:2013iv,
  Catani:2019hip,Behring:2019iiv,Czakon:2020qbd,
  Stremmer:2024ecl,Ablat:2023tiy,Meng:2022htg}.
Even though these fixed-order results are able to describe the
production cross sections in the majority of the phase space,
considerable corrections can emerge in particular kinematic limits from
all orders in the perturbative series underpinning these calculations,
calling for resummation techniques to improve the perturbative convergence
and, in turn, provide reliable theoretical predictions.
Existing research in this context comprises soft-gluon resummation in  
\ttbar\ production  
\cite{Kidonakis:2014pja,Kidonakis:2023juy,Kidonakis:2010dk,
  Kidonakis:2009ev,Kidonakis:2014isa,Kidonakis:2019yji,
  Ahrens:2010zv,Ferroglia:2012ku,Ferroglia:2013awa,Pecjak:2016nee,
  Czakon:2018nun,Almeida:2008ug,Pecjak:2018lif}, the Coulomb
resummation around $\Mttbar\to2m_t$ \cite{Hagiwara:2008df,Kiyo:2008bv,
  Ju:2020otc,Ju:2019mqc} with a generic transverse recoil against the
$\ttbar$ system, the combined resummation of Coulomb and soft-gluon
corrections \cite{Beneke:2009ye,Beneke:2010da,Beneke:2011mq,
  Cacciari:2011hy,Piclum:2018ndt},
and the resummation of soft and collinear parton emissions
\cite{Zhu:2012ts,Li:2013mia,Catani:2014qha,Catani:2017tuc,
  Catani:2018mei,Alioli:2021ggd,Ju:2022wia} in the small
transverse recoil region.
Parton-shower matched predictions at the highest fixed-order
precision can be found in \cite{Mazzitelli:2020jio,
  Mazzitelli:2021mmm,Jezo:2023rht}.
  
In this work we will continue to study the resummation of
\changed{QCD} logarithms in the process $pp\to\ttbar+X$ and put
particular emphasis on the asymptotic regions $\qT\to0$ and
$\dPhittbar\to\pi$.
The \changed{azimuthally averaged} distribution $\done\sigma_{\ttbar}/\done\qT$
\changed{is an observable that
is free of any azimuthally asymmetric divergences \cite{Catani:2017tuc}.}
It thus allows for a systematic resummation of the asymptotic behaviour
of each perturbative order by means of
exponentiating the logarithmic contributions in impact-parameter space,
akin to the corresponding procedure in the Drell-Yan process
\cite{Collins:1984kg,Becher:2010tm,Bozzi:2010xn,
  GarciaEchevarria:2011rb,Becher:2011xn,Banfi:2011dx,
  Banfi:2011dm,Banfi:2012du,Catani:2015vma,Scimemi:2017etj,
  Bizon:2018foh,Bacchetta:2019sam,Bizon:2019zgf,Becher:2020ugp,
  Ebert:2020dfc,Re:2021con,Camarda:2021ict,Ju:2021lah,
  Camarda:2023dqn,Neumann:2022lft,Moos:2023yfa} or Higgs
hadroproduction \cite{Belyaev:2005bs,Bozzi:2005wk,Bozzi:2007pn,
  Monni:2016ktx,Becher:2020ugp,Becher:2012yn,Neill:2015roa,
  Bizon:2017rah,Chen:2018pzu,Bizon:2018foh,
  Gutierrez-Reyes:2019rug,Harlander:2014hya,Billis:2021ecs,
  Cal:2023mib}.
In the existing literature, focusing on the leading singular contributions
\changed{in low \qT\ domain},
such a logarithmic exponentiation has been presented in
\cite{Zhu:2012ts,Li:2013mia}, through a combination
of the Soft-Collinear Effective Theory (SCET) \cite{Bauer:2001yt,
  Bauer:2001ct,Bauer:2000yr,Bauer:2000ew,Bauer:2002nz,
  Beneke:2002ph,Beneke:2002ni,Bauer:2002aj,Lange:2003pk,
  Beneke:2003pa} with the Heavy-Quark Effective
Theory (HQET)~\cite{Eichten:1989zv,Georgi:1990um,Grinstein:1990mj,
  Neubert:1993mb}, as well as using a generalised CSS approach
in \cite{Catani:2014qha,Catani:2017tuc,Catani:2018mei}.
\changed{
Moreover, a second class of observables, the projected transverse
momentum spectra  $\done\sigma_{\ttbar}/\done q_{\tau}$, were proposed
in \cite{Ju:2022wia} to remove any azimuthally asymmetric contributions,
where $q_{\tau}\equiv |\qTvec \cdot \vec{\tau}|$ represents the magnitude
of the projection of \qTvec\ onto a reference unit vector $\vec{\tau}$
in the transverse plane.
Choosing this reference vector $\vec{\tau}$ perpendicular to the top
quark spatial momentum, $\done\sigma_{\ttbar}/\done q_{\tau}$ can be
related to the azimuthal decorrelation
$\done\sigma_{\ttbar}/\done\dPhittbar$ between the top and antitop quarks.
This observable has been of particular interest in recent measurements
at the LHC \cite{CMS:2022uae,CMS:2024ybg}, where it was observed that
fixed-order calculations generally exhibited significant theoretical uncertainties in the vicinity of $\dPhittbar\to\pi$, demanding the
inclusion and resummation of the dominant higher-order corrections.
}

\changed{
In addition, Coulomb divergences appear in the physical region of
top-quark pair production in the vicinity of their production
threshold \cite{Hagiwara:2008df,Kiyo:2008bv,
  Ju:2020otc,Ju:2019mqc,Beneke:2009ye,Beneke:2010da,Beneke:2011mq,
  Cacciari:2011hy,Piclum:2018ndt},
characterised through $\betattbar=0$ or $\dEttbar=0$,
with the relative velocity $\betattbar\equiv\sqrt{1-4m_t^2/M^2_\ttbar}$
and the energy separation $ \dEttbar\equiv \Mttbar-2m_t$.
This Coulomb divergence has the potential to formally spoil the
perturbative convergence of the above SCET+HQET resummed calculation.
In \cite{Ju:2022wia}, in order to remove the thus afflicted region
from our consideration, a lower bound on the top-antitop-pair
invariant mass, $\Mttbar\ge400\,\text{GeV}$, was introduced.
In the following, we will explore methods to lift this kinematic
constraint, so as to extend our resummation to encompass the whole
$\ttbar$ production phase space.
Meanwhile, as the leading singular terms as $\qT\to0$ and
$\dPhittbar\to\pi$ are driven by the same dynamic modes, we will
also generalise the resummation formalism in \cite{Ju:2022wia}
through an adapted multipole expansion procedure to access
$\done\sigma_{\ttbar}/\done\qT$ itself.
}

\changed{
Ideally, the overlap of Coulomb divergences and divergences of
soft-collinear origin calls for a combined resummation to study
the $\qT$ and $\dPhittbar$ spectra near the threshold regime.
In terms of combining SCET and potential non-relativistic
QCD (pNRQCD) \cite{Pineda:1997bj,Brambilla:1999xf,Beneke:1999zr,
  Beneke:1999qg} (or, alternatively, velocity non-relativistic QCD,
vNRQCD \cite{Luke:1999kz,Manohar:1999xd,Manohar:2000hj,
  Manohar:2000kr,Hoang:2002ae}), such a resummation at
leading-logarithmic (LL) accuracy primarily entails the products
of LO Green function \cite{Beneke:1999qg,Beneke:1999zr,
  Pineda:2006ri,Beneke:2011mq}, the tree-level amplitudes for
the hard processes, and the Sudakov factor consisting of the cusp
anomalous dimensions at LO.
Even though the leading threshold enhancements induced by Coulomb
interactions can be resummed at such an accuracy, it limits
the perturbative corrections of soft and collinear radiation
to its lowest order at leading power, thereby neglecting crucial
known higher-order corrections that have been embedded in the
earlier works \cite{Zhu:2012ts,Li:2013mia,Ju:2022wia}.
Owing to the superrenormalisable nature of the Coulomb vertices
\cite{Beneke:2009ye,Pineda:2000gza}, beyond-LL precision
involves both leading and sub-leading power perturbative
contributions from the soft-collinear sector,
in analogy with the combination of soft and Coulomb resummation
\cite{Beneke:2010da,Beneke:2011mq,Piclum:2018ndt,Ju:2019lwp}.
In recent years, a lot of effort has been devoted to calculate
subleading power contributions to the \qT\ spectrum in colour-singlet
hadroproduction
\cite{Balitsky:2017gis,Balitsky:2020jzt,Balitsky:2021fer,
  Balitsky:2017flc,Vladimirov:2021hdn,Ebert:2021jhy,
  Gamberg:2022lju,Rodini:2022wic,Rodini:2023plb,
  Vladimirov:2023aot,Rodini:2023mnh,Ebert:2018gsn,
  Inglis-Whalen:2021bea,Inglis-Whalen:2022vyn,Ferrera:2023vsw,
  Campbell:2024hjq},
while results for coloured heavy partons processes have still to appear.
In addition, off-shell top quarks effects  will become relevant if the top-antitop quark
pair invariant mass is lowered further, $\Mttbar\le 2m_t$.
They have been investigated in lepton collider environments in the
vicinity of the \ttbar\ production threshold \cite{Beneke:2003xh,
  Beneke:2004km,Hoang:2004tg,Hoang:2006pd,Beneke:2010mp,
  Penin:2011gg,Jantzen:2013gpa,Beneke:2017rdn}.
Generalising these calculations to a hadron collider, however, involves
further complexity due to the jet-like nature of the final state,
necessitating the use of jet algorithms for event selection.
This introduces the non-global dynamics \cite{Becher:2023vrh,
  Becher:2023mtx,Becher:2024nqc} into the problem, structurally
changing the factorisation formulae derived in \cite{Ju:2022wia}.%
}

\changed{
Therefore, choosing a pragmatic approach aiming for a phenomenological
appraisal of both the \qT\ and \dPhittbar\ spectra, we will employ ad hoc
prescriptions to achieve a meaningful soft-collinear resummation in the
presence of Coulomb divergences.
The expectation is that for the process $pp\to\ttbar+X$ taking place
 at a colliding energy $\sqrt{s}=13\,\text{TeV}$, the bulk of
events are produced well above the \ttbar\ production threshold in the
domain $\Mttbar\ge400\,\text{GeV}$ where $\betattbar\sim\order(1)$ or
$\dEttbar\sim\order(m_t)$.
Here, the top and antitop quarks are kinematically well separated and
the dynamics behind the asymptotic behaviour are captured entirely by
SCET and HQET.
Then, extending the coverage of the QCD resummation established in this
domain to the whole phase space including the threshold region, using
such an ad hoc prescription, can serve as a rapid and reasonable
estimation for the single differential observables
$\done\sigma_{\ttbar}/\done\qT$ and $\done\sigma_{\ttbar}/\done\dPhittbar$
without its details having a major impact when integrated over the entire
invariant mass range.
Such a methodology has been extensively applied in existing calculations
on the total cross section of the processes
$pp\to t\bar{t}+X$ \cite{Kidonakis:1997gm,Bonciani:1998vc,
  Kidonakis:2001nj,Ahrens:2010zv,Czakon:2018nun},
$pp\to t\bar{t} B(B=H,Z,W^{\pm})+X$ \cite{Broggio:2015lya,
  Broggio:2016lfj,Broggio:2017kzi,Kulesza:2017ukk,
  Kulesza:2017jqv,Kulesza:2018tqz},
and $pp\to t\bar{t}t\bar{t}+X$~\cite{vanBeekveld:2022hty}.
}

\changed{
However, extending the resummation phase space to include the threshold
regime, $\Mttbar\ge 2m_t$, is not always straightforward in a SCET$+$HQET
based resummation.
To be precise, the resummation in the well-separated region
\cite{Ju:2022wia} features the products of the fixed-order sectors,
including hard, soft, and beam-collinear functions, as well as their
corresponding scale evolution kernels.
In this work, we will demonstrate that, starting from next-to-next-to
leading logarithmic order (\NNLL), taking the threshold limit
$\beta_{\ttbar}\to0$ of those ingredients will develop cubic and
quadratic divergences in the triple differential cross sections
$\done^3\sigma_{\ttbar}/(\done\betattbar\done\Yttbar\done\qT)$ and
$\done^3\sigma_{\ttbar}/(\done\betattbar\done\Yttbar\done\dPhittbar)$,
respectively.
This, in turn, leads to a diverging phase space integration in evaluating
$\done\sigma_{\ttbar}/\done\qT$ and $\done\sigma_{\ttbar}/\done\dPhittbar$
without further regularisation on the threshold enhancements.
}

\changed{To this end, in this paper, we introduce two ad hoc
prescriptions to treat this problem.}
Their derivation is based on the observation that in using the
expanded solution of the hard RGE~\cite{Buras:1991jm,Buchalla:1995vs}
the main driver for the threshold divergences are the
non-logarithmic products of the hard scale evolution kernels,
contributing $\mathcal{O}(\betattbar^{-4})$ in the limit of
$\betattbar\to0$.
Ideally, this behaviour can be mitigated by implementing the exact
solution of the hard RGE.
However, in presence of soft colour correlations, such an exact
solution necessitates the path-ordered integration over a set of
threshold-enhanced colour matrices.
Unfortunately, neither an analytically compact expression nor a
numerical approximation via Taylor expansion is straightforward.
Hence, \changed{akin to \cite{Kulesza:2017ukk,Alioli:2021ggd,
  Ahrens:2010zv,Zhu:2012ts,Li:2013mia}}, we first introduce
the ``decomposition (D) \changed{prescription}'', in
which the threshold-singular contributions at \NNLL\ are, in part,
shifted to a higher logarithmic accuracy at the cost of mild
corrections in the domain $\betattbar\sim\mathcal{O}(1)$.
\changed{It} allows a smooth and consistent extrapolation
to the threshold area $\betattbar\to0$.
On the other hand, we will also introduce the
``re-exponentiation (R) \changed{prescription}''.
In spite of the difficulties in determining a rigorous solution
of hard RGE for a generic \betattbar, we will demonstrate that
solving hard RGE can be substantially simplified in the vicinity
of $\betattbar=0$.
This is thanks to the fact that up to two-loop level the leading
threshold divergences all reside in the diagonal entries of the
hard anomalous dimensions~\cite{Ferroglia:2009ii,Ferroglia:2009ep}.
Consequently, the leading singular behaviour of the hard anomalous
dimensions can be exponentiated by solving an approximate hard RGE.
The resulting resummation kernels in the R-\changed{prescription} present intensively
oscillatory but integrable behaviour in the limit $\betattbar\to0$.
\changed{Due to the fact that in these prescriptions,  the threshold enhanced series are truncated in two radically different approaches, comparing their  outcome can  deliver a quantitative assessment on the dependence of $\done\sigma_{\ttbar}/\done\qT$ and
$\done\sigma_{\ttbar}/\done\dPhittbar$ on the ad hoc prescriptions and in turn unveil their sensitivity to the higher order Coulomb interactions.}

The paper is structured as follows.
In Sec.~\ref{sec:methods} we start with a brief review of the
soft-collinear resummation on the \qT\ and \dPhittbar\ spectra
for the well-separated region, thereby specifying the fixed-order
ingredients and anomalous dimensions comprised up to \NNLL.
Then, Sec.~\ref{sec:sigma:thr:hqet} is devoted to an analysis
of the asymptotic behaviour of \changed{HQET and SCET based resummation}
in the vicinity of $\betattbar\to0$, from which we raise the
concern over the integrability of the resummation kernel at \NNLL.
In turn, we propose the two \changed{ad hoc} prescriptions discussed above in
Sec.~\ref{sec:thr:div:extrap} to mitigate the arising threshold
singularities before we match the resummed \qT\ and \dPhittbar\
distributions to the exact fixed-order calculations in Sec. \ref{sec:mat}.
With our framework in place, we deliver a numeric evaluation
in Sec.~\ref{sec:results}.
Therein, we will at first validate the perturbative expansion of our
resummed results by comparing against the \qT\ and \dPhittbar\
distribution computed in the full theory in three different
\Mttbar\ slices, i.e.\ the threshold domain
$\Mttbar\in[2m_t, 360]\,\text{GeV}$, the transitional region
$\Mttbar\in[360,400]\,\text{GeV}$, and the well-separated realm
$\Mttbar\ge400\,\text{GeV}$.
Finally, we present our final resummation improved \qT\ and
\dPhittbar\ distributions at \NNLLNNLO\ accuracy using both
the D- and R-schemes before  concluding this work in
Sec.~\ref{sec:conclusions}.

\section{Theoretical details}
\label{sec:methods}
  
\subsection{Soft and collinear resummation in the domain \texorpdfstring{$\dEttbar\sim\mathcal{O}(m_t)$}{Delta E of O(mt)}}
\label{sec:methods:domains}

From the QCD factorisation theorem~\cite{Collins:1989gx},  the differential cross section of a generic observable $\mathcal{Q}$ for the process $pp\to \ttbar+X$ can be expressed as,
\begin{equation}\label{eq:QCDF}
  \begin{split}
    \frac{\done^3{\sigma_{\ttbar}}}
         {\done \Mttbar^2\,
          \done \Yttbar\,\done \mathcal{Q}}
    \,=\;&
      \sum_{\mathrm{sign}[P_{t}^{z}]}\frac{1}{16s\,(2\pi)^6 }
      \int \done^2\vec{P}_{t}^{\perp}\,\done^2 \qTvec\;
      \delta\Big[\mathcal{Q}-\mathcal{F}_{\mathcal{Q}}\Big]\;
      \frac{\Sigma_{\ttbar}}{\MTttbar\,|P_{t}^{z}|}\,,
  \end{split}
\end{equation}
where $s$ denotes collider energy and will be taken to be $13\,\text{TeV}$
throughout our investigation.
$\vec{P}_{t}^{\perp}$ stands for the transverse momentum of
the top quark measured in the laboratory reference frame (LRF),
while $ P_{t}^{z}$ marks its longitudinal components detected
from the $z$-direction rest frame ($z$RF) of the top-antitop pair.
Further, $\qTvec$, $\Mttbar$, and $\Yttbar$ represent the
transverse momentum, invariant mass and pseudo-rapidity of the
$\ttbar$ system in LRF, respectively, from which the transverse
mass of the top-antitop pair can be expressed as
\begin{align}
  \Mttbar^{\mathrm{T}}=\sqrt{\Mttbar^2+\qT^2}.
\end{align}
$\mathcal{Q}$ in Eq.~\eqref{eq:QCDF} refers to the observable,
which can be evaluated via its definition function
$\mathcal{F}_{\mathcal{Q}}$.
$\mathcal{F}_{\mathcal{Q}}$ takes the following form
for the observables of interest in the present paper,
\begin{align}
  \label{eq:meas:func:qT_dphi}
  \begin{split}
    \mathcal{Q}&=\qT\,,\quad\quad\quad\quad\quad\quad \quad\quad ~~~~\mathcal{F}_{\mathcal{Q}}=\changed{\left|\qTvec\right|}\,,\\
    \mathcal{Q}&= \Delta\phi_{\ttbar}\equiv\pi-\Delta\Phi_{\ttbar}  \,,\quad\quad\quad \mathcal{F}_{\mathcal{Q}}=\pi-\arccos \left[\frac{\vec{P}^{\bot}_t\cdot \vec{P}^{\bot}_{ \tbar}}{ |\vec{P}^{\bot}_t | |\vec{P}^{\bot}_{  \tbar} |}\right]\,.
  \end{split}
\end{align}
Here, $\vec{P}^{\bot}_{\tbar}$ stands for the transverse momenta
of the antitop quark in the LRF, satisfying
$\vec{P}^{\bot}_{\tbar}=\qTvec-\vec{P}^{\bot}_{t}$.
$\Delta\Phi_{\ttbar}$ measures the azimuthal separation of the
top and antitop quarks in the transverse plane.

At last, ${\Sigma_{\ttbar}}$ in Eq.~\eqref{eq:QCDF} collects the
contributions from all participating partonic processes,
\begin{align}\label{eq:def:QCD:parton}
  \begin{split}
    {\Sigma_{\ttbar}}
    \,=&\;
      \sum_{i,j} \int^1_0 \frac{\done x_n}{x_n}\,
                          \frac{\done x_{\nbar}}{x_{\nbar}}\,
      f_{i/N}(x_n)\,f_{j/\bar{N}}(x_{\nbar})\,
      \sum_{r\changed{=0}}^{\changed{\infty}}\,\int\,
      \left[\prod_{m\changed{=1}}^r \frac{\done^3 \vec{k}_m}{(2\pi)^3\,2E_{k_m}}\right]
      \overline{\sum_{\mathrm{hel, col}}}
            \left|\mathcal{M}(i+j\to t+\tbar+ X ) \right|^2\\
    &\hspace*{70mm}
      \times\;
      (2\pi)^4\;\delta^4\left(p_i+p_{j}-P_{t}-P_{\tbar}-\sum_{m\changed{=0}}^{\changed{r}} k_m\right)\,,
  \end{split}
\end{align}
where the $f_{i/N}(x)$ is the parton distribution function (PDF)
for parton $i$ with the momentum fraction $x$ from proton $N$, and
$E_{k_m}$ and $\vec{k}_{m}$ are the energy and spatial momentum
of the $m$-th emitted parton, respectively.
$\mathcal{M}$ evaluates the transition amplitude of the occurring
partonic scattering $i+j\to t+\tbar+X$, with
$\{i,j\}\in[u,\bar{u},d,\bar{d},s,\bar{s},c,\bar{c},b,\bar{b},g]$,
in line with the $5$ active flavour scheme.
 
Substituting Eq.\ \eqref{eq:def:QCD:parton} into
Eq.\ \eqref{eq:QCDF}, we can now appraise the $\qT$
and $\Delta\phi_{\ttbar}$ spectra on the fixed-order level.
Although such a calculation delivers satisfactory predictions
in most phase space regions, it converges poorly as
$\qT\to0$ or $\dphittbar\to0$ due to the occurrence of
large logarithmic corrections to all orders.
Thus, a resummation of this asymptotic behaviour is mandated.
 
In the domain where the top and antitop quarks are kinematically
well-separated, i.e.\
\begin{align}\label{eq:defdE}
  \dEttbar\equiv \Mttbar-2m_t\sim\order(m_t)\,,
\end{align}
the factorisation and resummation of the azimuthally averaged
distribution $\done{\sigma_{\ttbar}}/\done\qT$ have been
investigated in different approaches, including the EFT-based
analysis~\cite{Zhu:2012ts,Li:2013mia} and the generalized
CSS framework~\cite{Catani:2014qha,Bonciani:2015sha,
  Catani:2017tuc,Catani:2018mei}.
It is demonstrated that (at least) the leading singular behaviour
of the \qT\ distribution is predominantly driven by the hard,
soft and beam-collinear domains in the loop and phase space
integrations.
This conclusion has been extensively applied in fixed order calculations \cite{Bonciani:2015sha,Catani:2019iny,Catani:2019hip,Catani:2020tko,Garzelli:2023rvx,Catani:2021cbl,Catani:2022mfv,Buonocore:2022pqq,Catani:2020kkl,Buonocore:2023ljm} and also their combination with parton showers \cite{Mazzitelli:2020jio,Mazzitelli:2021mmm,Mazzitelli:2023znt,Mazzitelli:2024ura}.

Recently, to further investigate the top-antitop-pair dynamics,
the differential distribution of the projected transverse momentum
$\done{\sigma_{\ttbar}}/\done q_{\tau}$ was computed in
\cite{Ju:2022wia}, where $q_{\tau}$ signifies the projection of
$\qTvec$ onto a reference unit vector $\vec{\tau}$ on the azimuthal
plane, from which the $\Delta\phi_{\ttbar}$ spectrum can be derived
by choosing $\vec{\tau}$ perpendicular to the flight direction of
(anti)top quark.
At variance with the small \qT\ region, which imposes constraints
on both components of $\qTvec$, the asymptotic regime
$q_{\tau}\to0$ or $\Delta\phi_{\ttbar}\to0$ concerns only the
longitudinal projection $q_{\tau}=|\qTvec\cdot \vec{\tau}|$, leaving the
transverse part unresolved.
To probe the dynamic modes for the transverse component, in
\cite{Ju:2022wia}, employing the method of expansion of dynamic
regions \cite{Beneke:1997zp,Smirnov:2002pj,Smirnov:2012gma,
  Jantzen:2011nz} as well as the SCET formalism \cite{Bauer:2001yt,
  Bauer:2001ct,Bauer:2000yr,Bauer:2000ew,Bauer:2002nz,
  Beneke:2002ph,Beneke:2002ni,Bauer:2002aj,Lange:2003pk,
  Beneke:2003pa},
we enumerate the possible regions that can prompt energetic recoil
against the top-antitop system, finding that assigning the label
momenta to the transverse direction will incur an additional suppression
from the phase space by at least one power of
$\lambda_{\tau}\equiv q_{\tau}/\Mttbar$, such that the leading
singular behaviour of $\done{\sigma_{\ttbar}}/\done q_{\tau}$ is
also captured by the hard, soft and beam-collinear regions, akin
to the \qT\ resummation in \cite{Zhu:2012ts,Li:2013mia,Catani:2014qha}.

Given their common dynamic regions that preside over the leading
singular contributions \footnote{%
  As far as we know, this coincidence only takes place in the
  leading power factorisation and resummation, since without
  accidental cancellations the central collinear mode can be
  relevant for $\done\sigma_{\ttbar}/\done\Delta\phi_{\ttbar}$
  starting from the subleading power~\cite{Ju:2022wia},
  whereas its participation in $\done\sigma_{\ttbar}/\done\qT$
  is postponed to the sub-subleading power by its kinematics
  \cite{Collins:1980ui,Collins:1981uk,Ju:2022wia}.
  Analogously,  structural similarities between Eqs.~(\ref{eq:LP:res}-\ref{eq:LP:res:gg}) and those governing 
  resummation-improved azimuthal decorrelation of the
  jet-boson~\cite{Chien:2019gyf,Chen:2018fqu,Chien:2020hzh,
    Bouaziz:2022tik,Chien:2022wiq} and dijet
  \cite{delCastillo:2021znl,Banfi:2008qs,Zhang:2022wvs,
    Gao:2023ulg} processes
  may also be limited to leading power, especially when the jets therein
  are defined exclusively.
},
we can utilise a uniform framework to compute the resummed expressions
for both the $\qT$ and $\Delta\phi_{\ttbar}$ distributions.
Within the context of \SCETII~\cite{Bauer:2002aj,Lange:2003pk,
  Beneke:2003pa} and HQET~\cite{Eichten:1989zv,Georgi:1990um,
  Grinstein:1990mj,Neubert:1993mb}, both of them comprise the
resummed partonic function,
\begin{align}\label{eq:LP:res}
  \Sigma_{\ttbar }^{\mathrm{res}}
  =&\;
    \frac{8\pi^2}{\Mttbar^2}\,
    \sum_{\kappa}\,
    \int {\done^2\bTvec}\,\exp\left(\mathrm{i}\,\bTvec \cdot\qTvec \right)\,
    \widetilde{\Sigma}_{\ttbar}^{\mathrm{res}, [\kappa]}
    (\bTvec,\Yttbar, \Mttbar,\Omega_t)\,,
\end{align}
where $\kappa$ runs over
$\{g_{n}g_{\nbar},q^i_{n}\qbar^j_{\nbar},q^i_{\nbar}\qbar^j_{n}\}$,
enumerating the active initial-state parton-pairs contributing to
the hard kernels, with $i,j\in\{u,d,c,s,b\}$ specifying the flavour
of the quark fields.
$\widetilde{\Sigma}_{\ttbar}^{\mathrm{res},[\kappa]}$ collects the
partonic contribution after Fourier transforming it into impact-parameter
space, which is in general a function of the impact parameter $\bTvec$,
the pseudorapidity $\Yttbar$, the invariant mass $\Mttbar$,
and the solid angle $\Omega_t$ of the top quark measured in the
rest reference frame of $\ttbar$ system.
$\widetilde{\Sigma}_{\ttbar}^{\mathrm{res},[\kappa]}$ is formally
related to the choice of the scheme regularising the rapidity
divergences.
In the following, we will use the soft and beam functions
evaluated within the exponential regulator as proposed in
\cite{Li:2016axz,Li:2016ctv}.
Alternative choices can also be found in \cite{Zhu:2012ts,
  Li:2013mia,Angeles-Martinez:2018mqh,Gehrmann:2012ze,
  Gehrmann:2014yya} calculated via analytic rapidity regulator
\cite{Becher:2011dz}, and in \cite{Catani:2011kr,Catani:2021cbl,Catani:2022sgr,Catani:2023tby} using a generalised
CSS method~\cite{Catani:2014qha}.
It follows that,
\begin{align}
  \label{eq:LP:res:qqbar}
  \begin{split}
  &\hspace*{-5mm}
  \widetilde{\Sigma}_{\ttbar}^{\mathrm{res},[q^i_{n}\qbar^j_{\nbar}]}
  (\bTvec,\Yttbar, \Mttbar,\Omega_t)\\
  \,=&\;
  \left(\frac{1}{2N_c}\right)^2\,
  \mathcal{D}^{\mathrm{res}}_{[q^i_{n}\qbar^j_{\nbar}]}
  (\bT,\Mttbar,\mu_h,\mu_b,\mu_s,\nu_b,\nu_s)\,
  \mathcal{B}_{n}^{[q_n^i]}(\eta_n,\bT,\mu_b,\nu_b)\,
  \mathcal{B}_{\nbar}^{[\qbar_\nbar^j]}(\eta_{\nbar},\bT,\mu_b,\nu_b) \,
  \\
  &
  \sum_{\{\alpha,\beta,h\}}\Bigg\{
  \mathcal{S}^{\alpha_1\beta_1}_{[q_{n}\qbar_{\nbar}]}(\bTvec,v_{t},v_{\tbar},\mu_s,\nu_s)
  \left[\mathcal{V}^{[q_{n}\qbar_{\nbar}]}_{\alpha_1\alpha_2}(v_{t},v_{\tbar},\mu_s,\mu_h) \mathcal{C}^{[q^i_{n}\qbar^j_{\nbar}]}_{\alpha_2;h_nh_{\nbar}h_th_{\tbar}} \right]^*\, \mathcal{V}^{[q_{n}\qbar_{\nbar}]}_{\beta_1\beta_2}(v_{t},v_{\tbar},\mu_s,\mu_h)\,
  \\
  &   
  \mathcal{C}_{\beta_2;h_nh_{\nbar}h_th_{\tbar}}^{[q^i_{n}\qbar^j_{\nbar}]}\, \Bigg\}\,,
  \end{split}
\end{align}
and
\begin{align}
  \label{eq:LP:res:gg}
  \begin{split}
  &\hspace*{-5mm}
  \widetilde{\Sigma}_{\ttbar  }^{\mathrm{res},[g_{n}g_{\nbar}]}
  (\bTvec,\Yttbar, \Mttbar,\Omega_t)\\
  \,=&\;
  \left(\frac{1}{N^2_c-1}\right)^2\,
  \mathcal{D}^{\mathrm{res}}_{[g_{n}g_{\nbar}]}
  (\bT,\Mttbar,\mu_h,\mu_b,\mu_s,\nu_b,\nu_s)\,
  \sum_{\{\alpha,\beta, h,h'\}}
  \Bigg\{
      \mathcal{S}^{\alpha_1\beta_1}_{[g_ng_{\nbar}]}(\bTvec,v_{t},v_{\tbar},\mu_s,\nu_s)\,
    \\
  &\times\,
    \mathcal{B}^{[g_n]}_{n,h_n'{h_n}}(\eta_n,\bTvec,\mu_b,\nu_b)\,
    \mathcal{B}^{[g_{\nbar}]}_{\nbar,h_{\nbar}'{h_{\nbar}}}(\eta_{\nbar},\bTvec,\mu_b,\nu_b)  \,
\left[\mathcal{V}^{[g_{n}g_{\nbar}]}_{\alpha_1\alpha_2}(v_{t},v_{\tbar},\mu_s,\mu_h)\mathcal{C}^{[g_{n}g_{\nbar}]}_{\alpha_2;h'_nh'_{\nbar}h_th_{\tbar}}\right]^*\,
  \\
  &\times
\mathcal{V}^{[g_{n}g_{\nbar}]}_{\beta_1\beta_2}(v_{t},v_{\tbar},\mu_s,\mu_h)\, \mathcal{C}_{\beta_2;h_nh_{\nbar}h_th_{\tbar}}^{[g_ng_{\nbar}]}\,
  \Bigg\}\,,
  \end{split}
\end{align}
where the soft function is given by $\mathcal{S}^{\alpha\beta}_{[\kappa]}$
as a function of the impact parameter $\bTvec$, the velocity
$v_{t(\tbar)}$ of the (anti)top quark, and the soft
virtuality (rapidity) scale $\mu_s(\nu_s)$.
To facilitate our calculations, we have projected the colour
states of the soft function onto the orthonormal
bases $c_{\{a_i\}}^{qq}$ and $c_{\{a_i\}}^{gg}$ of
\cite{Beneke:2009rj}, leading to the colour indices
$\{\alpha,\beta\}$ emerging as superscripts.
\changed{Their expressions are presented in App.~\ref{app:def:col:spin}.}
It is important to note that, heretofore, while the azimuthally
averaged soft function have been calculated up to
\NNLO~\cite{Angeles-Martinez:2018mqh,Catani:2023tby},
its fully azimuthal-angle-dependent form that are essential to
compute the $\phi_{\ttbar}$ resummation are only available at
\NLO~\cite{Catani:2021cbl,Ju:2022wia}.

Furthermore, Eqs.~\eqsref{eq:LP:res:qqbar}{eq:LP:res:gg} include
the hard functions
$\mathcal{C}_{\alpha;h_nh_{\nbar}h_th_{\tbar}}^{[q^i_{n}\qbar^j_{\nbar}]}$ and
$\mathcal{C}_{\beta;h_nh_{\nbar}h_th_{\tbar}}^{[g_ng_{\nbar}]}$
which consist of the UV-renormalized and IRC-subtracted amplitudes
of the relevant hard partonic processes.
Again, the $\{\alpha,\beta\}$ encode the colour states as in the
soft function, while the tuple $\{h_{n},h_{\nbar},h_t,h_{\tbar}\}$
is introduced to specify the helicity states of the external particles.
Throughout this work, the helicity bases of
\cite{Actis:2012qn,Actis:2016mpe} are taken as our default choice
to evaluate the helicity projections
\changed{ and we present their expression in App.~\ref{app:def:col:spin}.}
In calculating
$\mathcal{C}_{\alpha;h_nh_{\nbar}h_th_{\tbar}}^{[q^i_{n}\qbar^j_{\nbar}]}$ and
$\mathcal{C}_{\beta;h_nh_{\nbar}h_th_{\tbar}}^{[g_ng_{\nbar}]}$,
the \MSbar\ scheme is utilised to renormalise the UV divergences
associated with the massless partons and the zero-momentum subtraction
prescription \cite{Beenakker:2002nc} is employed to cope with those
pertaining to the (anti)top quarks.
The remaining IRC singularities are removed following the procedures in
\cite{Ferroglia:2009ii}.
Up to \NLO, the automated program \Recola~\cite{Actis:2012qn,Actis:2016mpe}
is employed in this paper to extract the amplitudes of $\qqbar\to\ttbar$
and $gg\to\ttbar$ in all the helicity and colour configurations.
The \NNLO\ calculation are more involved.
For now, the grid-based numerical results have been presented in
\cite{Chen:2017jvi}, while the progress towards the full analytic
evaluations are made in \cite{DiVita:2019lpl,Badger:2021owl,Mandal:2022vju,Wang:2023qbf}.
  
Next, Eqs.~\eqsref{eq:LP:res:qqbar}{eq:LP:res:gg} also comprise
the beam functions $\mathcal{B}_{n(\nbar)}^{[q_{n(\nbar)}^i]}$ and
$\mathcal{B}^{[g_{n(\nbar)}]}_{n(\nbar)}$ governing the beam-collinear
contributions along the $n(\nbar)$-direction.
They are the functions of the virtuality (rapidity) scale $\mu_b(\nu_b)$
and the momentum fractions $\eta_n={\Mttbar\,e^{\Yttbar}}/{\sqrt{s}}$
and $\eta_\nbar={\Mttbar\,e^{-\Yttbar}}/{\sqrt{s}}$.
In comparison with the quark beam function
$\mathcal{B}_{n(\nbar)}^{[q_{n(\nbar)}^i ]}$, the gluon case additionally
depends on the gluon helicities $\{h_{n(\nbar)},h_{n(\nbar)}'\}\in\{+,-\}$
to accommodate the helicity-flipping and helicity-conserving contributions.
At present, the quark beam function,
$\mathcal{B}_{n(\nbar)}^{[q_{n(\nbar)}^i]}$, and the helicity-conserving
components of the gluon beam function,
$\mathcal{B}_{n(\nbar),++}^{[g_{n(\nbar)}]}$ and
$\mathcal{B}_{n(\nbar),--}^{[g_{n(\nbar)}]}$, have been calculated
up to \NNNLO\ \cite{Luo:2020epw,Luo:2019szz,Luo:2019bmw,Luo:2020epw},
while the helicity-flipping entries
$\mathcal{B}_{n(\nbar),+-}^{[g_{n(\nbar)}]}$ and
$\mathcal{B}_{n(\nbar),-+}^{[g_{n(\nbar)}]}$ are only know on the \NNLO\
level \cite{Luo:2019bmw,Gutierrez-Reyes:2019rug,Catani:2022sgr}.

Finally, in addition to the above fixed-order contributions,
Eqs.\ \eqsref{eq:LP:res:qqbar}{eq:LP:res:gg} contain  the
evolution kernels $\mathcal{D}^{\mathrm{res}}_{[\kappa]}$ and
$\mathcal{V}^{[\kappa]}_{\alpha\beta}$ as well.
They bridge the gap between the intrinsic scales in the hard,
soft, and beam-collinear contributions by resumming the occurring 
large logarithms and are derived by solving
the respective R(a)GEs of the corresponding constituents
\cite{Chiu:2011qc,Chiu:2012ir,Li:2016axz,Li:2016ctv}.
For instance, $ \mathcal{D}^{\mathrm{res}}_{[\kappa]}$ consists
of the solutions of the beam-collinear R(a)GEs and the diagonal
part of the hard RGEs, see~\cite{Ju:2022wia},
\begin{align} \label{eq:def:Dres}
  &\ln\mathcal{D}^{\mathrm{res}}_{[\kappa]}
    (\bT,\Mttbar,\mu_h,\mu_b,\mu_s,\nu_b,\nu_s)
    \nnb\\
  &\,=\;
    \int^{\mu^2_s}_{\mu^2_b} \,
    \frac{\done\bar{\mu}^2}{\bar{\mu}^2}\,
    \Bigg\{
      C_{[\kappa]}\, \Gamma_{\mathrm{cusp}}\big[\alpha_s(\bar{\mu}) \big]\,
      \ln\bigg[ \frac{\nu_b^2}{\Mttbar^2}\bigg]
      +2\,\gamma^{[\kappa]}_b\big[\alpha_s(\bar{\mu}) \big]\,
    \Bigg\}
    -\int^{\mu_s^2}_{\mu_h^2}\frac{\done \bar{\mu}^2}{\bar{\mu}^2}\,
    \bigg\{
    {C_{[\kappa]}}\,
    \Gamma_{\mathrm{cusp}}\big[\alpha_s(\bar{\mu}) \big]\,
    \ln\left[\frac{\bar{\mu}^2}{\Mttbar^2}\right]\,\bigg\}\nnb
    \\
  &\,\phantom{=}\;{}
    +C_{[\kappa]}\,
    \ln\bigg[\frac{\nu_s^2}{\nu_b^2}\bigg]\,
    \int_\frac{b_0^2}{\bT^2}^{\mu_s^2}\,
    \frac{\done\bar{\mu}^2}{\bar{\mu}^2}\,
    \Gamma_{\mathrm{cusp}}\big[\alpha_s(\bar{\mu}) \big]\,
    -C_{[\kappa]}\,
     \ln\bigg[\frac{\nu_s^2}{\nu_b^2}\bigg]\,
     \gamma_r
     \bigg[\alpha_s\left(\frac{b_0}{\bT}\right)\bigg]\,. 
\end{align}
Therein, $\Gamma_{\mathrm{cusp}}$, $\gamma^{[\kappa]}_b$, and
$\gamma_r$ denote the cusp anomalous dimension, the non-cusp anomalous
dimension associated with the virtuality divergences in the beam functions,
and the non-cusp anomalous dimension of the rapidity renormalisation.
All their expressions up to \NNNNLO\ are already available in the literature
\cite{Moch:2004pa,Henn:2019swt,vonManteuffel:2020vjv,Herzog:2018kwj}
and \cite{Li:2016axz,Luo:2019bmw,Luo:2019hmp,Li:2016ctv,
  Vladimirov:2016dll,Luo:2020epw,Ebert:2020yqt,Luo:2019szz,
  Das:2019btv,Duhr:2022cob,Duhr:2022yyp,Moult:2022xzt}, respectively.
In writing Eq.~\eqref{eq:def:Dres}, the following abbreviations are
employed for the corresponding colour factor in QCD,
\begin{align}
  \kappa\in\{g_{n}g_{\nbar}\}\,:\;
  C_{[\kappa]}=C_A\;,\qquad\qquad
  \kappa\in\{q^i_{n}\qbar^j_{\nbar},q^i_{\nbar}\qbar^j_{n}\}\,:\;
  C_{[\kappa]}=C_F\,,
\end{align} 
as well as the non-cusp anomalous dimensions,  
\begin{equation}
  \kappa\in\{g_{n}g_{\nbar}\}\,:\;
  \gamma_b^{[\kappa]}=\gamma_b^{[g]}\,,
  \qquad\qquad
  \kappa\in\{q^i_{n}\qbar^j_{\nbar},q^i_{\nbar}\qbar^j_{n}\}\,:\;
  \gamma_b^{[\kappa]}=\gamma_b^{[q]}\,.
\end{equation}
Complementarily, the $\mathcal{V}^{[\kappa]}_{\alpha\beta}$ are in
charge of the non-cusp hard anomalous dimension $\gamma^{[\kappa]}_h$
\cite{Ferroglia:2009ii,Ferroglia:2009ep}.
Up to \NLL, the $\mathcal{V}^{[\kappa]}_{\alpha\beta}$ can be derived
by solving the RGE of the hard function in the diagonal colour space
\cite{Buras:1991jm,Buchalla:1995vs,Ahrens:2010zv},
\begin{align}\label{eq:def:Vres:NLL}
  \mathbf{V}_h^{[\kappa]}(v_t,v_{\tbar},\mu_s,\mu_h)\Bigg|_{\mathrm{NLL} }
  \,=\;
  \mathbf{R}^{-1}_{[\kappa]}\,
  \exp\bigg\{
    \frac{\mathbf{r}^{[\kappa],(0)}_{h}}{2\beta_0}
    \ln\bigg[
      \frac{\alpha_s(\mu_h)}{\alpha_s(\mu_s)}
    \bigg]
  \bigg\}\,
  \mathbf{R}_{[\kappa]}\,,
\end{align}
where $\mathbf{V}_h^{[\kappa]}$ is the matrix representation of
$\mathcal{V}^{[\kappa]}_{\alpha\beta}$.
$\mathbf{r}^{[\kappa],(0)}_{h}$ stands for the diagonalised one-loop
non-cusp anomalous dimension of the hard function, by means of the
invertible transformation matrix $\mathbf{R}_{[\kappa]}$.
$\alpha_s$ denotes the strong coupling evaluated in the $N_F=5$
flavour scheme, with the according anomalous dimension $\beta_k$
at $(k+1)$-loop accuracy.

This approach can also been generalized to \NNLL\ by including the
off-diagonal entries of the two-loop hard anomalous dimensions as
appropriate \changed{\cite{Buras:1991jm,Buchalla:1995vs}}, i.e.,
\begin{align}\label{eq:def:Vres:N2LL}
  \begin{split}
  &\mathbf{V}_h^{[\kappa]}(v_t,v_{\tbar},\mu_s,\mu_h)\Bigg|_{\NNLL}
  \\
  &\,=\;
    \mathbf{R}^{-1}_{[\kappa]}\,
    \bigg[\mathbf{I}+\frac{\alpha_s(\mu_s)}{4\pi}\mathbf{J}^{[\kappa]}\bigg]\,
    \exp\bigg\{
      \frac{\mathbf{r}^{[\kappa],(0)}_{h}}{2\beta_0}
      \ln\bigg[
        \frac{\alpha_s(\mu_h)}{\alpha_s(\mu_s)}
      \bigg]
    \bigg\}\,
    \bigg[
      \mathbf{I}-\frac{\alpha_s(\mu_h)}{4\pi}\mathbf{J}^{[\kappa]}
    \bigg]\,
    \mathbf{R}_{[\kappa]}\,,
  \end{split}
\end{align}
where the matrix $\mathbf{J}^{[\kappa]}$ is introduced here to take in the two loop ingredients,
\begin{align}
\label{eq:def:Vres:N2LL:Jmax}
  \mathbf{J}^{[\kappa]}_{ij}
  \,=\;
    \mathbf{r}^{{[\kappa]},(0)}_{h,ii}\;
    \delta_{ij}\,\frac{\beta_1}{2\beta^2_0}
    -\frac{\mathbf{r}^{{[\kappa]},(1)}_{h,ij}}
          {2\beta_0+\mathbf{r}^{{[\kappa]},(0)}_{h,ii}
           -\mathbf{r}^{{[\kappa]},(0)}_{h,jj}}\,.
\end{align}
Herein, $\delta_{ij}$ represents the Kronecker delta function carrying the
indices $\{i,j\}\in\{1,2\}\;(\{1,2,3\}$) for the quark (gluon) channel.
$\mathbf{r}^{{[\kappa]},(1)}_{h}$ is defined analogously to
$\mathbf{r}^{{[\kappa]},(0)}_{h}$ in terms of the two-loop
non-cusp anomalous dimension $\gamma^{(1)}_h$ within the diagonal space
of $\gamma^{(0)}_h$.


Reinserting the results of Eqs.\ \eqsref{eq:LP:res:qqbar}{eq:LP:res:gg}
into Eq.~\eqref{eq:QCDF} and expanding the kinematic variables to
leading power, we arrive at the resummed \qT\ and \dphittbar\
spectra~\cite{Ju:2022wia}, 
\begin{align}\label{eq:methods:res:qT_dphi}
  \begin{split}
    \frac{\done{\sigma^{\mathrm{res}}_{\ttbar}}}
         {\done \qT}
    \,=\;&
      \changed{
        \frac{\qT}{64\pi^3\,s}
        \sum_{\kappa,\mathrm{sign}[\widetilde{P}^z_{t}]}
        \int\done\Mttbar^2\,\done\Yttbar\,\done^2\widetilde{P}_{t}^{\perp}\;
        \frac{\theta(M_\ttbar-\Mttbar^\text{min})}
             {\Mttbar^3\;|\widetilde{P}^z_{t}|}}
        \int{\done^2\bTvec}\,J_{0}(\bT\qT)\,
        \widetilde{\Sigma}_{\ttbar}^{\mathrm{res},[\kappa]}
          (\bTvec,\Yttbar,\Mttbar,\Omega_t)\,,\hspace*{-20mm}\\
    \frac{\done{\sigma^{\mathrm{res}}_{\ttbar}}}
         {\done \dphittbar}
    \,=\;&
      \changed{
        \frac{1}{32\pi^3\,s}
        \sum_{\kappa,\mathrm{sign}[\widetilde{P}^z_{t} ]}
        \int\done\Mttbar^2\,\done\Yttbar\,\done^2\widetilde{P}_{t}^{\perp}\;
        \frac{\theta(M_\ttbar-\Mttbar^\text{min})}{\Mttbar^3}\,
        \frac{|\widetilde{P}_{t}^{\perp}|}{|\widetilde{P}^z_{t}|}\;}\\
    &\hspace*{60mm}\times\;
        \int{\done b_{\tau}}\,
        \cos(b_{\tau}\,|\widetilde{P}_{t}^{\perp}|\,\dphittbar)\,
        \widetilde{\Sigma}_{\ttbar}^{\mathrm{res},[\kappa]}
          (\vec{b}^{\|}_{\tau},\Yttbar,\Mttbar,\Omega_t)\,,\hspace*{-20mm}
  \end{split}
\end{align}
where $\widetilde{P}^z_{t}$ and $\widetilde{P}_{t}^{\perp}$ are
the longitudinal and transverse momenta of the top quark measured
in the rest frame of the top and antitop pair.
\changed{A lower cutoff $\Mttbar^\text{min}$, which was chosen to be
$400\,\text{GeV}$ in~\cite{Ju:2022wia}, is introduced to avoid any
threshold enhanced contributions, thereby ensuring the applicability
of HQET. }
$J_0(x)$ represents the zeroth-rank Bessel function.
$\vec{b}_{\tau}$ refers to the projected component of the
impact parameter $\bTvec$,
\begin{align}
  \bTvec
  =\vec{b}_{\tau}^{\perp}+ \vec{b}_{\tau}^{\|}
  \equiv b_{\tau}^{\perp}\vec{\tau}\times\vec{n}+ b_{\tau}\vec{\tau}\,.
\end{align} 
Here $\vec{n}$ stands for a unit vector pointing to one of beam
directions in the laboratory reference frame, whilst in calculating
$\dphittbar$ distribution, $\vec{\tau}$ is always chosen to be
perpendicular to the flight direction of top quark.

Before closing this subsection, we want to discuss the choice of the
auxiliary scales in Eqs.\ \eqsref{eq:LP:res:qqbar}{eq:LP:res:gg}.
Therein, two sets of auxiliary scales $\{\mu_h,\mu_b,\mu_s\}$ and
$\{\nu_b,\nu_s\}$ are introduced during the virtuality and rapidity
renormalisation in the relevant sectors.
An appropriate choice of their values can minimise the logarithmic
dependences in the fixed-order functions, and in turn improve the
convergence of the resummation.
To this end, the following values will be taken by default
in this paper~\cite{Chiu:2012ir,Neill:2015roa,Ju:2022wia},
\begin{align}\label{eq:scale:nat:qT_dphi}
  \begin{split}
    \mathcal{Q}&=\qT\,,\quad\quad\quad\quad \mu^{\mathrm{def}}_h=\nu^{\mathrm{def}}_{b}=\Mttbar\,,\quad\quad\quad  \mu^{\mathrm{def}}_b= \mu^{\mathrm{def}}_s=\nu^{\mathrm{def}}_s= b_0/\bT\,,\\
    \mathcal{Q}&=\dphittbar\,,\quad\quad\quad \mu^{\mathrm{def}}_h=\nu^{\mathrm{def}}_{b}=\Mttbar\,,\quad\quad\quad  \mu^{\mathrm{def}}_b= \mu^{\mathrm{def}}_s=\nu^{\mathrm{def}}_s= b_0/b_{\tau}\,,
  \end{split}
\end{align}
where $b_0=2\exp(-\gamma_{\mathrm{E}})$ with $\gamma_{\mathrm{E}}$
being the Euler constant.
With the choice of Eq.~\eqref{eq:scale:nat:qT_dphi}, the
evaluation of Eq.~\eqref{eq:methods:res:qT_dphi} can encounter
the Landau singularity of the strong coupling $\alpha_s$ during
the impact parameter space integration, which we regularise using
the cutoff prescription proposed in \cite{Neill:2015roa}.

\subsection{\changed{Extending the resummation region --
                     properties and caveats}}
\label{sec:sigma:thr:hqet}

In the last subsection, we introduced the resummed \qT\ and
\dphittbar\ spectra in the domain where the top and antitop
quarks are kinematically well-separated, i.e.\
$\dEttbar\sim\mathcal{O}(m_t)$ or larger.
In this regime, thanks to HQET~\cite{Eichten:1989zv,
  Georgi:1990um,Grinstein:1990mj,Neubert:1993mb}, the
(anti)top quark field will not interact with the other particles
at leading power accuracy after applying the decoupling transformation
\cite{Bauer:2001yt,Beneke:2010da}.
In consequence, at least up to leading power, the hard, soft and
beam-collinear regions are sufficient to describe the asymptotic
behaviour of the $\qT$ and $\dphittbar$ spectra.

\changed{
We are now interested in exploring the possibility to lift
this kinematic restriction, $\theta(M_\ttbar-\Mttbar^\text{min})$,
in Eq.~\eqref{eq:methods:res:qT_dphi} and extend the single
differential observables $\done\sigma_\ttbar/\done\qT$ and
$\done\sigma_\ttbar/\done\Delta\phi_\ttbar$ over the full $\Mttbar$ range.
Such an approach is motivated by the anticipation that for
top-quark pair production at the LHC at 13\,TeV the kinematic
region $\Delta E_\ttbar\sim\mathcal{O}(m_t)$ accounts for the bulk
of the total production cross section.
Hence, the asymptotic behaviour in the $\qT\to0$ and $\dphittbar\to0$
limits of the unconstrained cross section is expected to be mostly
governed by the dynamics in HQET and SCET, in analogy to the
methodology used in \cite{Kidonakis:1997gm,Bonciani:1998vc,
  Kidonakis:2001nj,Ahrens:2010zv,Czakon:2018nun,
  Broggio:2015lya,Broggio:2016lfj,Broggio:2017kzi,
  Kulesza:2017ukk,Kulesza:2017jqv,Kulesza:2018tqz,
  vanBeekveld:2022hty}.
}

\changed{
Nonetheless, we will discuss in the following the implications of
removing the phase space restriction $\Mttbar^\text{min}$ in Eq.~\eqref{eq:methods:res:qT_dphi}.
In particular, we will show that the integrand
$\widetilde{\Sigma}_{\ttbar}^{\mathrm{res},[\kappa]}$, in
particular the amplitudes $\mathcal{C}^{[\kappa]}_{\alpha;\{h_i\}}$
of the hard sector as well as the evolution kernels
$\mathbf{V}_{h}^{[\kappa]}$ it comprises, develops
power-like divergences $\sim(\alpha_s/\betattbar)^n$ as the
Coulomb interactions manifest themselves in the threshold limit
as $\dEttbar\to0$, or more conventionally
}
\begin{equation}\label{eq:defbeta}
  \betattbar\equiv\sqrt{1-\frac{4m_t^2}{\Mttbar^2}}\to0\,.
\end{equation}

\subsection*{Beam function and the evolution kernel
             \texorpdfstring{$\mathcal{D}^{\mathrm{res}}_{[\kappa]}$}
             {Dres}}

As illustrated in Eqs.~\eqsref{eq:LP:res:qqbar}{eq:LP:res:gg},
$\widetilde{\Sigma}_{\ttbar}^{\mathrm{res},[\kappa]}$ contains
the fixed-order contribution functions
$\mathcal{B}_{n(\nbar)}^{[\kappa]}$,
$\mathcal{S}^{\alpha\beta}_{[\kappa]}$, and
$\mathcal{C}^{[\kappa]}_{\alpha,\{h\}}$ as well as the evolution  kernels
$\mathcal{D}^{\mathrm{res}}_{[\kappa]}$ and
$\mathcal{V}^{[\kappa]}_{\alpha\beta}$.
\changed{
Here, we start with an analysis of the threshold limit of the
beam sector $\mathcal{B}_{n(\nbar)}^{[\kappa]}$
and the evolution kernel $\mathcal{D}^{\mathrm{res}}_{[\kappa]}$.
Given the fact that both are functions of $\Mttbar$, $\Yttbar$,
and the magnitude of impact parameters $\bT$ and $b_{\tau}$, 
taking the threshold limit is straightforward and does not incur
any singular behaviour at any perturbative order.%
}
It follows that,
\begin{align}
\label{eq:bf:asy:thr}
 \mathcal{B}_{n(\nbar)}^{[\kappa]} & \xrightarrow[]{\betattbar\to0}
 \sum_{m=0}^{\infty}\left(\frac{\alpha_s(\mu_b)}{4\pi}\right)^m
 \underbrace{\mathcal{B}_{n(\nbar),\mathrm{thr}}^{[\kappa],{(m)}}}_{\mathcal{O}(\betattbar^0)}
 +\,\mathcal{O}(\betattbar)\,,\\
 \label{eq:Dexp:asy:thr}
 \mathcal{D}^{\mathrm{res}}_{[\kappa]}&  \xrightarrow[]{\betattbar\to0}
 \underbrace{\mathcal{D}^{\mathrm{res}}_{\mathrm{thr},[\kappa]}}_{\mathcal{O}(\betattbar^0)}
 +\,\mathcal{O}(\betattbar)\,.
\end{align} 
Herein, to facilitate the later discussion, the functions
$\mathcal{B}_{n(\nbar),\mathrm{thr}}^{[\kappa],{(m)}} $ and
$\mathcal{D}^{\mathrm{res}}_{\mathrm{thr},[\kappa]}$, that
represent leading contributions of the beam-collinear sector
and the cusp evolution kernel in the vicinity of $\Mttbar=2m_t$,
respectively, are introduced, with the corresponding scalings
indicated in the underbraces.

\subsection*{Hard function}

Approaching the limit $\betattbar\to0$ can induce
a distinct asymptotic behaviour in the hard function
$\mathcal{C}^{[\kappa]}_{\alpha,\{h\}}$.
Within the context of the expansion by regions
\cite{Beneke:1997zp,Smirnov:2002pj,Smirnov:2012gma,Jantzen:2011nz},
we can perform the asymptotic expansion of
$\mathcal{C}^{[\kappa]}_{\alpha,\{h\}}$ in $\betattbar$ via a
set of dynamic regions in the loop integrals, which in general
includes the hard, collinear, soft, ultrasoft, and Coulomb
regions~\cite{Beneke:2010da}.
In the following, we will use the soft-collinear effective field
theory (SCET) \cite{Bauer:2000ew,Bauer:2000yr,Bauer:2001yt,
  Beneke:2002ph,Beneke:2002ni} and potential non-relativistic
QCD (pNRQCD) \cite{Pineda:1997bj,Brambilla:1999xf,Beneke:1999zr,
  Beneke:1999qg} frameworks to capture their contributions.

At leading power, the SCET and pNRQCD effective Lagrangians
can be expressed as \cite{Beneke:2002ph,Beneke:2002ni,Beneke:1999zr,
  Beneke:1999qg,Kniehl:2002br}
\begin{align}
  \label{eq:L0SCET}
  \mathcal{L}_{\text{SCET}}
  &=
    \bar{\varphi}_n
    \Big(
      i n \cdot D_n
      + i \slashed{D}_{n\perp} \frac{1}{i \nbar \cdot D_n} i \slashed{D}_{n\perp}
    \Big)
    \frac{\slashed{\nbar}}{2} \varphi_n
    -\frac{1}{2} \mathrm{Tr} \Big\{ F^{\mu\nu}_n F_{\mu\nu}^n \Big\}
    + ( n \leftrightarrow \nbar )
    - \frac{1}{2} \mathrm{Tr} \Big\{F^{\mu\nu}_{\text{us}} F_{\mu\nu}^{\text{us}}\Big\}\,,
  \\
  \label{eq:L0pNRQCD}
  \mathcal{L}_{\text{pNR}}
  &=
    \psi^{\dag} \left( i \partial^0 +\frac{\vec{\partial}^2}{2m_t} \right) \psi
    + \chi^{\dag} \left( i \partial^0 - \frac{\vec{\partial}^2 }{2m_t} \right) \chi
    - \int \mathrm{d}^3\vec{r} \, \psi^{\dag} T^{a}
      \psi \big( x^0,\vec{x}+\vec{r} \big)
      \left( \frac{\alpha_s}{r} \right)
      \chi^{\dag} T^{a}\chi \big(x^0,\vec{x}\big) \,,
\end{align}
where $\varphi_{n}$ denotes the collinear quark field, while
$F_{n}^{\mu\nu}$ is the collinear gluon field strength tensor.
Likewise, $F_{\text{us}}^{\mu\nu}$ represents the field strength
tensor for the ultrasoft gluons.
$\psi^{\dagger}(\chi)$ stands for the \changed{heavy quark} field creating
the (anti)top quark.
The $T^a$ are the usual generators of QCD.
In writing Eqs.~\eqref{eq:L0SCET} and~\eqref{eq:L0pNRQCD}, the
decoupling transformation~\cite{Beneke:2010da} has been carried
out on the collinear and heavy quark fields so as to remove all
the ultrasoft-collinear and ultrasoft-heavy-quark interactions at
leading power, respectively.

We are now ready to appraise the leading contribution of
$\mathcal{C}^{[\kappa]}_{\alpha,\{h\}}$ at each perturbative order.
On the tree level, the leading terms of
$\mathcal{C}^{[\kappa]}_{\alpha,\{h\}}$ are determined by the
effective Hamiltonian constructed out of the SCET and pNRQCD fields
above.
To evaluate the amplitudes induced by this Hamiltonian, we match
the QCD amplitudes evaluated at the threshold $\Mttbar=2m_t$ onto
the effective field theories.
During the calculation, we  make use of the \texttt{Mathematica}
packages \texttt{FeynArts} \cite{Hahn:2000kx},
\texttt{FeynCalc} \cite{Mertig:1990an,Shtabovenko:2016sxi,
  Shtabovenko:2020gxv},
and \texttt{FeynHelpers} \cite{Shtabovenko:2016whf} to generate
the amplitudes for the individual partonic channels and then employ
\texttt{FeynOnium} \cite{Brambilla:2020fla} to recast the Dirac
spinors of the heavy quarks in terms of Pauli spinors.
It follows that,
\begin{align}
  \mathcal{C}^{[\kappa]}_{\alpha,\{h\}}\xrightarrow[]{\betattbar\to0}
  \sum^{\infty}_{n=0}\,\left(\frac{\alpha_s}{4\pi}\right)^{n+1}\mathcal{C}^{[\kappa],(n)}_{\mathrm{thr},\alpha,\{h\}}+\dots\,,
\end{align}
where $\mathcal{C}^{[\kappa],(n)}_{\mathrm{thr},\alpha,\{h\}}$
characterises the leading contribution in the threshold domain
$\betattbar\to0$ at the $n$-th order.
The LO results read,
\begin{align}
  \label{eq:def:Chard:thr:qqb_qbq_gg:LO}
  \begin{split}
  \mathcal{C}^{[q_{n}\qbar_{\nbar}],(0)}_{\mathrm{thr},\{h\}}&=
\begin{bmatrix}
 0& \frac{4\mathrm{i}\sqrt{2}\pi^2}{m^2_t}\left(\xi_{t}^{\dagger}\vec{\sigma}\eta_{\tbar}\right)\!\cdot\!\left(\bar{v}_{\nbar}\vec{\gamma}_{\perp}u_{n}\right)\, \\
\end{bmatrix}
^{\mathbf{T}}\,,
\\ 
\mathcal{C}^{[q_{\nbar}\qbar_{n}],(0)}_{\mathrm{thr},\{h\}}&= 
 \begin{bmatrix}
 0& \frac{4\mathrm{i}\sqrt{2}\pi^2}{m^2_t}\left(\xi_{t}^{\dagger}\vec{\sigma}\eta_{\tbar}\right)\!\cdot\!\left(\bar{v}_{{n}}\vec{\gamma}_{\perp}u_{\nbar}\right)\, \\
\end{bmatrix}
^{\mathbf{T}}\,,
  \\ 
  \mathcal{C}^{[g_{n}g_{\nbar}],(0)}_{\mathrm{thr},\{h\}}&= 
 \begin{bmatrix}
 -\frac{8\pi^2}{m_t}\sqrt{\frac{2}{3}}\,\xi_{t}^{\dagger}\eta_{\tbar}\,\varepsilon_{\perp}^{\epsilon_n\epsilon_{\nbar}}&0&-\frac{8\pi^2}{m_t}\sqrt{\frac{5}{3}}\,\xi_{t}^{\dagger}\eta_{\tbar}\,\varepsilon_{\perp}^{\epsilon_n\epsilon_{\nbar}}  \\
\end{bmatrix}
^{\mathbf{T}}\,.
  \end{split}
\end{align}
Here, $\xi^{\changed{\dagger}}_{t}$ and $\eta_{\tbar}$ denote the Pauli spinors for
the top and antitop quarks, respectively,
\changed{resulting from the quantum field operators $\psi^{\dagger}$ and
$\chi$ in Eq.~\eqref{eq:L0pNRQCD} acting on the external states
$\langle\ttbar|$, with $\vec{\sigma}$ being a spatial vector consisting
of the Pauli matrices.}
Similarly, $u_{n(\nbar)}$ and $v_{n(\nbar)}$ denote the Dirac
spinors of the incoming massless quark and antiquarks
\changed{induced by the collinear field operator $\varphi_n$ in
Eq.~\eqref{eq:L0SCET}}, while
$\vec{\gamma}_\perp$ is the transverse component of the Dirac matrices.
The contraction of the totally antisymmetric tensor and the
polarisation vectors is abbreviated to
$\varepsilon_{\perp}^{\epsilon_n\epsilon_{\nbar}}\equiv
 \varepsilon^{\mu\nu\rho\sigma}n_{\mu}\nbar_{\nu}
 \epsilon_{n,\rho}\epsilon_{\nbar,\sigma}$.
\changed{
In writing Eq.~\eqref{eq:def:Chard:thr:qqb_qbq_gg:LO}, the bases
$c_{\{a_i\}}^{qq}$ and $c_{\{a_i\}}^{gg}$ in
Eq.~\eqref{eq:def:color:basis:qq_gg} are employed
to project out the colour states of the hard amplitudes.
As shown in Eq.~\eqref{eq:def:Chard:thr:qqb_qbq_gg:LO},
$\mathcal{C}^{[q_{n}\qbar_{\nbar}],(0)}_{\mathrm{thr},\{h\}}$
and $\mathcal{C}^{[q_{\nbar}\qbar_{n}],(0)}_{\mathrm{thr},\{h\}}$
only include the colour-octet contributions as the LO partonic
process $q\bar{q}\to\ttbar$ only contains colour-octet $s$-channel
diagrams.
This differs for $gg\to\ttbar$, where the $s$-channel and $u$-channel
diagrams both contribute, leading to the presence of both colour-singlet
and color-octet configurations in
$\mathcal{C}^{[g_{n}g_{\nbar}],(0)}_{\mathrm{thr},\{h\}}$.
}

The leading contribution of $\mathcal{C}^{[\kappa]}_{\alpha,\{h\}}$
on the one-loop level is calculated with the amplitudes induced by the
time product of the Coulomb vertex in Eq.~\eqref{eq:L0pNRQCD} and the
tree-level Hamiltonian.
To evaluate the ensuing loop integral, following the method in
\cite{Smirnov:2002pj}, the residue theorem is first applied to
integrate out the temporal component of the loop momentum, and the
integration of the remaining spatial components can be completed via
Feynman parameterisation.
After removing the IRC poles within the \MSbar scheme
\cite{Ferroglia:2009ii}, it yields,
\begin{align}
  \label{eq:def:Chard:thr:qqb_qbq_gg:NLO}
  \begin{split}
\mathcal{C}^{[q_{n}\qbar_{\nbar}],(1)}_{\mathrm{thr},\{h\}}&= 
 \begin{bmatrix}
0
&
 \frac{4\mathrm{i}\sqrt{2}\pi^2}{m^2_t}\left(\xi_{t}^{\dagger}\vec{\sigma}\eta_{\tbar}\right)\!\cdot\!\left(\bar{v}_{\nbar}\vec{\gamma}_{\perp}u_{n}\right) \left( -\frac{\pi ^2}{6 \betattbar}+\frac{i \pi \Lttbar  }{3 \betattbar}\right)  \\
\end{bmatrix}
^{\mathbf{T}}\,,
\\ 
\mathcal{C}^{[q_{\nbar}\qbar_{n}],(1)}_{\mathrm{thr},\{h\}}&= 
 \begin{bmatrix} 
0
&
 \frac{4\mathrm{i}\sqrt{2}\pi^2}{m^2_t}\left(\xi_{t}^{\dagger}\vec{\sigma}\eta_{\tbar}\right)\!\cdot\!\left(\bar{v}_{{n}}\vec{\gamma}_{\perp}u_{\nbar}\right)\left( -\frac{\pi ^2}{6 \betattbar}+\frac{i \pi \Lttbar  }{3 \betattbar}\right)  \\
\end{bmatrix}
^{\mathbf{T}}\,,
\\ 
\mathcal{C}^{[g_{n}g_{\nbar}],(1)}_{\mathrm{thr},\{h\}}&= 
 \begin{bmatrix}
-\frac{8\pi^2}{m_t}\sqrt{\frac{2}{3}}\,\xi_{t}^{\dagger}\eta_{\tbar}\,\varepsilon_{\perp}^{\epsilon_n\epsilon_{\nbar}}\left(\frac{4 \pi ^2}{3 \betattbar}-\frac{8 i \pi \Lttbar  }{3 \betattbar} \right)
&
0
&
-\frac{8\pi^2}{m_t}\sqrt{\frac{5}{3}}\,\xi_{t}^{\dagger}\eta_{\tbar}\,\varepsilon_{\perp}^{\epsilon_n\epsilon_{\nbar}}
\left( -\frac{\pi ^2}{6 \betattbar}+\frac{i \pi \Lttbar  }{3 \betattbar}\right) \\
\end{bmatrix}
^{\mathbf{T}}\,,
  \end{split}
\end{align}
where $\Lttbar\equiv\ln\left(\frac{\mu}{2\betattbar m_t}\right)$%
\changed{, using an analogous notation to
Eq.~\eqref{eq:def:Chard:thr:qqb_qbq_gg:LO}.}
We have verified that the logarithmic dependences in
Eq.~\eqref{eq:def:Chard:thr:qqb_qbq_gg:NLO} indeed satisfy
the RGE suggested in \cite{Becher:2009cu,Ferroglia:2009ii} up
to the power corrections \changed{of} $\mathcal{O}(\betattbar^0)$ and also that
the non-logarithmic terms of Eq.\ \eqref{eq:def:Chard:thr:qqb_qbq_gg:NLO}
reproduce the NLO correction of the imaginary part of the pNRQCD
Green function \cite{Beneke:1999qg,Beneke:1999zr,Pineda:2006ri,
  Beneke:2011mq}.
\changed{
Comparing with the LO results of
Eq.~\eqref{eq:def:Chard:thr:qqb_qbq_gg:LO}, we observe that the
leading-power one-loop corrections of
Eq.~\eqref{eq:def:Chard:thr:qqb_qbq_gg:NLO}
contain the same colour configurations only.
This echoes the colour conservation in the leading-power
Coulomb-gluon exchanges but might be broken when adding
ultrasoft radiation at subleading power.
}

At last, it is worth noting that aside from the Coulomb exchanges,
it is also possible to consider the collinear and hard contribution
to the one-loop amplitude $\mathcal{C}^{[\kappa]}_{\alpha,\{h\}}$.
However, while the hard loop momenta can not generate any threshold
enhanced contributions, according to Eq.~\eqref{eq:L0SCET}, the
internal collinear propagators can only result in scaleless and thus
vanishing loop integrals for on-shell amplitudes.
Therefore, in deriving Eq.~\eqref{eq:def:Chard:thr:qqb_qbq_gg:NLO},
we are only concerned with the contributions induced by the Coulomb
potential.

From Eq.~\eqref{eq:def:Chard:thr:qqb_qbq_gg:LO} and
Eq.~\eqref{eq:def:Chard:thr:qqb_qbq_gg:NLO}, we can determine the
asymptotic behaviour of $\mathcal{C}^{[\kappa]}_{\alpha,\{h\}}$ in the
threshold regime,
\begin{align} \label{eq:chard:asy:thr:qq:gg}
 \mathcal{C}^{[\kappa]}_{\alpha,\{h\}}\xrightarrow[]{\betattbar\to0}  \,\left(\frac{\alpha_s}{4\pi}\right)\Bigg\{\underbrace{\mathcal{C}^{[\kappa],(0)}_{\mathrm{thr},\alpha,\{h\}}}_{\sim\mathcal{O}(\betattbar^0)}+ \mathcal{O}(\betattbar)\Bigg\}+\left(\frac{\alpha_s}{4\pi}\right)^2\Bigg\{\underbrace{\mathcal{C}^{[\kappa],(1)}_{\mathrm{thr},\alpha,\{h\}}}_{\sim\mathcal{O}(\beta^{-1}_{\ttbar})}+ \mathcal{O}(\betattbar^0)\Bigg\}\dots\,.
\end{align}
Here we only present the results up to the one-loop level, which is
sufficient for us to analyse the \NNLL\ resummation in
$\widetilde{\Sigma}_{\ttbar}^{\mathrm{res},[\kappa]}$.
The asymptotic expansion of $\mathcal{C}^{[\kappa]}_{\alpha,\{h\}}$
at the two loop accuracy and beyond can be carried out in an
analogous manner, even including higher power correction in $\betattbar$.
Further discussion can be found in \cite{Beneke:2009ye,Barnreuther:2013qvf}.

\subsection*{Soft function}

We now move onto the investigation of the behaviour of the soft
function $\mathcal{S}^{\alpha\beta}_{[\kappa]}$ in the limit
$\betattbar\to0$.
In principle, the threshold limit of the HQET-based soft
function could be extracted by comparison with the soft
function in pNRQCD.
However, due to the fact that HQET and pNRQCD follow a different
sequence in performing the UV renormalisation and the asymptotic
expansion---the threshold expansion of the soft function in HQET
prioritises the UV renormalisation, whilst the soft sector in
pNRQCD is derived by the $\betattbar$ expansion in the first
place---this kind of comparison has to be delivered on the
differential cross section level, rather than mapping the soft
sectors between the two directly.
One example to demonstrate the non-commutativity can be found in
the inclusive soft functions~\cite{Czakon:2013hxa,Wang:2018vgu}
for the threshold resummation.

With this in mind, we will directly expand the analytic results
for $\mathcal{S}^{\alpha\beta}_{[\kappa]}$ in the limit $\betattbar\to0$.
Remaining at the \NNLL\ level in
$\widetilde{\Sigma}_{\ttbar}^{\mathrm{res},[\kappa]}$, using
Eqs.~(\ref{eq:LP:res:qqbar}-\ref{eq:LP:res:gg}), we only require
the soft contribution up to the one-loop level, for which the
analytic expression have been derived in \cite{Ju:2022wia}
with the help of a Mellin-Barnes transformation
\cite{Smirnov:1999gc,Tausk:1999vh}.
Expanding those renormalised results in the small parameter
$\betattbar$, it yields that
\begin{align} \label{eq:sf:asy:thr}
  \mathcal{S}^{\alpha\beta}_{[\kappa]}  \xrightarrow[]{\betattbar\to0} \delta^{\alpha\beta}+\left(\frac{\alpha_s}{4\pi}\right)\, \left[\mathcal{S}^{(1),\alpha\beta}_{\mathrm{thr},[\kappa]}+\mathcal{O}(\betattbar)\right]+\mathcal{O}(\alpha^2_s)\dots\,,
\end{align} 
where 
\begin{align}  
  \begin{split}
 \mathcal{S}^{(1),\alpha\beta}_{\mathrm{thr},[q_n\qbar_{\nbar}]}&= \mathcal{S}^{(1),\alpha\beta}_{\mathrm{thr},[q_{\nbar}\qbar_n]}=
 \begin{bmatrix}
\frac{16 L_{\nu } L_{\mathrm{T}}}{3}-\frac{8 L_{\mathrm{T}}^2}{3}-\frac{4 \pi ^2}{9} & 0 \\
0& \frac{16 L_{\nu } L_{\mathrm{T}}}{3}-\frac{8 L_{\mathrm{T}}^2}{3}+6 L_{\mathrm{T}}-\frac{4 \pi ^2}{9}
\end{bmatrix}\,,
\\
\mathcal{S}^{(1),\alpha\beta}_{\mathrm{thr},[gg]}&=
 \begin{bmatrix}
12 L_{\nu } L_{\mathrm{T}}-6 L_{\mathrm{T}}^2-\pi ^2 & 0 &0 \\
0& 12 L_{\nu } L_{\mathrm{T}}-6 L_{\mathrm{T}}^2+6 L_{\mathrm{T}}-\pi ^2 &0\\
0& 0 &12 L_{\nu } L_{\mathrm{T}}-6 L_{\mathrm{T}}^2+6 L_{\mathrm{T}}-\pi ^2\\
\end{bmatrix}\,.
  \end{split}
\end{align}
Herein, we use the notations $L_{\nu}=\ln[\mu^2/\nu^2]$,
$L_{\mathrm{T}}=\ln[\bT^2\mu^2/b_0^2]$, and
$b_0=2\exp(-\gamma_{\mathrm{E}})$ with $\gamma_{\mathrm{E}}$
being again the Euler constant.
From the results above, it is seen that no threshold enhanced
behaviour emerges from the NLO soft function.
We can therefore establish,
\begin{align}  \label{eq:soft:asy:thr:qq:gg}
\mathcal{S}^{(1),\alpha\beta}_{\mathrm{thr},[\kappa]}\sim\mathcal{O}(1)\,.
\end{align}

\subsection*{Evolution kernel $\mathcal{V}^{[\kappa]}_{\alpha\beta}$}
  
Finally, we investigate the behaviour of non-cusp resummation
kernel $\mathcal{V}^{[\kappa]}_{\alpha\beta}$ in the vicinity of
the threshold.
According to the definitions in Eqs.~\eqref{eq:def:Vres:NLL} and
\eqref{eq:def:Vres:N2LL}, $\mathcal{V}^{[\kappa]}_{\alpha\beta}$
comprises the exponential of the $\mathbf{r}^{{[\kappa]},(0)}_{h}$
matrices up to \NLL\ accuracy.
Starting at \NNLL, however, they are supplemented with additional
perturbative correction matrices, $\mathbf{J}^{[\kappa]}$, to
accommodate the two-loop non-cusp anomalous dimension
\cite{Ferroglia:2009ii,Ferroglia:2009ep}.
Hence, the analysis of the threshold behaviour of
$\mathcal{V}^{[\kappa]}_{\alpha\beta}$ reduces to the expansion of
$\mathbf{r}^{{[\kappa]},(0)}_{h}$, $\mathbf{J}^{[\kappa]}$, and the
transformation matrices $\mathbf{R}_{[\kappa]}$ in $\betattbar$.
 
The $\mathbf{r}^{{[\kappa]},(0)}_{h}$ matrices can be constructed
from the eigenvalues of the one-loop non-cusp anomalous dimensions
$\gamma_h^{[\kappa],(0)}$~\cite{Ferroglia:2009ii,Ferroglia:2009ep},
for which we solve the characteristic equations for the contributing
partonic processes using \texttt{Mathematica}.
Expanding in $\betattbar$, the leading and subleading
power contributions read,
\begin{align}
  \label{eq:rh:thr:exp}
\mathbf{r}^{{[\kappa]},(0)}_{h} \xrightarrow[]{\betattbar\to0}\mathbf{r}^{{[\kappa]},(0)}_{h,\mathrm{thr}}+\mathcal{O}(\betattbar)\,,
\end{align}
where
\begin{align}
  \label{eq:rh:thr:exp:qq_gg}
  \begin{split}
\mathbf{r}^{{[q_n\qbar_{\nbar},\qbar_n{q}_{\nbar}]},(0)}_{h,\mathrm{thr}}=&
 \begin{bmatrix} 
-8-\frac{8 i \pi }{3 \betattbar}&0\\
0&\frac{i \pi }{3 \betattbar}+6 i \pi -14\,
\end{bmatrix}  \,,\\ 
\mathbf{r}^{{[g_ng_{\nbar}]},(0)}_{h,\mathrm{thr}}=&
 \begin{bmatrix} 
-\frac{46}{3}-\frac{8 i \pi }{3 \betattbar}&0&0\\
0&-\frac{64}{3}+\frac{i\pi }{3 \betattbar}+6i \pi&0\\
0&0&-\frac{64}{3}+\frac{i\pi }{3 \betattbar}+6i \pi\\
\end{bmatrix}  \,.
  \end{split}
\end{align}
Here, all terms suppressed by positive powers of $\betattbar$
are omitted as they are not related to the leading behaviour of
the exponential function of Eq.~\eqref{eq:def:Vres:NLL} in
the limit $\betattbar\to0$.
Of the remaining expression, the threshold-enhanced imaginary parts echo
the $\Lttbar$-dependences in Eq.~\eqref{eq:def:Chard:thr:qqb_qbq_gg:NLO},
driven by Coulomb vertex in Eq.~\eqref{eq:L0pNRQCD}.

To derive the diagonalisation matrix $\mathbf{R}_{[\kappa]}$,
we solve for the eigenvectors of $\gamma_h^{[\kappa],(0)}$ with
the diagonal entries of $\mathbf{r}^{{[\kappa]},(0)}_{h}$ and then
fill the columns of $\mathbf{R}_{[\kappa]}$ with the resulting
eigenvectors in line with the positions of their eigenvalues.
There is, however, some arbitrariness involved in the solutions for
the eigenvectors themselves.
In this work, we require the eigenvectors constructing
$\mathbf{R}_{[\kappa]}$ to, at most, be of $\mathcal{O}(\betattbar^0)$
in the threshold domain.
Alternative choices of eigenvectors will lead to distinct expressions
of $\mathbf{R}_{[\kappa]}$ as well as $\mathbf{J}^{[\kappa]}$, but
do not alter the resulting $\mathcal{V}^{[\kappa]}_{\alpha\beta}$.
To confirm this, we have compared the non-cusp kernel
$\mathcal{V}^{[\kappa]}_{\alpha\beta}$ evaluated by our
$\mathbf{R}_{[\kappa]}$ and its inverse matrix with those generated
by the program~\texttt{Diag}~\cite{Hahn:2006hr} and the built-in
functions in \texttt{Mathematica}, finding numerical agreements in
all three partonic channels at both \NLL\ and \NNLL\ accuracy.
After carrying out the expansion in $\betattbar$, the leading terms
from $\mathbf{R}_{[\kappa]}$ read,
\begin{align}
\mathbf{R}_{[\kappa]} \xrightarrow[]{\betattbar\to0}
\mathbf{R}^{\mathrm{thr}}_{[\kappa]}+\mathcal{O}(\betattbar)\,,
\end{align}
where
\begin{align}
  \begin{split}
\mathbf{R}^{\mathrm{thr}}_{{[q_n\qbar_{\nbar},\qbar_n{q}_{\nbar}]}}
=
&
 \begin{bmatrix} 
1&0\\
0&1\,
\end{bmatrix}  \,,\\
\mathbf{R}_{{[g_ng_{\nbar}]}}^{\mathrm{thr}}=&
 \begin{bmatrix} 
1&0&0\\
0&1&\mathrm{sign}[\cos(\theta_t)]\\
0&-\mathrm{sign}[\cos(\theta_t)]&1\\
\end{bmatrix}  \,.
  \end{split}
\end{align}
Herein, the transformation matrices take diagonal form for the
$\kappa=q_n\qbar_{\nbar}$ and $\kappa=\qbar_n{q}_{\nbar}$ channels
in the threshold limit, while
$\mathbf{R}_{{[g_ng_{\nbar}]}}^{\mathrm{thr}}$ comprises additional
off-diagonal entries $\pm\mathrm{sign}[\cos(\theta_t)]$ in the
colour-octet blocks.
The reason for this phenomenon is that in the quark-antiquark
initiated process, the eigenvalues for the one-loop anomalous
dimensions differ from each other by $\mathcal{O}(\betattbar^{-1})$,
but as for the $\kappa=g_ng_{\nbar}$ case, the eigenvalues
accounting for colour-octet projections overlap with each other
until $\mathcal{O}(\betattbar)$, which, in solving for their
eigenvectors, can bring in additional contributions from the
colour-octet blocks and in turn result in the appearances of
$\pm\mathrm{sign}[\cos(\theta_t)]$ in
$\mathbf{R}_{{[g_ng_{\nbar}]}}^{\mathrm{thr}}$.
When applying $\mathbf{R}_{{[g_ng_{\nbar}]}}^{\mathrm{thr}}$
onto the diagonalisation, one encounters a change in sign when
the scattering angle $\theta_t$ crosses $\pi/2$.
This is caused by the small-$\betattbar$ expansion of the square
root operation in the eigenvalues and is associated with the
branch cuts therein.

Equipped with the above transformation matrices and the two-loop
anomalous dimensions~\cite{Ferroglia:2009ii,Ferroglia:2009ep},
we are now able to evaluate and expand the matrix
$\mathbf{J}^{[\kappa]}$ via Eq.~\eqref{eq:def:Vres:N2LL:Jmax},
\begin{align}
\mathbf{J}^{[\kappa]} \xrightarrow[]{\betattbar\to0}
\mathbf{J}^{[\kappa]}_{\mathrm{thr}}+\mathcal{O}(\betattbar^0)\,,
\end{align}
where
\begin{align}
  \label{eq:Jh:thr:exp:qq_gg}
  \begin{split}
\mathbf{J}_{\mathrm{thr}}^{{[q_n\qbar_{\nbar},\qbar_n{q}_{\nbar}]}}
=
&
 \begin{bmatrix} 
-\frac{220 i \pi }{4761 \beta_\ttbar}&0\\
0&\frac{55 i \pi }{9522\beta_\ttbar}\,
\end{bmatrix}  \,,\\ 
\mathbf{J}_{\mathrm{thr}}^{{[g_ng_{\nbar}]}}=&
 \begin{bmatrix} 
-\frac{220 i \pi }{4761 \beta_\ttbar}&0&0\\
0&\frac{55 i \pi }{9522\beta_\ttbar}&0\\
0&0&\frac{55 i \pi }{9522\beta_\ttbar}\\
\end{bmatrix}  \,.
  \end{split}
\end{align}
Akin to Eq.~\eqref{eq:rh:thr:exp:qq_gg}, the expressions for
$\mathbf{J}_{\mathrm{thr}}^{[\kappa]}$ contain the power-like
divergence in the imaginary parts.
Here, we only need to retain the leading singular terms.

Substituting the expressions of Eq.~\eqref{eq:rh:thr:exp:qq_gg}
into Eqs.~(\ref{eq:def:Vres:NLL}-\ref{eq:def:Vres:N2LL}), we
arrive at the leading behaviour of the evolution kernel
$\mathbf{V}_h^{[\kappa]}$ in the threshold domain,
\begin{align}
&   \mathbf{V}_h^{[\kappa]} \Bigg|_{\mathrm{NLL}(')}
  \,   \xrightarrow[]{\betattbar\to0}\;
\mathbf{V}_{h,\mathrm{thr}}^{[\kappa],(0)} +\mathcal{O}(\betattbar^0)  \,,
  \\ 
  & \mathbf{V}_h^{[\kappa]} \Bigg|_{\mathrm{N}^2\mathrm{LL}(')}
  \, \xrightarrow[]{\betattbar\to0}\;   \mathbf{V}_{h,\mathrm{thr}}^{[\kappa],(1)}+\mathcal{O}(\beta^{-1}_{\ttbar})\,,
\end{align}
where 
\begin{align}
  \label{eq:def:Vres:NLL:thr:qq_gg}
  \begin{split}
&\mathbf{V}_{h,\mathrm{thr}}^{[q_n\qbar_{\nbar},\qbar_n{q}_{\nbar}],(0)}
=
 \begin{bmatrix} 
\displaystyle\left[\frac{\alpha_s(\mu_h)}{\alpha_s(\mu_s)}
\right]^{-\frac{12}{23}-\frac{4 \mathrm{i} \pi }{23  \betattbar}}\,
&0\\
0&\displaystyle\left[
\frac{\alpha_s(\mu_h)}{\alpha_s(\mu_s)}
\right]^{\frac{\mathrm{i} \pi }{46 \betattbar}+\frac{9 \mathrm{i} \pi }{23}-\frac{21}{23}}
\end{bmatrix}  \,,
  \\ 
&\mathbf{V}_{h,\mathrm{thr}}^{[g_ng_{\nbar}],(0)}
=
 \begin{bmatrix}
\displaystyle\left[\frac{\alpha_s(\mu_h)}{\alpha_s(\mu_s)}
\right]^{-1-\frac{4\mathrm{i} \pi }{23 \betattbar}  }\,
&0&0\\
0&\displaystyle\left[
\frac{\alpha_s(\mu_h)}{\alpha_s(\mu_s)}
\right]^{\frac{\mathrm{ i} \pi }{46 \betattbar}+\frac{9 \mathrm{ i} \pi }{23}-\frac{32}{23}}&0\\
0&0&\displaystyle\left[
\frac{\alpha_s(\mu_h)}{\alpha_s(\mu_s)}
\right]^{\frac{\mathrm{ i} \pi }{46 \betattbar}+\frac{9 \mathrm{ i} \pi }{23}-\frac{32}{23}}\\
\end{bmatrix}  \,,
  \end{split}
\end{align}
and
\begin{align}
  \label{eq:def:Vres:N2LL:thr:qq_gg}
  \begin{split}
&\mathbf{V}_{h,\mathrm{thr}}^{[q_n\qbar_{\nbar},\qbar_n{q}_{\nbar}],(1)}
=
\frac{\alpha_s(\mu_s)\alpha_s(\mu_h)}{\beta^2_{\ttbar}}\,\\
&
 \begin{bmatrix}
\displaystyle
 \frac{3025}{22667121 } \,
 \left[\frac{\alpha_s(\mu_h)}{\alpha_s(\mu_s)}
\right]^{-\frac{12}{23}-\frac{4 \mathrm{i} \pi }{23  \betattbar}}\,
&0\\
0&
\displaystyle
\frac{3025}{1450695744 } \,
\left[
\frac{\alpha_s(\mu_h)}{\alpha_s(\mu_s)}
\right]^{\frac{\mathrm{i} \pi }{46 \betattbar}+\frac{9 \mathrm{i} \pi }{23}-\frac{21}{23}}
\end{bmatrix}\,,
  \\ 
&\mathbf{V}_{h,\mathrm{thr}}^{[g_ng_{\nbar}],(1)}
=
\frac{\alpha_s(\mu_s)\alpha_s(\mu_h)}{\beta^2_{\ttbar}}\,\\
&
 \begin{bmatrix} 
 \displaystyle
  \frac{3025}{22667121 } 
\left[\frac{\alpha_s(\mu_h)}{\alpha_s(\mu_s)}
\right]^{-1-\frac{4\mathrm{i} \pi }{23 \betattbar}  }\,
&0&0\\
0& \displaystyle \frac{3025}{1450695744 } \left[
\frac{\alpha_s(\mu_h)}{\alpha_s(\mu_s)}
\right]^{\frac{\mathrm{ i} \pi }{46 \betattbar}+\frac{9 \mathrm{ i} \pi }{23}-\frac{32}{23}}&0\\
 0&0& \displaystyle \frac{3025}{1450695744 }\left[
\frac{\alpha_s(\mu_h)}{\alpha_s(\mu_s)}
\right]^{\frac{\mathrm{ i} \pi }{46 \betattbar}+\frac{9 \mathrm{ i} \pi }{23}-\frac{32}{23}}\\
\end{bmatrix}
.\hspace*{-20mm}
  \end{split}
\end{align}
Examining the above evolution kernels in detail, we observe
an intensely oscillating behaviour in the diagonal entries at
\NLL\ as $\betattbar\to0$, which is always bounded from above
though and, thus, remains finite.
The results at \NNLL\ accuracy, however, exhibit  quadratic
divergences that factorise from the matrix structure of the
evolution kernel.
These divergences are induced by the product of pairs of
$\mathbf{J}_{\mathrm{thr}}^{[\kappa]}$ matrices, detailed in
Eq.\ \eqref{eq:Jh:thr:exp:qq_gg}, when assembled according to
Eq.~\eqref{eq:def:Vres:N2LL}.
Comparing this result to the exact evolution function of
Eqs.\ \eqsref{eq:def:Vres:NLL}{eq:def:Vres:N2LL}, we find that
the expressions in
Eqs.~\eqsref{eq:def:Vres:NLL:thr:qq_gg}{eq:def:Vres:N2LL:thr:qq_gg}
can indeed replicate the desired asymptotic behaviour in the
vicinity of $\beta_\ttbar=0$.
More details on this numerical assessment can be found in
App.~\ref{app:vali:Vthr}.

\subsection*{Combined resummation}

Summarising the scaling laws in
Eqs.~\eqsref{eq:bf:asy:thr}{eq:Dexp:asy:thr},
Eq.~\eqref{eq:chard:asy:thr:qq:gg},
Eq.~\eqref{eq:soft:asy:thr:qq:gg},
and Eqs.~\eqsref{eq:def:Vres:NLL:thr:qq_gg}{eq:def:Vres:N2LL:thr:qq_gg},
we can determine the asymptotic behaviour of
$\widetilde{\Sigma}_{\ttbar}^{\mathrm{res},[\kappa]}$ with the help
of Eqs.~\eqsref{eq:LP:res:qqbar}{eq:LP:res:gg},\footnote{
  Please note that the coefficient functions at the given orders
  will have to be expanded for the appropriate order counting
  of the resummed cross section. In particular,
  $(\mathcal{C}^{[\kappa],(0)}_{\alpha,\{h'\}}+\frac{\alpha_s}{4\pi}\mathcal{C}^{[\kappa],(1)}_{\alpha,\{h'\}})^{\dagger}
   (\mathcal{C}^{[\kappa],(0)}_{\beta,\{h\}}+\frac{\alpha_s}{4\pi}\mathcal{C}^{[\kappa],(1)}_{\beta,\{h\}}) 
   =\mathcal{C}^{[\kappa],(0),\dagger}_{\alpha,\{h'\}}\mathcal{C}^{[\kappa],(0)}_{\beta,\{h\}}
   +\frac{\alpha_s}{4\pi}(\mathcal{C}^{[\kappa],(0),\dagger}_{\alpha,\{h'\}}\mathcal{C}^{[\kappa],(1)}_{\beta,\{h\}}+\mathcal{C}^{[\kappa],(1),\dagger}_{\alpha,\{h'\}}\mathcal{C}^{[\kappa],(0)}_{\beta,\{h\}})
   +\order(\alpha_s^2)$, etc.
}
\begin{align}
  \label{eq:def:sigma:res:NLL:N2LL:thr}
  \begin{split}
    \widetilde{\Sigma}_{\ttbar}^{\mathrm{res},[\kappa]}\Bigg|_{\NLL}\,
    \xrightarrow[]{\betattbar\to0}&\;
    \underbrace{\widetilde{\Sigma}_{\ttbar,\mathrm{thr}}^{\mathrm{res},[\kappa]}\Bigg|_{\NLL}}_{\mathcal{O}(\betattbar^{0})}
    +\,\mathcal{O}(\betattbar)\,,\\ 
    \widetilde{\Sigma}_{\ttbar}^{\mathrm{res},[\kappa]}\Bigg|_{\NNLL}\,
    \xrightarrow[]{\betattbar\to0}&\;
    \underbrace{\widetilde{\Sigma}_{\ttbar,\mathrm{thr}}^{\mathrm{res},[\kappa]}\Bigg|_{\NNLL}}_{\mathcal{O}(\betattbar^{-5})}
    +\,\mathcal{O}(\betattbar^{-4})
    \,,
  \end{split}
\end{align}
where
\begin{align}
  \label{eq:NLL:res:qq_gg:thr}
  \begin{split}
  \widetilde{\Sigma}_{\ttbar,\mathrm{thr}}^{\mathrm{res},[q^i_{n}\qbar^j_{\nbar}]}\Bigg|_{\mathrm{N}\mathrm{LL} }=&\;
 \frac{64 \pi ^2  \alpha^2_s(\mu_h) }{9} \,
  \left[\frac{\alpha_s(\mu_h)}{\alpha_s(\mu_s)}\right]^{-\frac{42}{23}}\, 
  \mathcal{D}^{\mathrm{res},(1)}_{\mathrm{thr},[q^i_{n}\qbar^j_{\nbar}]} \,
 f_{q_n^i/N}(\tilde{\eta}_n,\mu_b) \,f_{\qbar_\nbar^j/\bar{N}}(\tilde{\eta}_\nbar,\mu_b) \,,
\\[2mm]
  \widetilde{\Sigma}_{\ttbar,\mathrm{thr}}^{\mathrm{res},[g_{n}g_{\nbar}]}\Bigg|_{\mathrm{N}\mathrm{LL} }=&\;
  \Bigg\{
  \frac{2 \pi ^2\alpha_s^2(\mu_h)}{3}
  \left[\frac{\alpha_s(\mu_h)}{\alpha_s(\mu_s)}\right]^{-2}
  +
  \frac{5\pi ^2 \alpha_s^2(\mu_h)}{3}\left[
\frac{\alpha_s(\mu_h)}{\alpha_s(\mu_s)}
\right]^{-\frac{64}{23}}
  \Bigg\}
      \\&\times\,
      \mathcal{D}^{\mathrm{res},(1)}_{\mathrm{thr},[g_{n}g_{\nbar}]}\,
       f_{g/N}(\tilde{\eta}_n,\mu_b) \, f_{g/\bar{N}}(\tilde{\eta}_\nbar,\mu_b) \,,
  \end{split}
\end{align}
and
\begin{align}
  \label{eq:N2LL:res:qq_gg:thr}
  \begin{split}
  \widetilde{\Sigma}_{\ttbar,\mathrm{thr}}^{\mathrm{res},[q^i_{n}\qbar^j_{\nbar}]}\Bigg|_{\mathrm{N}^2\mathrm{LL} }=&\;
-\frac{\alpha_s^5(\mu_h) \alpha_s^2(\mu_s) }{\betattbar^5}\,
 \frac{9150625\,\pi^3}{3551374364050766592}
   \left[\frac{\alpha_s(\mu_h)}{\alpha_s(\mu_s)}\right]^{-\frac{42}{23}}
  \\&\times\,
  \mathcal{D}^{\mathrm{res},(2)}_{\mathrm{thr},[q^i_{n}\qbar^j_{\nbar}]}\,
 f_{q_i/N}(\tilde{\eta}_n,\mu_b) \,f_{\qbar_j/\bar{N}}(\tilde{\eta}_\nbar,\mu_b)\,,\hspace*{-20mm}
 \\[4mm]
  \widetilde{\Sigma}_{\ttbar,\mathrm{thr}}^{\mathrm{res},[g_{n}g_{\nbar}]} \Bigg|_{\mathrm{N}^2\mathrm{LL} }=&\;
   \frac{\alpha_s^5(\mu_h)\alpha_s^2(\mu_s)}{ \betattbar^5}\,
  \Bigg\{
  \frac{36602500\,\pi^3}{4624185369857769 }
  \left[\frac{\alpha_s(\mu_h)}{\alpha_s(\mu_s)}\right]^{-2} \!\!\!\!\!\!
  -
  \frac{45753125\,\pi^3}{75762653099749687296}
  \left[\frac{\alpha_s(\mu_h)}{\alpha_s(\mu_s)}\right]^{-\frac{64}{23}} 
  \!\Bigg\}\hspace*{-20mm}\\
&\times\,  \mathcal{D}^{\mathrm{res},(2)}_{\mathrm{thr},[g_{n}g_{\nbar}]}\,
       f_{g/N}(\tilde{\eta}_n,\mu_b) \, f_{g/\bar{N}}(\tilde{\eta}_\nbar,\mu_b) \,.
  \end{split}
\end{align}
Once again, we omit the expression for the
$\kappa=\qbar^i_{n}q^j_{\nbar}$ case, for which the results at
\NLL\ and \NNLL\ can be derived from the $\kappa=q^i_{n}\qbar^j_{\nbar}$
case by appropriately swapping the labels $n\leftrightarrow \nbar$.
In Eqs.~\eqsref{eq:NLL:res:qq_gg:thr}{eq:N2LL:res:qq_gg:thr},
we have introduced the resummation kernels
$\mathcal{D}^{\mathrm{res},(1,2)}_{\mathrm{thr},[\kappa]}$ to encode
the contribution of Eq.~\eqref{eq:def:Dres} evaluated at threshold,
$\Mttbar=2m_t$, with the superscripts $\{1,2\}$ denoting the logarithmic
precision.
For the \NLL\ results in Eqs.~\eqref{eq:NLL:res:qq_gg:thr}, due to the
lack of perturbative corrections to the fixed-order ingredients, the
soft function is equal to a unit matrix and the beam functions are
reduced to the PDFs with the momentum fractions
$\tilde{\eta}_n= {2m_te^{\Yttbar}}/{\sqrt{s}}$ and
$\tilde{\eta}_\nbar={2m_te^{-\Yttbar}}/{\sqrt{s}}$.
Conversely, evaluating the \NNLL\ expressions of
Eq.~\eqref{eq:N2LL:res:qq_gg:thr}, we emphasise that the perturbative
corrections, which comprise the hard contributions of
Eq.~\eqref{eq:def:Chard:thr:qqb_qbq_gg:NLO} and its complex conjugate
as well as their non-cusp evaluations in
Eq.~\eqref{eq:def:Vres:N2LL:thr:qq_gg}, account for the leading singular
behaviour of $\widetilde{\Sigma}_{\ttbar}^{\mathrm{res},[\kappa]}$.

Using the results of Eqs.~\eqsref{eq:NLL:res:qq_gg:thr}{eq:N2LL:res:qq_gg:thr},
we note that while the \NLL\ resummation approaches a constant as
$\betattbar\to0$,  
the \NNLL\ results display quintic divergences.
To be precise,
$\widetilde{\Sigma}_{\ttbar}^{\mathrm{res},[q^i_{n}\qbar^j_{\nbar}]}$
approaches negative infinity in the limit $\betattbar\to0$, whereas
the sign of the threshold limit of
$\widetilde{\Sigma}_{\ttbar}^{\mathrm{res},[g_{n}g_{\nbar}]}$
is subject to the competition between
colour-singlet and colour-octet contributions, as shown in the first and
second term in the curly
brackets of Eq.~\eqref{eq:N2LL:res:qq_gg:thr}, respectively.
Under regular LHC conditions and conventional scale
definitions, the singlet term is by far dominant, though,
inducing a positive overall sign.\footnote{
  The difference in magnitude of the prefactors of the singlet and
  octet coefficients would have to be overcome by an extreme ratio
  of the strong couplings at the soft and hard scales, necessitating
  a soft scale choice extremely close to the $\Lambda_\text{QCD}$.
}

Combining the scalings of Eq.~\eqref{eq:def:sigma:res:NLL:N2LL:thr}
with Eqs.~\eqref{eq:methods:res:qT_dphi}, we are able to establish
the asymptotic properties of the resummed \qT\ and \dphittbar\ spectra
in the threshold domain.
We note that the kinematic variables introduce an additional suppression
in the limit $\betattbar\to0$,
\begin{align}\label{eq:kin:asy:diff:pTt_Mtt}
  \begin{split}
 \done^2 \widetilde{P}_{t}^{\perp} &\sim\mathcal{O}(\beta^2_{\ttbar})\,,\qquad\qquad\qquad | \widetilde{P}_{t}^{\perp}|\sim|\widetilde{P}_{t}^{z} |\sim\mathcal{O}(\beta_\ttbar)\,,\\
 \done M^2_{\ttbar} &=2M^2_{\ttbar}\,\left(\frac{ \betattbar }{1-\beta^2_\ttbar}\right)\done \beta_\ttbar=8\, m_t^2\,\beta_\ttbar\,\done \beta_\ttbar +\dots\,.
  \end{split}
\end{align}
This yields,
\begin{align}\label{eq:asy:obs:qT_dphi}
  \begin{split}
%
  \frac{\done^3 \sigma^{\rm{res}}_{\ttbar}}{\done \betattbar\done \Yttbar \done\qT}
  \,\xrightarrow[]{\betattbar\to0} &\,
\underbrace{\beta^2_{\ttbar}}_{\rm{kin}}\,
\otimes
\,
\underbrace{
\Big\{ 
\overbrace{\mathcal{O}(\betattbar^0) }^{\rm{NLL}}
+
\overbrace{\mathcal{O}(\beta^{-1}_{\ttbar}) }^{\rm{N}^2\rm{LL}} 
+
\dots
\Big\}}_{\mathcal{C}^{[\kappa]}_{\alpha,\{h\}}}\,
\otimes
\,
\underbrace{
\Big\{\overbrace{\mathcal{O}(\betattbar^0) }^{\rm{NLL} }
+
\overbrace{\mathcal{O}(\beta^{-4}_{\ttbar}) }^{\rm{N}^2\rm{LL} }
+\dots
\Big\}}_{\mathbf{V}^{[\kappa]}_{h}}\,
\otimes\dots\\
\sim&\,
\underbrace{\mathcal{O}(\beta^2_{\ttbar}) }_{\rm{NLL}} 
+
\underbrace{\mathcal{O}(\beta^{-3}_{\ttbar}) }_{\rm{N}^2\rm{LL}}
+
\dots\,,\\
%
  \frac{\done^3 \sigma^{\rm{res}}_{\ttbar}}{\done \betattbar\done \Yttbar \done\dphi_\ttbar}
  \,\xrightarrow[]{\betattbar\to0} &\,
\underbrace{\beta^3_{\ttbar}}_{\rm{kin}}\,
\otimes
\,
\underbrace{
\Big\{ 
\overbrace{\mathcal{O}(\betattbar^0) }^{\rm{NLL}}
+
\overbrace{\mathcal{O}(\beta^{-1}_{\ttbar}) }^{\rm{N}^2\rm{LL}} 
+
\dots
\Big\}}_{\mathcal{C}^{[\kappa]}_{\alpha,\{h\}}}\,
\otimes
\,
\underbrace{
\Big\{\overbrace{\mathcal{O}(\betattbar^0) }^{\rm{NLL} }
+
\overbrace{\mathcal{O}(\beta^{-4}_{\ttbar}) }^{\rm{N}^2\rm{LL} }
+\dots
\Big\}}_{\mathbf{V}^{[\kappa]}_{h}}\,
\otimes\dots\\
\sim&\,
\underbrace{\mathcal{O}(\beta^3_{\ttbar}) }_{\rm{NLL}} 
+
\underbrace{\mathcal{O}(\beta^{-2}_{\ttbar}) }_{\rm{N}^2\rm{LL}}
+
\dots\,.
  \end{split}
\end{align}
In the first line of each of the equations in Eq.~\eqref{eq:asy:obs:qT_dphi},
the scalings for the kinematic prefactor, the hard sector, and the
non-cusp evolution kernel are spelt out, capturing the asymptotic
behaviour of the differential spectra
$\done^3\sigma^{\rm{res}}_{\ttbar}/(\done\betattbar\done\Yttbar\done\mathcal{Q})$
up to \NNLL\ accuracy.
For simplicity, we omit the scalings from the beam functions, the
soft sector, and the diagonal resummation kernel, since (at least) up
to \NNLL\ all of them approach a constant in the vicinity of the
threshold $\betattbar=0$.
The second lines then present the resulting asymptotic behaviour of
the \qT\ and \dphittbar\ differential distributions at the logarithmic
accuracies of our concern.

We observe that both the \qT\ and the \dphittbar\ differential spectra
at \NLL\ experience significant kinematic suppression near threshold,
whereas at \NNLL, thanks to the Coulomb enhancement
from the hard sector and the non-cusp evolution kernel, see
Eq.~\eqref{eq:def:Chard:thr:qqb_qbq_gg:NLO} and
Eq.~\eqref{eq:def:Vres:N2LL:thr:qq_gg}, the behaviour of
$\done^3\sigma^{\rm{res}}_{\ttbar}/(\done\betattbar\done\Yttbar\done\qT)$ and
$\done^3\sigma^{\rm{res}}_{\ttbar}/(\done\betattbar\done\Yttbar\done\dphittbar)$
reverses and they instead develop cubic and quadratic divergences,
respectively.
\changed{
Although the existence of such a divergence has been implied in
earlier calculations \cite{Kulesza:2017ukk,Zhu:2012ts,
  Li:2013mia,Ahrens:2010zv,Alioli:2021ggd}, where additional
perturbative expansions were applied to the non-cusp kernel
$\mathbf{V}^{[\kappa]}_{h}$ in place of the original solution
derived in \cite{Buras:1991jm,Buchalla:1995vs},
Eq.~\eqref{eq:asy:obs:qT_dphi} for the first time presents
their specific threshold behaviour.
In turn, this finding will help to interpret the limitation of a HQET
and SCET based analysis, thereby paving the way for the future
combined resummation of Coulomb, soft and collinear corrections.%
}

\changed{
Further, we want to emphasise that for lack of a resummation of the
Coulomb interactions, Eq.~\eqref{eq:asy:obs:qT_dphi} should be
interpreted as the threshold limit of a HQET and SCET based resummation,
testing the integrability in Eq.~\eqref{eq:methods:res:qT_dphi}
over the entire $\ttbar$ production phase space, rather than an
implication on a full QCD calculation in this limit.
An analysis of the threshold behaviour of a full QCD result invariably
necessitates the use of pNRQCD or vNRQCD in its derivation.
Finally,}
we conclude that the resummation formalism
in Eq.~\eqref{eq:methods:res:qT_dphi} \textit{cannot} be
straightforwardly applied to evaluate the single differential
observables $\done\sigma^{\rm{res}}_{\ttbar}/\done\qT$ and
$\done\sigma^{\rm{res}}_{\ttbar}/\done\dphittbar$ beyond \NLL,
unless a kinematic constraint on $\betattbar$, or equivalently
$\Delta E_\ttbar$ or $\Mttbar$, is put in place to remove the
threshold regime from the $\betattbar$ integration.
Instead, in the following, we will propose two \changed{\textit{ad hoc}}
prescriptions to
smoothly and consistently match the well-separated domain
$\dEttbar\sim\mathcal{O}(m_t)$ to the threshold region
$\dEttbar\ll\Mttbar$ for a generic observable
$\done^3\sigma^{\rm{res}}_{\ttbar}/({\done\betattbar\done\Yttbar\done\mathcal{Q}})$.

 \subsection{\changed{Prescriptions to extend the resummation region}}
\label{sec:thr:div:extrap}

\subsubsection{D-\changed{prescription}: Resummation with a decomposed Sudakov factor}

In order to \changed{sufficiently weaken} the threshold divergences in the evolution kernel
$\mathbf{V}_h^{[\kappa]}$, which, according to
Eqs.~\eqref{eq:asy:obs:qT_dphi}, constitute the main singular
contribution at \NNLL, we introduce a first prescription, dubbed
D-\changed{prescription} in the following.
  
We start by analysing the elements of $\mathbf{V}_h^{[\kappa]}$ in
the well-separated domain, i.e.\ $\dEttbar\sim\mathcal{O}(m_t)$.
As defined in Eq.~\eqref{eq:def:Vres:N2LL}, at \NNLL\ accuracy,
$\mathbf{V}_h^{[\kappa]}$ includes the \NLL\ resummation kernel
sandwiched between the perturbative corrections
$\big(1+\alpha_s(\mu_s)\mathbf{J}^{[\kappa]}/(4\pi)\big)$ and
$\big(1-\alpha_s(\mu_h)\mathbf{J}^{[\kappa]}/(4\pi)\big)$.
In the region $\dEttbar\sim\mathcal{O}(m_t)$,
both correction terms are of a similar magnitude to the
non-logarithmic contributions in the hard and soft functions.
Therefore, the product of them is expected to be numerically
comparable with the \NNNLL\ coefficients.
In consequence, during our phenomenological investigation, we can
truncate all terms proportional to $\alpha_s(\mu_s)\alpha_s(\mu_h)$
in Eq.~\eqref{eq:def:Vres:N2LL}, at the cost of additional
non-logarithmic corrections in the well-separated domain, yielding
\begin{align}\label{eq:def:Vres:N2LL:Dec}
 & \mathbf{V}_h^{[\kappa]} (v_t,v_{\tbar}, \mu_s,\mu_h)\Bigg|_{\mathrm{N}^2\mathrm{LL}_{\mathrm{D}}}= \sum_{n_s,n_h=0}^{n_s+n_h=1}\,\left(\frac{\alpha_s(\mu_s)}{4\pi}\right)^{n_s}\left(\frac{\alpha_s(\mu_h)}{4\pi}\right)^{n_h}\mathbf{V}_h^{[\kappa],(n_s,n_h)} (v_t,v_{\tbar}, \mu_s,\mu_h)\Bigg|_{\mathrm{N}^2\mathrm{LL}_{\mathrm{D}}}\,,
\end{align}
where
\begin{align}
  \label{eq:def:Vres:N2LL:Dec:00:10:01}
  \begin{split}
    \mathbf{V}_h^{[\kappa],(0,0)} (v_t,v_{\tbar}, \mu_s,\mu_h)\Bigg|_{\mathrm{N}^2\mathrm{LL}_{\mathrm{D}}} \,=\,&\mathbf{R}^{-1}_{[\kappa]}\,
    \exp\bigg\{
      \frac{\mathbf{r}^{[\kappa],(0)}_{h}}{2\beta_0}
      \ln\bigg[
        \frac{\alpha_s(\mu_h)}{\alpha_s(\mu_s)}
      \bigg]
    \bigg\}    \mathbf{R}_{[\kappa]}\,=\,\mathbf{V}_h^{[\kappa]} (v_t,v_{\tbar}, \mu_s,\mu_h)\Bigg|_{\mathrm{NLL}},\\
     \mathbf{V}_h^{[\kappa],(1,0)} (v_t,v_{\tbar}, \mu_s,\mu_h)\Bigg|_{\mathrm{N}^2\mathrm{LL}_{\mathrm{D}}} \,=\,& \mathbf{R}^{-1}_{[\kappa]}\,
   \mathbf{J}^{[\kappa]}\,
    \exp\bigg\{
      \frac{\mathbf{r}^{[\kappa],(0)}_{h}}{2\beta_0}
      \ln\bigg[
        \frac{\alpha_s(\mu_h)}{\alpha_s(\mu_s)}
      \bigg]
    \bigg\}\,    
    \mathbf{R}_{[\kappa]}\,,  \\
     \mathbf{V}_h^{[\kappa],(0,1)} (v_t,v_{\tbar}, \mu_s,\mu_h)\Bigg|_{\mathrm{N}^2\mathrm{LL}_{\mathrm{D}}} \,=\, &\,-\,
    \mathbf{R}^{-1}_{[\kappa]}\, 
    \exp\bigg\{
      \frac{\mathbf{r}^{[\kappa],(0)}_{h}}{2\beta_0}
      \ln\bigg[
        \frac{\alpha_s(\mu_h)}{\alpha_s(\mu_s)}
      \bigg]
    \bigg\}\,
   \mathbf{J}^{[\kappa]}\,
    \mathbf{R}_{[\kappa]}\,.
  \end{split}
\end{align}
Herein, we have decomposed the original evolution kernel of
Eq.~\eqref{eq:def:Vres:N2LL} according to their $\alpha_s(\mu_s)$
and $\alpha_s(\mu_h)$ powers.
The leading order contribution $\mathbf{V}_h^{[\kappa],(0,0)}$ contains
no perturbative corrections and thus coincides with the \NLL\ Sudakov
factor in Eq.~\eqref{eq:def:Vres:NLL}, while starting from
$\mathbf{V}_h^{[\kappa],(1,0)}$ and $\mathbf{V}_h^{[\kappa],(0,1)}$
perturbative corrections encoded in the $\mathbf{J}^{[\kappa]}$ enter.
To facilitate our discussion and comparison below, we will refer to the
results in Eq.~\eqref{eq:def:Vres:N2LL:Dec} as the non-cusp evolution
kernel evaluated in the decomposed \changed{prescription} (D-\changed{prescription}), i.e.\ \NNLLD.

It is worth noting that the non-cusp evolution evaluated in the D-\changed{prescription}
cannot precisely satisfy the hard RGE as its original form in
Eq.~\eqref{eq:def:Vres:N2LL} did
\changed{at \NNLL.
This can be confirmed by taking the derivative of
Eq.~\eqref{eq:def:Vres:N2LL:Dec} with respect to $\ln\mu_s$.
It follows that
\begin{equation}\label{eq:diff:Vh:N2LLD}
  \begin{split}
    \lefteqn{
      \frac{\partial}{\partial \ln\mu_s}\mathbf{V}_h^{[\kappa]}
      (v_t,v_{\tbar},\mu_s,\mu_h)\Bigg|_{\NNLLD}
      =
      \beta\big(\alpha_s(\mu_s)\big)\,\frac{\partial}{\partial \alpha_s(\mu_s)}
      \mathbf{V}_h^{[\kappa]}(v_t,v_{\tbar},\mu_s,\mu_h)\Bigg|_{\NNLLD}
    }\\
    =&\,
    \beta\big(\alpha_s(\mu_s)\big)\,
    \mathbf{R}^{-1}_{[\kappa]}
    \bigg[  -\frac{1}{\alpha_s(\mu_s)}\frac{\mathbf{r}^{[\kappa],(0)}_{h}}{2  \beta_0}+\frac{ \mathbf{J}^{[\kappa]}}{4\pi}-\frac{ \mathbf{J}^{[\kappa]}}{4\pi}\frac{\mathbf{r}^{[\kappa],(0)}_{h}}{2  \beta_0}\bigg]
            \exp\bigg\{
          \frac{\mathbf{r}^{[\kappa],(0)}_{h}}{2\beta_0}
          \ln\bigg[
            \frac{\alpha_s(\mu_h)}{\alpha_s(\mu_s)}
          \bigg]
        \bigg\}\,
    \mathbf{R}_{[\kappa]}\\
    &
    +\,\beta\big(\alpha_s(\mu_s)\big)\,
    \mathbf{R}^{-1}_{[\kappa]}
    \bigg[ \frac{1}{\alpha_s(\mu_s)}\frac{\mathbf{r}^{[\kappa],(0)}_{h}}{2  \beta_0}\bigg]
            \exp\bigg\{
          \frac{\mathbf{r}^{[\kappa],(0)}_{h}}{2\beta_0}
          \ln\bigg[
            \frac{\alpha_s(\mu_h)}{\alpha_s(\mu_s)}
          \bigg]
        \bigg\}\,
        \frac{\alpha_s(\mu_h)}{4\pi}\mathbf{J}^{[\kappa]}\,
    \mathbf{R}_{[\kappa]}\,.
  \end{split}
\end{equation}
According to the definition in Eq.~\eqref{eq:def:Vres:N2LL:Jmax},
$\mathbf{J}^{[\kappa]}$ fulfills the identity,
\begin{align}
\label{eq:id:Jh:N2LL}
\mathbf{J}^{[\kappa]}\left(\mathbf{I}- \frac{\mathbf{r}^{[\kappa],(0)}_{h}}{2  \beta_0}\right)=\frac{\beta_1   }{2\beta_0^2}\mathbf{r}^{[\kappa],(0)}_{h}-\frac{\mathbf{r}^{[\kappa],(1)}_{h}}{2\beta_0}-\frac{\mathbf{r}^{[\kappa],(0)}_{h}}{2  \beta_0}\mathbf{J}^{[\kappa]}\,.
\end{align}
Reinserting this relation back to Eq.~\eqref{eq:diff:Vh:N2LLD} gives
\begin{align}\label{eq:diff:Vh:N2LLD2}
&\frac{\partial}{\partial \ln\mu_s}\mathbf{V}_h^{[\kappa]} (v_t,v_{\tbar}, \mu_s,\mu_h)\Bigg|_{\NNLLD}\nnb\\
=&\,
\mathbf{R}^{-1}_{[\kappa]}
\Bigg\{
\bigg[\frac{\alpha_s(\mu_s)}{4\pi}\mathbf{r}^{[\kappa],(0)}_{h}+\left(\frac{\alpha_s(\mu_s)}{4\pi}\right)^{\!2}\!\!\mathbf{r}^{[\kappa],(1)}_{h}+
\mathcal{O}(\alpha_s^3)\bigg]
\bigg[\mathbf{I}+\frac{\alpha_s(\mu_s)}{4\pi}\mathbf{J}^{[\kappa]}\bigg]
\Bigg\}\,
  \exp\bigg\{
      \frac{\mathbf{r}^{[\kappa],(0)}_{h}}{2\beta_0}
      \ln\bigg[
        \frac{\alpha_s(\mu_h)}{\alpha_s(\mu_s)}
      \bigg]
    \bigg\}\,
\mathbf{R}_{[\kappa]}\nnb\\
&
-\mathbf{R}^{-1}_{[\kappa]}
\Bigg\{
 \frac{\alpha_s(\mu_s)}{4\pi}\mathbf{r}^{[\kappa],(0)}_{h}
+
\mathcal{O}(\alpha_s^2)
\Bigg\}\,
  \exp\bigg\{
      \frac{\mathbf{r}^{[\kappa],(0)}_{h}}{2\beta_0}
      \ln\bigg[
        \frac{\alpha_s(\mu_h)}{\alpha_s(\mu_s)}
      \bigg]
    \bigg\}\,
     \frac{\alpha_s(\mu_h)}{4\pi}\mathbf{J}^{[\kappa]}\,
\mathbf{R}_{[\kappa]}\,,
\end{align}
where we have expanded the QCD beta function in $\alpha_s(\mu_s)$.
In this result we observe that while the expression in the second
line meets the \NNLL\ RGE of the hard sector, up to perturbative
corrections of $\sim\mathcal{O}(\alpha_s^3)$, which is given by
differentiating Eq.~\eqref{eq:def:Vres:N2LL} with respect to
$\ln\mu_s$, yielding
\begin{align}
\label{eq:diff:Vh:N2LL}
&\frac{\partial}{\partial \ln\mu_s}\mathbf{V}_h^{[\kappa]}
=
\mathbf{R}^{-1}_{[\kappa]}
\bigg[\frac{\alpha_s(\mu_s)}{4\pi}\mathbf{r}^{[\kappa],(0)}_{h}+\left(\frac{\alpha_s(\mu_s)}{4\pi}\right)^2\mathbf{r}^{[\kappa],(1)}_{h}+
\mathcal{O}(\alpha_s^3)\bigg]
\mathbf{R}_{[\kappa]}\,
\mathbf{V}_h^{[\kappa]} \,,
\end{align}
an additional
contribution enters in the third line.
This additional term takes a form similar to the \NLL\ evolution
equation, spoiling the \NNLL\ accuracy of
$\mathbf{V}_h^{[\kappa]}|_{\NNLLD}$.
In view of this difference, the decomposition of
Eq.~\eqref{eq:def:Vres:N2LL:Dec} should be interpreted as an
\textit{ad hoc} prescription to diminish the divergence in the
threshold limit rather than an exact solution of the hard RGE.
}
  

An analogous decomposition should also be applied to the other partonic
functions in Eqs.~\eqref{eq:LP:res:qqbar} and \eqref{eq:LP:res:gg}
to remove the combined contributions from different fixed-order ingredients,
giving
\begin{align}
  \label{eq:LP:res:qqbar:gg:Dec}
  \begin{split}
  &\hspace*{-5mm}
    \widetilde{\Sigma}_{\ttbar}^{\mathrm{res},[q^i_{n}\qbar^j_{\nbar}]}
    (\bTvec,\Yttbar, \Mttbar,\Omega_t)\Bigg|_{\mathrm{N}^2\mathrm{LL}_{\mathrm{D}}} \,\\
  \,=&\;
    \left(\frac{1}{2N_c}\right)^2\,
    \mathcal{D}^{\mathrm{res}}_{[q^i_{n}\qbar^j_{\nbar}]}
    (\bT,\Mttbar,\mu_h,\mu_b,\mu_s,\nu_b,\nu_s)\\
  &\times
    \sum_{\{n_i,n_i',n_i'',n_i'''\}}\,
    \left(\frac{\alpha_s(\mu_s)}{4\pi}\right)^{n_s+n'_s+n''_s}\,
    \left(\frac{\alpha_s(\mu_b)}{4\pi}\right)^{n_b+n_b'}\,
    \left(\frac{\alpha_s(\mu_h)}{4\pi}\right)^{n_h+n'_h+n''_h+n'''_h}\\
  &\hspace*{20mm}\times
    \mathcal{B}_{n}^{[q_n^i],(n_b)}(\eta_n,\bT,\mu_b,\nu_b)\,
    \mathcal{B}_{\nbar}^{[\qbar_\nbar^j],(n'_b)}(\eta_{\nbar},\bT,\mu_b,\nu_b) \,\\
  &\hspace*{20mm}\times
    \sum_{\{\alpha,\beta,h\}}
    \Bigg\{
      \mathcal{S}^{\alpha_1\beta_1}_{[q_{n}\qbar_{\nbar}],(n_s)}
      (\bTvec,v_{t},v_{\tbar},\mu_s,\nu_s)
      \left[
        \mathcal{V}^{[q_{n}\qbar_{\nbar}],(n'_s,n'_h)}_{\alpha_1\alpha_2}(v_{t},v_{\tbar},\mu_s,\mu_h)\,
        \mathcal{C}^{[q^i_{n}\qbar^j_{\nbar}],(n_h)}_{\alpha_2;h_nh_{\nbar}h_th_{\tbar}}
      \right]^*\\[-4mm]
  &\hspace*{80mm}\times
      \left[
        \mathcal{V}^{[q_{n}\qbar_{\nbar}],(n''_s,n''_h)}_{\beta_1\beta_2}(v_{t},v_{\tbar},\mu_s,\mu_h)\,
        \mathcal{C}_{\beta_2;h_nh_{\nbar}h_th_{\tbar}}^{[q^i_{n}\qbar^j_{\nbar}],(n_h''') }\,
      \right]
    \Bigg\}\\
  &\hspace*{20mm}\times
    \theta\Big[1-\left(n_h+n'_h+n''_h+n'''_h\right)-\left(n_s+n'_s+n''_s\right)-\left(n_b+n_b'\right)\Big]\,,
  \\[2mm]
  &\hspace*{-5mm}
    \widetilde{\Sigma}_{\ttbar }^{\mathrm{res},[g_{n}g_{\nbar}]}
    (\bTvec,\Yttbar, \Mttbar,\Omega_t)\Bigg|_{\mathrm{N}^2\mathrm{LL}_{\mathrm{D}}} \,\\
  \,=&\;
    \left(\frac{1}{N^2_c-1}\right)^2\,
    \mathcal{D}^{\mathrm{res}}_{[g_{n}g_{\nbar}]}
    (\bT,\Mttbar,\mu_h,\mu_b,\mu_s,\nu_b,\nu_s)\\
  &\times
    \sum_{\{n_i,n_i',n_i'',n_i'''\}}\,
    \left(\frac{\alpha_s(\mu_s)}{4\pi}\right)^{n_s+n'_s+n''_s}\,
    \left(\frac{\alpha_s(\mu_b)}{4\pi}\right)^{n_b+n_b'}\,
    \left(\frac{\alpha_s(\mu_h)}{4\pi}\right)^{n_h+n'_h+n''_h+n'''_h}\\
  &\hspace*{20mm}\times
    \sum_{\{\alpha,\beta, h,h'\}}
    \Bigg\{
      \mathcal{S}^{\alpha_1\beta_1}_{[g_ng_{\nbar}],(n_s)}(\bTvec,v_{t},v_{\tbar},\mu_s,\nu_s)\,
      \mathcal{B}^{[g_n],(n_b)}_{n,h_n'{h_n}}(\eta_n,\bTvec,\mu_b,\nu_b)\,
      \mathcal{B}^{[g_{\nbar}],(n'_b)}_{\nbar,h_{\nbar}'{h_{\nbar}}}(\eta_{\nbar},\bTvec,\mu_b,\nu_b)\\
  &\hspace*{40mm}\times
      \left[
        \mathcal{V}^{[g_{n}g_{\nbar}],(n'_s,n'_h)}_{\alpha_1\alpha_2}(v_{t},v_{\tbar},\mu_s,\mu_h)\,
        \mathcal{C}^{[g_{n}g_{\nbar}],(n_h)}_{\alpha_2;h'_nh'_{\nbar}h_th_{\tbar}}
      \right]^*\\
  &\hspace*{40mm}\times
      \left[
        \mathcal{V}^{[g_{n}g_{\nbar}],(n''_s,n''_h)}_{\beta_1\beta_2}(v_{t},v_{\tbar},\mu_s,\mu_h)\,
        \mathcal{C}_{\beta_2;h_nh_{\nbar}h_th_{\tbar}}^{[g_ng_{\nbar}],(n'''_h)}
      \right]
    \Bigg\}\\
  &\hspace*{20mm}\times
    \theta\Big[1-\left(n_h+n'_h+n''_h+n'''_h\right)-\left(n_s+n'_s+n''_s\right)-\left(n_b+n_b'\right)\Big],
  \end{split}\hspace*{-20mm}
\end{align}
where the Heaviside function $\theta(x)$ is introduced with $\theta(x)=1$
for $x\ge0$ and $\theta(x)=0$ otherwise.
The $\mathcal{V}^{[\kappa],(n_s,n_h)}_{\alpha\beta}$ refers to the element
in the non-cusp resummation kernel of Eq.~\eqref{eq:def:Vres:N2LL:Dec}
at index $\{\alpha,\beta\}$.
The perturbative expansion of the fixed-order coefficient functions is
defined as,
\begin{align}
  \begin{split}
\mathcal{S}^{\alpha\beta}_{[\kappa]}=\sum_{n_s=0}^{\infty}\left(\frac{\alpha_s(\mu_s)}{4\pi}\right)^{n_s}\mathcal{S}^{\alpha\beta}_{[\kappa],(n_s)}\,,\\
\mathcal{C}^{[\kappa]}_{\alpha,\{h\}}=\sum_{n_h=0}^{\infty}\left(\frac{\alpha_s(\mu_h)}{4\pi}\right)^{n_h+1}\mathcal{C}^{[\kappa],(n_h)}_{\alpha,\{h\}}\,,\\
\mathcal{B}_{n(\nbar)}^{[\kappa]}=\sum_{n_b=0}^{\infty}\left(\frac{\alpha_s(\mu_b)}{4\pi}\right)^{n_b}\mathcal{B}_{n(\nbar)}^{[\kappa],(n_b)}\,.
  \end{split}
\end{align}
The asymptotic behaviour of Eqs.~\eqref{eq:LP:res:qqbar:gg:Dec}
in the threshold limit $\betattbar\to0$ can be obtained by repeating
the expansion procedure of Sec.~\ref{sec:sigma:thr:hqet}.
Analysing the fixed order constituents, as demonstrated in
Eq.~\eqref{eq:bf:asy:thr}, \eqref{eq:chard:asy:thr:qq:gg},
and \eqref{eq:soft:asy:thr:qq:gg}, the soft and beam-collinear
functions approach a constant in the limit $\betattbar\to0$ up to \NLO,
while a power like divergence of $\order(\betattbar^{-1})$ still emerges
from the \NLO\ hard sector as a result of the Coulomb interaction.
As for the evolution kernels, the diagonal entries
$\mathcal{D}^{\mathrm{res}}_{[\kappa]}$ continue to be regular
in the threshold domain, see Eq.~\eqref{eq:Dexp:asy:thr},
while the singular behaviour of the non-cusp kernel $\mathbf{V}_h$ is
now reduced by one power of $\betattbar$ after the decomposition in
Eq.~\eqref{eq:def:Vres:N2LL:Dec}, according to the scaling rules
of Eq.~\eqsref{eq:rh:thr:exp}{eq:Jh:thr:exp:qq_gg}, i.e.\
\begin{align} 
\label{eq:LP:res:Dec:thr:betam1}
\mathbf{V}_h^{[\kappa]}\Bigg|_{\mathrm{N}^2\mathrm{LL}_{\mathrm{D}}}\sim \mathcal{O}(\beta^{-1}_{\ttbar})\,.
\end{align}
In summary, we arrive at,
\begin{align}\label{eq:LP:res:Dec:thr}
\lim_{\betattbar\to0}\widetilde{\Sigma}_{\ttbar }^{\mathrm{res},[\kappa]}\Bigg|_{\mathrm{N}^2\mathrm{LL}_{\mathrm{D}}}\sim\mathcal{O}\left(\betattbar^{-1}\right)\,.
\end{align}
In comparison with Eq.~\eqref{eq:def:sigma:res:NLL:N2LL:thr},
the thus defined D-\changed{prescription} reduces the degree of the divergence
to $\mathcal{O}(\beta^{-1}_{\ttbar})$, pushing all terms of higher
divergence to \NNLLp\ and beyond.
Although our resummed cross section still diverges as $\betattbar\to0$,
this singularity can be well contained by the kinematical suppression
introduced through the phase space element and the observable definition,
see Eq.~\eqref{eq:kin:asy:diff:pTt_Mtt}.
\changed{Therefore, we can now safely lift the kinematic constraint in Eq.~\eqref{eq:methods:res:qT_dphi}
and take into account the full $M_\ttbar$ range in the phase space integral for
the single  differential observables $\done\sigma^{\rm{res}}_{\ttbar}/\done\qT$ and $\done\sigma^{\rm{res}}_{\ttbar}/\done\Delta\phi_\ttbar$. 
}

 At last, we would like to stress that in the previous calculations
on the soft \cite{Kulesza:2017ukk} and zero-jettiness
\cite{Alioli:2021ggd} resummations, the expansion of the product
of the fixed-order contributions and the hard evolution kernels in
the strong coupling $\alpha_s$ was already used to remove the
Coulomb divergence and thereby accomplish \NNLL\ accurate results.
Eqs.~\eqref{eq:LP:res:qqbar:gg:Dec} in our formulation is in fact
equivalent to their solution, with the only exception that the soft
and beam-collinear sectors were adapted as appropriate to the
observables of interest.
An analogous \changed{prescription} was also implemented in the \qT\ \cite{Zhu:2012ts,
  Li:2013mia} and threshold \cite{Ahrens:2010zv} resummation, where,
in place of $\alpha_s$, the expansion therein proceeded in the scaling
$\lambda_{N}\sim \alpha_s\sim1/\ln(\mu_h/\mu_s)$.
This method is equivalent to Eqs.~\eqref{eq:LP:res:qqbar:gg:Dec}
of our formulation as well, since, up to \NNLL, the result of the
$\lambda_{N}$ expansion can be absorbed into the running of $\alpha_s$.

\subsubsection{R-\changed{prescription}: Resummation with a re-exponentiated anomalous dimension} 

Alternatively, we can also mitigate the threshold singularity
of $\mathbf{V}_h^{[\kappa]}$ at \NNLL\ by
re-exponentiating the divergent contributions in the anomalous
dimension $\gamma^{[\kappa]}_h$.
We will call this the R-\changed{prescription} in the following.
To accomplish this it is worth noting that, to accommodate the
Coulomb enhancement in the threshold domain, it is convenient
to organise the perturbative contributions using the parameter
$\lambda_{\ttbar}\sim \betattbar\sim\alpha_s$
\cite{Beneke:1999qg,Beneke:1999zr,Pineda:2006ri,Beneke:2011mq}.
Even though a systematic resummation of Coulomb, soft, and
beam-collinear singularities is not the focus of this paper,
in the following we will show that this scaling rule can facilitate
the regularisation of the threshold divergence of
Eq.~\eqref{eq:def:Vres:N2LL:thr:qq_gg}.

Expanding the anomalous dimension $\gamma^{[\kappa]}_h$ up to
two-loop level~\cite{Ferroglia:2009ii,Ferroglia:2009ep} in the
parameter $\lambda_{\ttbar}$, we arrive at the following power series,
\begin{align}
 \gamma_h^{[\kappa]} \xrightarrow[]{\betattbar\to0} \left(\frac{\alpha_s(\mu)}{4\pi}\right)\gamma_{h,\mathrm{thr}}^{[\kappa],(0)}+\left(\frac{\alpha_s(\mu)}{4\pi}\right)^2\gamma_{h,\mathrm{thr}}^{[\kappa],(1)}+\mathcal{O}(\lambda^2_{\ttbar})\,,
\end{align}
where
\begin{align}
\label{eq:def:ad:hard:thr:lo_nlo:qq_gg}
\begin{split}
\gamma_{h,\mathrm{thr}}^{[q_n\qbar_{\nbar}],(0)}= &\;\gamma_{h,\mathrm{thr}}^{[\qbar_n{q}_{\nbar}],(0)}=
\begin{bmatrix}
 -\frac{8 i \pi }{3  \betattbar} -8& 0 \\
0& \frac{i \pi }{3 \betattbar}+6 i \pi -14
\end{bmatrix}
\,,
\\
 \gamma_{h,\mathrm{thr}}^{[g_{n}g_{\nbar}],(0)}
 =&
  \begin{bmatrix}
 -\frac{46}{3}-\frac{8 i \pi }{3  \betattbar} & 0&0 \\
0& \frac{i \pi }{3  \betattbar}+6 i \pi -\frac{64}{3}&0\\
0&0&\frac{i \pi }{3  \betattbar}+6 i \pi -\frac{64}{3}
\end{bmatrix}
\,,
\\
\gamma_{h,\mathrm{thr}}^{[q_n\qbar_{\nbar}],(1)}= &\;\gamma_{h,\mathrm{thr}}^{[\qbar_n{q}_{\nbar}],(1)}=
 \begin{bmatrix}
 -\frac{344 i \pi }{27  \betattbar}& 0 \\
0& \frac{43 i \pi }{27  \betattbar}
\end{bmatrix}
\,,
\\
 \gamma_{h,\mathrm{thr}}^{g_{n}g_{\nbar},(1)}
=&
  \begin{bmatrix}
 -\frac{344 i \pi }{27 \betattbar} & 0&0 \\
0&  \frac{43 i \pi }{27  \betattbar}&0\\
0&0& \frac{43 i \pi }{27  \betattbar}
\end{bmatrix}
\,.
\end{split}
\end{align}
Here, while we retain the leading and subleading singular contributions
in the one-loop anomalous dimension, only the leading terms are needed
in the two loop results, in accordance with our scaling rule.
At this point, it is important to note that in the threshold limit
all $\gamma^{[\kappa]}_h$ are diagonal up to two-loop order.
This allows us to solve the hard RGE for the evolution kernel in the
low $\betattbar$ region exactly,
\begin{align}
\label{eq:def:RGE:ReExp}
\frac{\done}{\done\ln\mu}\ln \widetilde{\mathbf{V}}_{h,\mathrm{thr}}^{[\kappa]}
= \gamma_{h,\mathrm{thr}}^{[\kappa]}\,,
\end{align}
which leads to the results at \NLL\ and \NNLL\ accuracy,
\begin{align}
\label{eq:def:evo:ReExp:nll_n2ll}
\begin{split}
 \widetilde{\mathbf{V}}_{h,\mathrm{thr}}^{[\kappa],(0)} 
     \,   =&\;  
    \exp\bigg\{
      \frac{\gamma^{[\kappa],(0)}_{\mathrm{thr},h}}{2\beta_0}
      \ln\bigg[
        \frac{\alpha_s(\mu_h)}{\alpha_s(\mu_s)}
      \bigg]
    \bigg\} \,,
    \\
 \widetilde{\mathbf{V}}_{h,\mathrm{thr}}^{[\kappa],(1) } 
     \,   =&\;  
    \exp\bigg\{
      \frac{\gamma^{[\kappa],(0)}_{\mathrm{thr},h}}{2\beta_0}
      \ln\bigg[
        \frac{\alpha_s(\mu_h)}{\alpha_s(\mu_s)}
      \bigg]
      -\frac{\alpha_s(\mu_h)-\alpha_s(\mu_s)}{4\pi}    
     \underbrace{ \left(\gamma^{{[\kappa]},(0)}_{h,\mathrm{thr}}\;
     \frac{\beta_1}{2\beta^2_0}
    -\frac{ \gamma^{{[\kappa]},(1)}_{h,\mathrm{thr}}}
          {2\beta_0 }\right)}_{\to\mathbf{J}^{[\kappa]}_{\mathrm{thr}}+\mathcal{O}(\betattbar^0)}
    \bigg\} \,,
\end{split}
\end{align}
where $\beta_{i}$ stands for the QCD beta function
\cite{vanRitbergen:1997va,Czakon:2004bu}.
With this result we find that the \NLL\ evolutions here can exactly
reproduce the leading contributions in
Eq.~\eqref{eq:def:Vres:NLL:thr:qq_gg} which are derived by
expanding the analytic expression of Eq.~\eqref{eq:def:Vres:NLL}
in the limit $\betattbar\to0$.

In particular, at finite \betattbar, where the $\gamma^{[\kappa]}_h$
are in general not diagonal at \NNLL, no closed solutions are available.
Hence, approximate solutions are used, for example in
Eq.~\eqref{eq:def:Vres:N2LL}, where only the one-loop anomalous
dimensions are exponentiated and the logarithmic corrections relevant
at \NNLL\ are applied by multiplying
$(\mathbf{I}+\frac{\alpha_s(\mu_s)}{4\pi}\mathbf{J}^{[\kappa]})$ and
$(\mathbf{I}-\frac{\alpha_s(\mu_h)}{4\pi}\mathbf{J}^{[\kappa]})$,
respectively. 
This structural difference can lead to differences in the asymptotic
behaviour between the solutions in Eq.~\eqref{eq:def:evo:ReExp:nll_n2ll}
and Eq.~\eqref{eq:def:Vres:N2LL:thr:qq_gg}, respectively, in the threshold
limit.
For instance, the ${\mathbf{V}}_{h,\mathrm{thr}}^{[\kappa],(1)}$ of
Eq.~\eqref{eq:def:Vres:N2LL:thr:qq_gg} are directly proportional
to the product
$\alpha_s(\mu_h)\alpha_s(\mu_s)\big(\mathbf{J}^{[\kappa]}_{\mathrm{thr}}\big)^2$
which, according to Eq.~\eqref{eq:Jh:thr:exp:qq_gg}, develops
divergences of $\mathcal{O}(\betattbar^{-2})$ as $\betattbar\to0$.
However, as we have now moved all anomalous dimensions into the
exponent and owing to the fact that their singular terms reside in the
imaginary part only, see Eq.~\eqref{eq:def:ad:hard:thr:lo_nlo:qq_gg},
the $\widetilde{\mathbf{V}}_{h,\mathrm{thr}}^{[\kappa],(1)}$
exhibits oscillatory but finite behaviour in the limit $\betattbar\to0$.
We can exploit this improved behaviour 
to remove the threshold divergences of Eq.~\eqref{eq:def:Vres:N2LL}.

Noting that the RGE of Eq.~\eqref{eq:def:RGE:ReExp} is subject
to the counting rule $\lambda_{\ttbar}\sim\betattbar\sim\alpha_s$,
which is appropriate in the threshold domain but can receive
significant power corrections in the well-separated region
at large \betattbar, we introduce the
following matching procedure,
\begin{align}
\label{eq:def:Vres:N2LL:Reexp}
\mathbf{V}_{h }^{[\kappa] } \Bigg|_{\mathrm{N}^2\mathrm{LL}_{\mathrm{R}} }
     \, =&\;  
{f}_{\rm{tran}}(\betattbar,c_{\mathrm{thr}} ,r_{\mathrm{thr}})\widetilde{\mathbf{V}}_{h,\mathrm{thr}}^{[\kappa],(1)}
\left[  \widetilde{\mathbf{V}}_{h,\mathrm{exp}}^{[\kappa],(1)} 
\right]^{-1} 
{ \mathbf{V}}_h^{[\kappa]} \Bigg|_{\mathrm{N}^2\mathrm{LL} }
+\Big[1-{f}_{\rm{tran}}(\betattbar,c_{\mathrm{thr}} ,r_{\mathrm{thr}})\Big]{ \mathbf{V}}_h^{[\kappa]} \Bigg|_{\mathrm{N}^2\mathrm{LL} }\hspace*{-5mm},
\end{align}
where in the first term the matrix
$\widetilde{\mathbf{V}}_{h,\mathrm{exp}}^{[\kappa],(1)}$ is used to remove
the overlap between
$\widetilde{\mathbf{V}}_{h,\mathrm{thr}}^{[\kappa],(1)}$ and
$\mathbf{V}_h^{[\kappa]}$.
It can be extracted by expanding
$\widetilde{\mathbf{V}}_{h,\mathrm{thr}}^{[\kappa],(1)}$
in $\alpha_s(\mu_s)$ and $\alpha_s(\mu_h)$ and retaining all
contributions up to \NLO, yielding
\begin{align}
\label{eq:def:Vres:N2LL:Reexp:exp}
\begin{split}
\widetilde{\mathbf{V}}_{h,\mathrm{exp}}^{[\kappa],(1)}\,=&\;
      \bigg[1
      +\frac{ \alpha_s(\mu_s)}{4\pi}    
 \left(\gamma^{{[\kappa]},(0)}_{h,\mathrm{thr}}\;
     \frac{\beta_1}{2\beta^2_0}
    -\frac{ \gamma^{{[\kappa]},(1)}_{h,\mathrm{thr}}}
          {2\beta_0 }\right)   \bigg]\\
  &\hspace*{20mm}\times\,
 \exp\bigg\{
      \frac{\gamma^{[\kappa],(0)}_{\mathrm{thr},h}}{2\beta_0}
      \ln\bigg[
        \frac{\alpha_s(\mu_h)}{\alpha_s(\mu_s)}
      \bigg]
      \bigg\}
      \bigg[1
      -\frac{\alpha_s(\mu_h) }{4\pi}    
 \left(\gamma^{{[\kappa]},(0)}_{h,\mathrm{thr}}\;
     \frac{\beta_1}{2\beta^2_0}
    -\frac{ \gamma^{{[\kappa]},(1)}_{h,\mathrm{thr}}}
          {2\beta_0 }\right)   \bigg]\,.
\end{split}
\end{align}
Multiplying $\widetilde{\mathbf{V}}_{h,\mathrm{thr}}^{[\kappa],(1)}$
by $[\widetilde{\mathbf{V}}_{h,\mathrm{exp}}^{[\kappa],(1)}]^{-1}$
removes terms of the same perturbative order as those already present in
$\mathbf{V}_h^{[\kappa]}$ in Eq.~\eqref{eq:def:Vres:N2LL},
thereby eliminating any double-counting in the matched result.
In the limit $\betattbar\to0$, both
$\widetilde{\mathbf{V}}_{h,\mathrm{exp}}^{[\kappa],(1)}$ and
$\mathbf{V}_h^{[\kappa]}$ approach
$\mathbf{V}_{h,\mathrm{thr}}^{[\kappa],(1)}$ of
Eq.~\eqref{eq:def:Vres:N2LL:thr:qq_gg}, such that the first term
of Eq.~\eqref{eq:def:Vres:N2LL:Reexp} is actually dictated by
$\widetilde{\mathbf{V}}_{h,\mathrm{thr}}^{[\kappa],(1)}$.
Away from the threshold regime, power corrections to
Eq.~\eqref{eq:def:RGE:ReExp} become relevant and its solution
$\widetilde{\mathbf{V}}_{h,\mathrm{thr}}^{[\kappa],(1)}$ gradually
loses its accuracy.
Here, we introduce the transition function ${f}_{\rm{tran}}$ to
switch off their contribution in the well-separated regime, i.e.,
\begin{equation}
\label{eq:def:ftran}
  f_{\rm{tran}}({\betattbar},c_{\mathrm{thr}},r_{\mathrm{thr}}) =
    \begin{cases}
      1\,, & \betattbar\le c_{\rm{thr}}-r_{\mathrm{thr}}\,;\\
    \displaystyle 1-\frac{(\betattbar-c_{\rm{thr}}+r_{\mathrm{thr}})^2}{2r_{\mathrm{thr}}^2} \,,& c_{\rm{m}}-r_{\mathrm{thr}}<\betattbar\le c_{\rm{thr}}\,;\\
         \displaystyle \frac{ (\betattbar-c_{\rm{thr}}-r_{\mathrm{thr}})^2 }{2r_{\mathrm{thr}}^2}\,,& c_{\rm{m}}<\betattbar\le c_{\rm{thr}}+r_{\mathrm{thr}}\,;\\
      0 \,,&  c_{\rm{thr}}+r_{\mathrm{thr}}\le \betattbar \,,
    \end{cases} 
\end{equation}
where the parameters $c_{\rm{thr}}$ and $r_{\mathrm{thr}}$
are introduced to characterize the focal point and the
transition radius, respectively.
To determine their central values and ranges for the uncertainty
estimation for our numerical evaluation in Sec.\ \ref{sec:results},
we compare the numeric
values of $\widetilde{\mathbf{V}}_{h,\mathrm{exp}}^{[\kappa],(1)}$ and
$\mathbf{V}_h^{[\kappa]}$ to determine the range of validity
for the RGE in Eq.~\eqref{eq:def:RGE:ReExp}, for details
see App. \ref{app:vali:Vthr}.
In consequence, we choose
\begin{align}\label{eq:scale:ftran:beta_TT}
  c^{\mathrm{def}}_{\rm{thr}}&=0.4\,,\quad\quad
  r^{\mathrm{def}}_{\mathrm{thr}}  =0.1\,,
\end{align}
as our default choices and use the sets
\begin{align}\label{eq:scale:ftran:beta_TT:var}
\{c_{\rm{thr}},r_{\mathrm{thr}}\}=\{0.35, 0.05\}, \{0.45, 0.15\}\,
\end{align}
to estimate the theoretical uncertainty associated with
our matching procedure.

At variance with the D-\changed{prescription} of Eq.~\eqref{eq:def:Vres:N2LL:Dec},
where the Coulomb singular terms are pushed to a higher
logarithmic order, Eq.~\eqref{eq:def:Vres:N2LL:Reexp} reduces
the threshold divergence by re-exponentiating the \NNLL\
corrections that have been abandoned in the formalism of
Eq.~\eqref{eq:def:Vres:N2LL}.
To this end, we will call the hereby defined \changed{prescription} the R-\changed{prescription}
in the following and label the evolution kernel evaluated via
Eq.~\eqref{eq:def:Vres:N2LL:Reexp} with
$\mathrm{N}^2\mathrm{LL}_{\mathrm{R}}$.
Incorporating Eq.~\eqref{eq:def:Vres:N2LL:Reexp} into
Eqs.~\eqref{eq:LP:res:qqbar} and \eqref{eq:LP:res:gg},
we derive the resummed partonic cross section in the
R-\changed{prescription},
\begin{align}
  \label{eq:LP:res:qqbar:gg:Reexp}
  \begin{split}
  &\hspace*{-5mm}
  \widetilde{\Sigma}_{\ttbar}^{\mathrm{res},[q^i_{n}\qbar^j_{\nbar}]}
  (\bTvec,\Yttbar, \Mttbar,\Omega_t)\Bigg|_{\mathrm{N}^2\mathrm{LL}_{\mathrm{R}}} \,\\
  \,=&\;
  \left(\frac{1}{2N_c}\right)^2\,
  \mathcal{D}^{\mathrm{res}}_{[q^i_{n}\qbar^j_{\nbar}]}
  (\bT,\Mttbar,\mu_h,\mu_b,\mu_s,\nu_b,\nu_s)\\
  &\times
  \sum_{\{n_i,n_i'\}}\,
  \left(\frac{\alpha_s(\mu_s)}{4\pi}\right)^{n_s}\,
  \left(\frac{\alpha_s(\mu_b)}{4\pi}\right)^{n_b+n_b'}\,
  \left(\frac{\alpha_s(\mu_h)}{4\pi}\right)^{n_h+n'_h}\\
  &\hspace*{20mm}\times
  \mathcal{B}_{n}^{[q_n^i],(n_b)}(\eta_n,\bT,\mu_b,\nu_b)\,
  \mathcal{B}_{\nbar}^{[\qbar_\nbar^j],(n'_b)}(\eta_{\nbar},\bT,\mu_b,\nu_b) \, \\
  &\hspace*{20mm}\times
  \sum_{\{\alpha,\beta,h\}}\Bigg\{
  \mathcal{S}^{\alpha_1\beta_1}_{[q_{n}\qbar_{\nbar}],(n_s)}(\bTvec,v_{t},v_{\tbar},\mu_s,\nu_s)
  \left[\mathcal{V}^{[q_{n}\qbar_{\nbar}]}_{\alpha_1\alpha_2}(v_{t},v_{\tbar},\mu_s,\mu_h)\,
  \mathcal{C}^{[q^i_{n}\qbar^j_{\nbar}],(n_h)}_{\alpha_2;h_nh_{\nbar}h_th_{\tbar}} \right]^*\,\\[-4mm]
  &\hspace*{80mm}{}\times
  \left[\mathcal{V}^{[q_{n}\qbar_{\nbar}]}_{\beta_1\beta_2}(v_{t},v_{\tbar},\mu_s,\mu_h)\,
  \mathcal{C}_{\beta_2;h_nh_{\nbar}h_th_{\tbar}}^{[q^i_{n}\qbar^j_{\nbar}],(n_h')}\right]
  \Bigg\}\\
  &\hspace*{20mm}\times
  \theta\Big(1-n_h\Big)\theta\Big(1-n'_h\Big)\theta\Big(1-n_s\Big)\theta\Big(1-n_b\Big)\theta\Big(1-n_b'\Big)\,,
  \\[2mm]
  &\hspace*{-5mm}
  \widetilde{\Sigma}_{\ttbar }^{\mathrm{res},[g_{n}g_{\nbar}]}
  (\bTvec,\Yttbar, \Mttbar,\Omega_t)\Bigg|_{\mathrm{N}^2\mathrm{LL}_{\mathrm{R}}} \,\\
  \,=&\;
  \left(\frac{1}{N^2_c-1}\right)^2\,
  \mathcal{D}^{\mathrm{res}}_{[g_{n}g_{\nbar}]}
  (\bT,\Mttbar,\mu_h,\mu_b,\mu_s,\nu_b,\nu_s)\\
  &\times
  \sum_{\{n_i,n_i'\}}\,
  \left(\frac{\alpha_s(\mu_s)}{4\pi}\right)^{n_s}\,
  \left(\frac{\alpha_s(\mu_b)}{4\pi}\right)^{n_b+n_b'}\,
  \left(\frac{\alpha_s(\mu_h)}{4\pi}\right)^{n_h+n'_h}\\
  &\hspace*{20mm}\times
  \sum_{\{\alpha,\beta, h,h'\}}
  \Bigg\{
      \mathcal{S}^{\alpha_1\beta_1}_{[g_ng_{\nbar}],(n_s)}(\bTvec,v_{t},v_{\tbar},\mu_s,\nu_s)\,
       \mathcal{B}^{[g_n],(n_b)}_{n,h_n'{h_n}}(\eta_n,\bTvec,\mu_b,\nu_b)\,
      \mathcal{B}^{[g_{\nbar}],(n'_b)}_{\nbar,h_{\nbar}'{h_{\nbar}}}(\eta_{\nbar},\bTvec,\mu_b,\nu_b)\hspace*{-20mm}\\[-2mm]
  &\hspace*{40mm}\times\,
      \left[\mathcal{V}^{[g_{n}g_{\nbar}]}_{\alpha_1\alpha_2}(v_{t},v_{\tbar},\mu_s,\mu_h)
      \mathcal{C}^{[g_{n}g_{\nbar}],(n_h)}_{\alpha_2;h'_nh'_{\nbar}h_th_{\tbar}}\right]^*
      \left[\mathcal{V}^{[g_{n}g_{\nbar}]}_{\beta_1\beta_2}(v_{t},v_{\tbar},\mu_s,\mu_h)\,
      \mathcal{C}_{\beta_2;h_nh_{\nbar}h_th_{\tbar}}^{[g_ng_{\nbar}],(n'_h)}\,
      \right]\Bigg\}\hspace*{-20mm}\\
  &\hspace*{20mm}\times
  \theta\Big(1-n_h\Big)\theta\Big(1-n'_h\Big)\theta\Big(1-n_s\Big)\theta\Big(1-n_b\Big)\theta\Big(1-n_b'\Big)\,.
  \end{split}
\end{align}
Here the perturbative correction in each contribution has been
included independently up to NLO, for which the product of the
Heaviside step functions $\theta(1-n_i)$ is introduced to impose
the boundary condition of the \NNLL-level resummation.
Again, $\mathcal{V}^{[\kappa]}_{\alpha\beta}$ denotes the element
in the non-cusp resummation kernel of Eq.~\eqref{eq:def:Vres:N2LL:Reexp}
at the $\alpha$-th row and $\beta$-th column.
Differing from the D-prescription in Eqs.~\eqref{eq:LP:res:qqbar:gg:Dec},
where the product of the NLO fixed-order contributions are pushed to
terms of higher logarithmic order, all of those contributions are
taken into account in the R-\changed{prescription} of Eq.~\eqref{eq:LP:res:qqbar:gg:Reexp}.
In principle, if there were no threshold divergences emerging
from the NLO non-logarithmic terms, these products could be
categorised into the higher logarithmic corrections and should play
a numerically minor role in the resummation.
However, in light of the Coulomb singularity in
Eq.~\eqref{eq:chard:asy:thr:qq:gg} and the threshold enhancement
in Eq.~\eqref{eq:def:Vres:NLL:thr:qq_gg}, the differences in
organising the fixed-order correction between the D- and R-\changed{prescription}s
can \changed{impact} the \qT\ and \dphittbar\ spectra
in the vicinity of $\betattbar=0$ \changed{non-trivially}.
\changed{In Sec.~\ref{sec:res:num}, we will make use of this property
to test the sensitivity of  $\done\sigma_\ttbar/\done\dphi_\ttbar$ and
$\done\sigma_\ttbar/\done\qT$ to the treatment of the Coulomb interactions. }

Taking the threshold limit $\betattbar\to0$, the only singular
contribution in Eq.~\eqref{eq:LP:res:qqbar:gg:Reexp} comes from
the perturbative correction to the hard function,
Eq.~\eqref{eq:chard:asy:thr:qq:gg}, giving rise to a quadratic
divergence in the partonic function,
\begin{align}
\label{eq:LP:res:Reexp:thr:betam2}
  \widetilde{\Sigma}_{\ttbar}^{\mathrm{res},[\kappa]} \Bigg|_{\mathrm{N}^2\mathrm{LL}_{\mathrm{R}}}\xrightarrow[]{\betattbar\to0} \mathcal{O}(\beta^{-2}_{\ttbar})\,.
\end{align}
In comparison with the corresponding expression in the D-\changed{prescription},
Eq.~\eqref{eq:LP:res:Dec:thr:betam1}, the result in
Eq.~\eqref{eq:LP:res:Reexp:thr:betam2} exhibits a stronger
divergence in the threshold limit.
Nevertheless, this divergence can still be accommodated
by the kinematic suppression factors of
Eq.~\eqref{eq:kin:asy:diff:pTt_Mtt},
\changed{allowing us to remove the phase space restriction
$\Mttbar^\text{min}$ in Eq.~\eqref{eq:methods:res:qT_dphi}
when evaluating the single differential observables
$\done\sigma^{\rm{res}}_{\ttbar}/\done\qT$ and
$\done\sigma^{\rm{res}}_{\ttbar}/\done\dphittbar$.
}



\subsection{Matching to fixed-order QCD}
\label{sec:mat}

In the past subsections, we have taken the soft-collinear resummation
\changed{for the $\qT$ and $\Delta\phi_\ttbar$ spectra within
the well-separated domain $\dEttbar\sim\mathcal{O}(m_t)$
and then extended its coverage over the full $M_\ttbar$ range via two ad hoc prescriptions.
During this analysis, we worked at leading-power accuracy,
embedding the most singular behaviour as $\qT\to0$ or $\dphittbar\to0$,
but systematically neglected higher-power corrections.
In the followings, we will restore them in part by matching resummation
to a dedicated fixed-order QCD calculation.%
}
To this end, we introduce a matching procedure between the resummation
and the fixed-order QCD calculation, defined through
\cite{Banfi:2012jm,Banfi:2012yh,Bizon:2018foh}
\begin{equation}\label{eq:def:mat}
  \begin{split}
    \cfrac{\done \sigma_{\ttbar}^{\rm{mat}}}{\done \mathcal{Q}}
    \,\equiv&\;
      \Bigg\{
        \left[
          \cfrac{\done\sigma_{\ttbar}^{\rm{res}}}{\done \mathcal{Q}}
          -\cfrac{\done \sigma_{\ttbar}^{\rm{s}}(\mu_\text{f.o.})}
                 {\done \mathcal{Q}}
        \right]\,
        f_{\rm{tran}}({\mathcal{Q}},c_{\mathrm{m}},r_{\mathrm{m}})
        +\cfrac{\done \sigma_{\ttbar}^{\rm{s}}(\mu_\text{f.o.})}
               {\done \mathcal{Q}}
      \Bigg\}\;
      \mathcal{R}_{\mathrm{fs}}(\mu_\text{f.o.})\\
    =&\;
      f_{\rm{tran}}({\mathcal{Q}},c_{\mathrm{m}} ,r_{\mathrm{m}})\,
      \left(\cfrac{\done\sigma_{\ttbar}^{\rm{res}}}{\done \mathcal{Q}}\right)\,
      \mathcal{R}_{\mathrm{fs}}(\mu_\text{f.o.})\Bigg|_{\mathrm{exp}}
      +\Big\{
         1-{f}_{\rm{tran}}({\mathcal{Q}},c_{\mathrm{m}} ,r_{\mathrm{m}})
       \Big\}\,
       \cfrac{\done \sigma_{\ttbar}^{\rm{f.o.}}(\mu_\text{f.o.})}
             {\done \mathcal{Q}}
      +\dots\,,
\end{split}
\end{equation}
where
$\mathcal{Q}\in\{\qT,\Delta\phi_{\ttbar}\}$ stands for the
observables of our concern.
${\done\sigma_{\ttbar}^{\rm{res}}}/{\done\mathcal{Q}}$ and
${\done\sigma_{\ttbar}^{\rm{s}}}/{\done\mathcal{Q}}$ represent
the resummed differential cross section and its perturbative
expansion evaluated at the fixed-order scale $\mu_\text{f.o.}$.
The modification factor $\mathcal{R}_{\mathrm{fs}}$ is
introduced here to supply the power suppressed contributions
that have been discarded in deriving the resummation in
Eqs.~\eqref{eq:LP:res:qqbar} and \eqref{eq:LP:res:gg}.
It is defined as,
\begin{align}\label{eq:def:mat:R}
  \mathcal{R}_{\mathrm{fs}}(\mu_\text{f.o.})
  =
        \cfrac{{\done\sigma_{\ttbar}^{\mathrm{f.o.}}(\mu_\text{f.o.})}/
           {\done\mathcal{Q}}}
          {{\done\sigma_{\ttbar}^{\mathrm{s}}(\mu_\text{f.o.})}/
           {\done\mathcal{Q}}}\,.  
           \end{align}
Herein, $\done\sigma_{\ttbar}^{\mathrm{f.o.}}/{\done\mathcal{Q}}$
denotes the fixed-order QCD results at the fixed-order scale
$\mu_\text{f.o.}$, which will be appraised by means of
the program \Sherpa~\cite{Gleisberg:2003xi,Gleisberg:2008ta,
  Sherpa:2019gpd,Bothmann:2024wqs}.
In calculating $ \mathcal{R}_{\mathrm{fs}}$, it is worth
noting that starting from \NNLO, the denominator is not positive
definite and exhibits zeros.
In this case, we expand $\mathcal{R}_{\mathrm{fs}}$ in
$\alpha_s(\mu_\text{f.o.})$ in the second step of
Eq.~\eqref{eq:def:mat} following the methodology in
\cite{Bizon:2018foh}.
Throughout our calculation, we will utilise
\begin{align}
\label{eq:scale:mat}
\mu_\text{f.o.}^{\mathrm{def}}=M_\ttbar
\end{align} 
as our default choice but employ the interval $\mu_\text{f.o.}\in[1/2,2]M_\ttbar$ to estimate the theoretical uncertainty.

Again,  the transition function $f_{\rm{tran}}$ is employed
here to progressively fade out the resummation away from the
singular region.
$f_{\rm{tran}}$ in Eq.~\eqref{eq:def:mat} formally takes
the identical form as in Eq.~\eqref{eq:def:ftran}, only being
governed by different arguments $\mathcal{Q}$, $c_{\mathrm{m}}$,
and $r_{\mathrm{m}}$ here.
The latter two parameters are subject to the range of validity
of the leading power approximation, which we determine in Sec.~\ref{sec:results:validation} by comparing
${\done\sigma_{\ttbar}^{\rm{s}}}/{\done\mathcal{Q}}$ and
${\done\sigma_{\ttbar}^{\rm{f.o.}}}/{\done\mathcal{Q}}$.

\section{Numerical Results}
\label{sec:results}

\subsection{Input parameters}
\label{sec:results:inputparams}

In order to validate and evaluate the expressions for the
resummed cross sections of the \qT\ and \dphittbar\ spectra
derived in the last section, we need to specify the following
input parameters, the top quark mass $m_t$, strong coupling
constant $\alpha_s$, and the PDFs.
We define the top quark mass in the pole mass scheme, using
a value of $m_t=173.4\,\text{GeV}$.
This is in line with our adopted UV renormalisation scheme for
the hard sector.
The strong coupling and the PDFs are evaluated by the \LHAPDF
package \cite{Buckley:2014ana,Bothmann:2022thx}, using
the \texttt{NNPDF31\_nnlo\_as\_0118}~\cite{NNPDF:2017mvq}
PDF set with $\alpha_s(m_Z)=0.118$ in the $n_f=5$ light
flavour scheme.

The colour- and helicity-dependent amplitudes inherent in the
partonic functions of Eqs.~\eqsref{eq:LP:res:qqbar}{eq:LP:res:gg},
Eq.~\eqref{eq:LP:res:qqbar:gg:Dec}, and
Eq.~\eqref{eq:LP:res:qqbar:gg:Reexp}, are evaluated using
\Recola~\cite{Actis:2012qn,Actis:2016mpe}, up to \NLO\ accuracy.
After their combination with the soft and beam-collinear
functions as well as the scale evolution kernels, the resulting
resummed cross sections $\widetilde{\Sigma}_{\ttbar}^{\mathrm{res},[\kappa]}$
are integrated over the relevant momentum and impact-parameter
spaces using \texttt{Cuba} \cite{Hahn:2004fe,Hahn:2014fua} to give our
resummed differential spectra
$\done\sigma^{\mathrm{res}}_{\ttbar}/\done\qT$ and
$\done\sigma^{\mathrm{res}}_{\ttbar}/\done\dphittbar$.

Eventually, we match the resummation onto the fixed-order QCD
calculations via Eq.~\eqref{eq:def:mat}.
At \NLO, the fixed order contributions comprise only the
tree-level amplitudes in the domain $\qT>0$ and $\dphittbar>0$,
which can be automatically generated by \Sherpa's
\cite{Gleisberg:2003xi,Gleisberg:2008ta,Sherpa:2019gpd,Bothmann:2024wqs}
built-in matrix element generator \Amegic~\cite{Krauss:2001iv}.
We process its output using \Rivet~\cite{Buckley:2010ar,
  Bierlich:2019rhm,Bierlich:2024vqo} to extract the
observables $\done\sigma^{\mathrm{f.o.}}_{\ttbar}/\done\qT$
and $\done\sigma^{\mathrm{f.o.}}_{\ttbar}/\done\dphittbar$.
To calculate the \NNLO\ contributions, \OpenLoops \cite{Cascioli:2011va,
  Kallweit:2014xda,Buccioni:2017yxi,Buccioni:2019sur} is
interfaced with \Sherpa to calculate the renormalised one-loop
corrections, which is then combined with the real-emission
contribution generated again by~\Amegic within the dipole
subtraction framework for single-parton divergences
\cite{Catani:1996vz,Catani:2002hc,Gleisberg:2007md,
  Schonherr:2017qcj}.

\subsection{Validation}
\label{sec:results:validation}

In the following, we confront the fixed-order expansion of
the resummation in Eq.~\eqref{eq:methods:res:qT_dphi} with
those evaluated in full QCD in order to establish the ability
of our approximate calculation to reproduce the exact result
in the relevant soft-collinear limits.
Before analysing our numerical results in detail, it is
worth noting that the expressions in
Eq.~\eqref{eq:methods:res:qT_dphi} are applicable in the
domain where the top and antitop quarks are well separated
from their threshold production region.
In this domain we are able to apply SCET$+$HQET to extract
the soft and beam-collinear approximation in the low \qT\
and \dphittbar\ regime and thereby exploit the decoupling transformation~\cite{Bauer:2001yt,Beneke:2010da} to
accomplish the factorisation in
Eqs.~\eqsref{eq:LP:res:qqbar}{eq:LP:res:gg}.
However, the situation is different in the threshold region
where \betattbar\ and $\Delta E_{\ttbar}\to0$.
Here, the (Coulomb) potential mode~\cite{Beneke:1997zp}
comes into play via virtual gluon exchanges between the
heavy partons and therefore Eq.~\eqref{eq:methods:res:qT_dphi}
is not directly applicable.
To this end, in the analysis below, we divide the phase space
into three intervals, the threshold region
$\Mttbar\le360\,\text{GeV}$, the transitional region
$\Mttbar\in[360,400]\,\text{GeV}$, and the well-separated region
$\Mttbar\ge400\,\text{GeV}$.
We will use these three regions to examine the quality of the
approximate result in the \qT\ and \dphittbar\ spectra,
probing into the applicability and limitations of
Eq.~\eqref{eq:methods:res:qT_dphi}.

\begin{figure}[t!]
  \centering
  \begin{subfigure}{0.32\textwidth}
    \centering
    \includegraphics[width=.9\linewidth, height=0.98\linewidth]{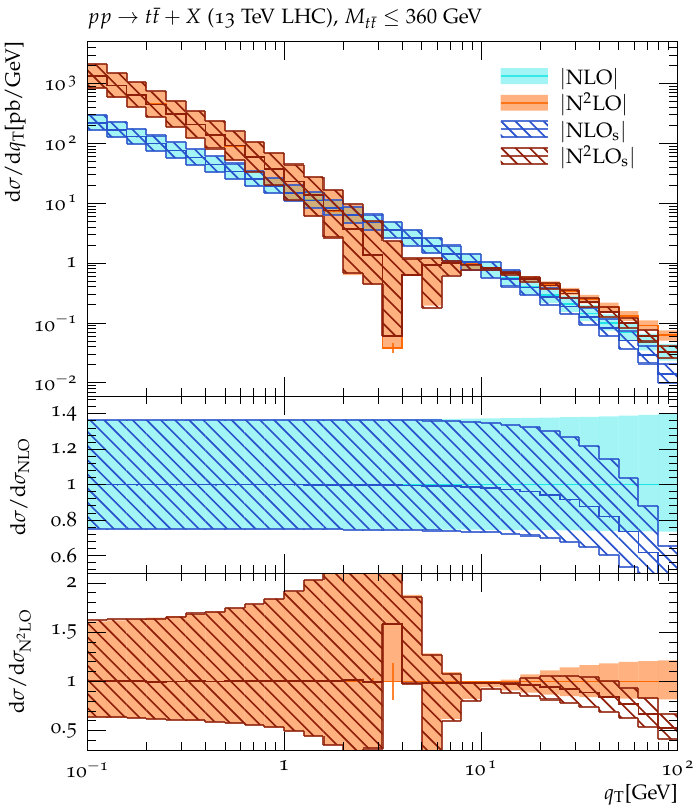}
    \caption{  }
    \label{fig:results:val:qT:a}
  \end{subfigure}
  \begin{subfigure}{0.32\textwidth}
    \centering
    \includegraphics[width=.9\linewidth, height=0.98\linewidth]{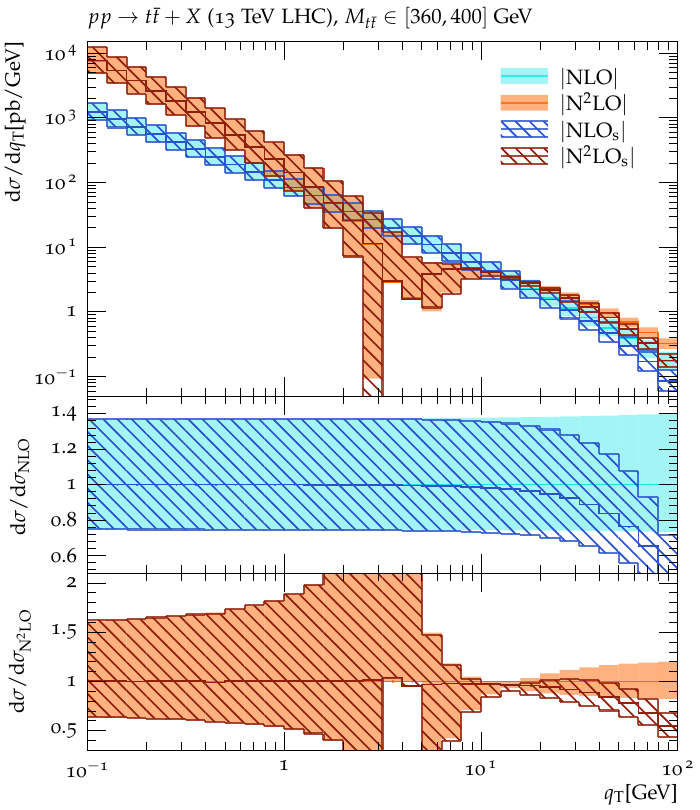}
    \caption{  }
    \label{fig:results:val:qT:b}
  \end{subfigure} 
  \begin{subfigure}{0.32\textwidth}
    \centering
    \includegraphics[width=.9\linewidth, height=0.98\linewidth]{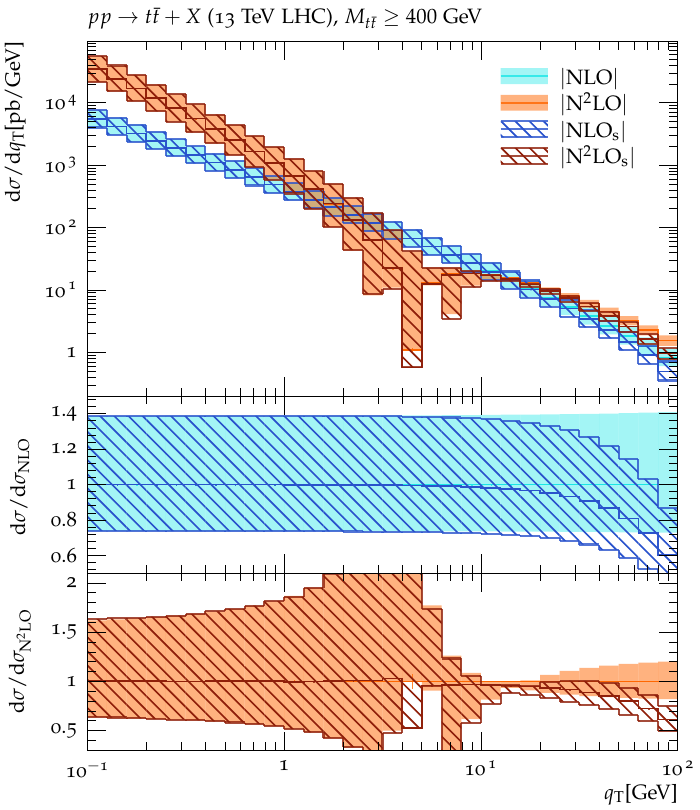}
    \caption{   }
    \label{fig:results:val:qT:c}
  \end{subfigure}
  \caption{
    The transverse momentum spectrum of the \ttbar-pair at fixed-order QCD
    at \NLO\ and \NNLO\ accuracy
    in the process $pp\to\ttbar+X$ at $\sqrt{s}=13\,\text{TeV}$ within the
    intervals $\Mttbar\le 360\,\text{GeV}$ (left),
    $\Mttbar\in[360,400]\,\text{GeV}$ (centre),
    and $\Mttbar\ge 400\,\text{GeV}$ (right).
    The \NLOs\ and \NNLOs\ results
    encode the leading singular behaviour derived
    from SCET$+$HQET.
  }
  \label{fig:results:val:qT}
\end{figure}

We begin our analysis with an examination of the transverse momentum
spectra of the \ttbar-pair in Fig.~\ref{fig:results:val:qT}, where
the differential distributions $\done\sigma_{\ttbar}/\done\qT$ of the
SCET$+$HQET approximation are compared to those derived in full QCD.
Therein, using cyan and apricot, we show the exact fixed-order full QCD
results at \NLO\ and \NNLO, respectively, while the approximations are
illustrated in the blue and red, labeled \NLOs\ and \NNLOs\ likewise.
During their evaluation, we set the renormalisation and factorisation
scales to $\mu_R=\mu_F=\Mttbar$ as our central scale choice and use
the interval $\mu_R=\mu_F\in[2,0.5]\,\Mttbar$ to estimate the theoretical
uncertainties.
We represent the scale uncertainties using corresponding coloured solid
and hatched bands.
In computing \NNLO\ results, we invariably encounter zeros in both the
full QCD and approximate calculations around
$\qT\approx3\,\text{to}\,5\,\text{GeV}$,
inducing significant Monte-Carlo statistical uncertainties shown via
vertical error bars.
Please note, that the distributions to the left of the respective
zero-crossings are negative and we are therefore showing the absolute
values.
 
From the main plots in  Fig.~\ref{fig:results:val:qT}, we observe
that, up to \NNLO, the asymptotic behaviour of the full QCD
calculation is well captured by the leading singular contributions
derived using SCET$+$HQET in the low \qT\ domain for all three
\Mttbar\ slices, including the scale variations. 
As \qT\ increases, power corrections progressively corrupt the
leading singular approximation and enlarge the discrepancies
between \NLO\ and \NLOs, and \NNLO\ and \NNLOs, respectively,
with the deviations becoming appreciable only around
$\qT\ge 20\,\text{GeV}$.
 This phenomenon suggests that, up to \NNLO, \changed{the
(Coulomb) potential region~\cite{Beneke:1997zp} near the \ttbar\
production threshold does not} incur additional
leading singular terms as $\qT\to0$.

To make a more quantitative assessment of the leading power approximation,
the first (second) subplots of  Fig.~\ref{fig:results:val:qT} are
devoted to the ratio between \NLOs\ (\NNLOs) and \NLO\ (\NNLO).
We observe that with only percent level deviations up to $\qT=20\,\text{GeV}$,
the \qT\ spectra derived through SCET$+$HQET manage to describe the
asymptotic behaviour of the full QCD results at both \NLO\ and \NNLO\
precisions.
Further increase in \qT\ increases the deviations between both approaches
as higher-power corrections become increasingly important.
Nevertheless, for $\qT\approx50\,\text{GeV}$, the leading power
approximation can still account for $80\%$ contribution of the exact
differential distributions $\done\sigma_{\ttbar}/\done\qT$.

\begin{figure}[t!]
  \centering
  \begin{subfigure}{0.32\textwidth}
    \centering
    \includegraphics[width=.9\linewidth, height=0.98\linewidth]{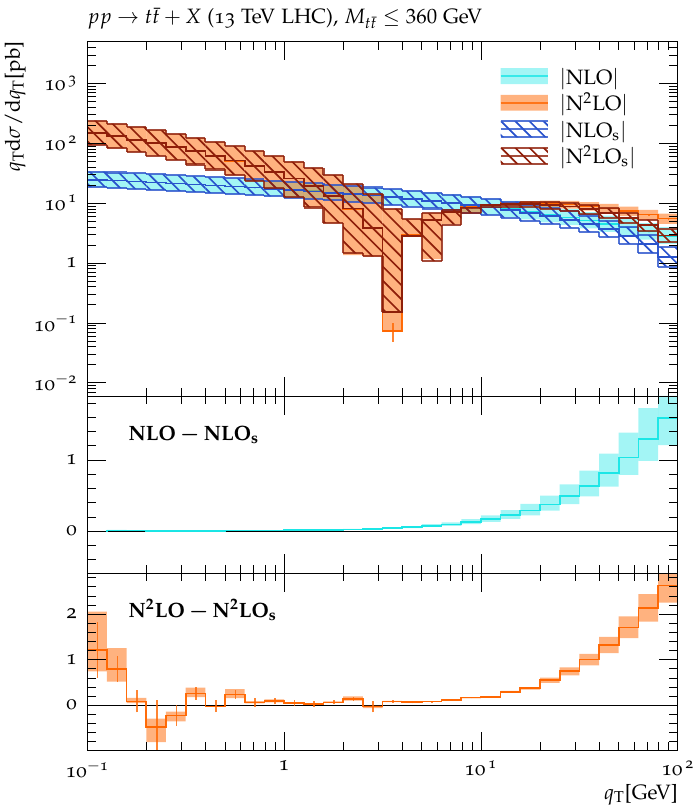}
    \caption{  }
    \label{fig:results:val:qTwt:a}
  \end{subfigure}
  \begin{subfigure}{0.32\textwidth}
    \centering
    \includegraphics[width=.9\linewidth, height=0.98\linewidth]{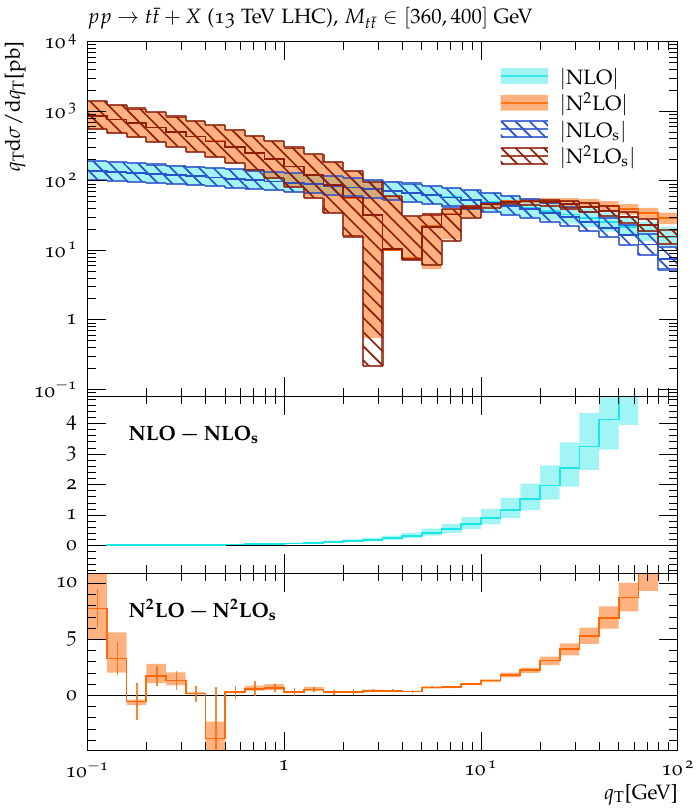}
    \caption{   }
    \label{fig:results:val:qTwt:b}
  \end{subfigure} 
  \begin{subfigure}{0.32\textwidth}
    \centering
    \includegraphics[width=.9\linewidth, height=0.98\linewidth]{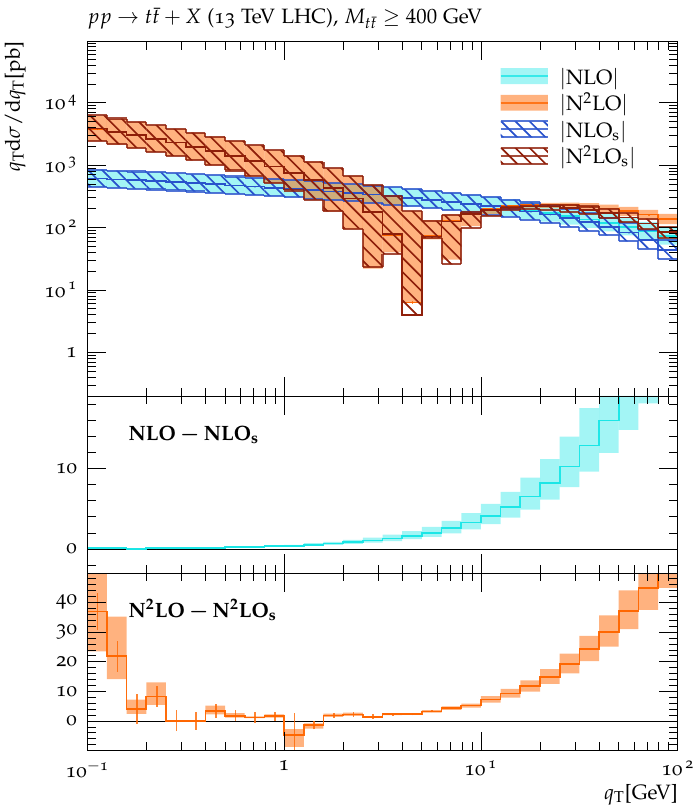}
    \caption{   }
    \label{fig:results:val:qTwt:c}
  \end{subfigure}
  \caption{
    The weighted transverse momentum spectrum of the \ttbar-pair
    at fixed-order QCD at \NLO\ and \NNLO\ accuracy
    in the process $pp\to\ttbar+X$ at $\sqrt{s}=13\,\text{TeV}$ within the
    intervals $\Mttbar\le 360\,\text{GeV}$ (left),
    $\Mttbar\in[360,400]\,\text{GeV}$ (centre),
    and $\Mttbar\ge 400\,\text{GeV}$ (right).
    The \NLOs\ and \NNLOs\ results
    encode the leading singular behaviour derived
    from SCET$+$HQET.
  }
  \label{fig:results:val:qTwt}
\end{figure}

In pursuit of further ascertaining the SCET$+$HQET prediction, we
investigate the weighted differential distributions
$\done\sigma_{\ttbar}/\done \ln \qT$ in the low $\qT\to 0$ regime,
which are expected to observe the power series,
\begin{align}\label{eq:AsyExp:qTweighted}
  \frac{\done{\sigma_{\ttbar}}}{\done \ln \qT}
  =
  \qT\,\frac{\done{\sigma_{\ttbar}}}{\done\qT}
  \,\sim\;
    \sigma_\ttbar^\text{LO}\,
    \sum_{m,n}\,
    \left(\frac{\alpha_s }{4\pi}\right)^m\,
    \left[\,
      \underbrace{c^{(0)}_{m,n}\,{\ln^{n}(\qT)}}_{\mathrm{LP}}
      +\,\underbrace{c^{(1)}_{m,n}\,\qT\ln^{n}(\qT)}_{\mathrm{NLP}}
      +\,\underbrace{c^{(2)}_{m,n}\,\qT^2 \ln^{n}(\qT)}_{\mathrm{N}^2\mathrm{LP}}
      +\dots
    \right]\,,
\end{align}
where $\sigma_\ttbar^\text{LO}$ is the LO cross section
and the $c_{m,n}^{(i)}$ are the coefficients at the respective
order of the expansion.
The numerical results for $\done\sigma_{\ttbar}/\done\ln\qT$ are
presented in Fig.~\ref{fig:results:val:qTwt}.
Differing from the findings of Fig.~\ref{fig:results:val:qT},
where acute enhancements are showcased in the low \qT\ region,
the asymptotic behaviour in $\done\sigma_{\ttbar}/\done\ln\qT$
is alleviated as compared to that of $\done\sigma_{\ttbar}/\done\qT$
by the application of the weighting factor $\qT$.
To examine whether the leading power behaviour can be entirely
replicated by Eq.~\eqref{eq:methods:res:qT_dphi}, we exhibit the
difference between the full QCD results and the EFT ones in the
first and second subgraphs of Fig.~\ref{fig:results:val:qTwt}.
At \NLO, their difference declines monotonously as \qT\ decreases
in all three \Mttbar\ intervals, demonstrating that the leading
power contributions are indeed subtracted by the fixed-order
expansion of Eq.~\eqref{eq:methods:res:qT_dphi}.
Regarding the interval $\qT\in[0.3,100]~$GeV, analogous scenarios
can also be found at \NNLO\ from the bottom subgraphs in
Fig.~\ref{fig:results:val:qTwt}, thereby justifying the EFT
results from Eq.~\eqref{eq:methods:res:qT_dphi}.
However, further decreasing \qT\ incurs non-negligible Monte-Carlo
statistical uncertainties, giving rise to deviations between the
exact and approximate calculations of up to $\sim\!3$ times the
variance estimated there, but still within (sub)percent level
relative accuracy w.r.t.\ the magnitude of
$\done\sigma_{\ttbar}/\done \ln \qT$.

\begin{figure}[t!]
  \centering
  \begin{subfigure}{0.42\textwidth}
    \centering
    \includegraphics[width=.9\linewidth, height=0.98\linewidth]{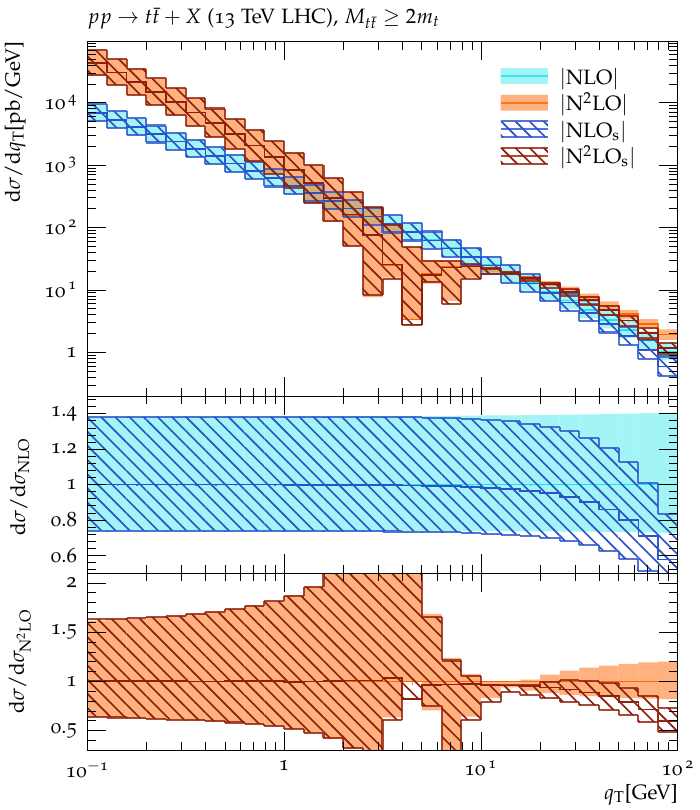}
    \caption{   }
    \label{fig:results:val:qT:FPS}
  \end{subfigure}
  \begin{subfigure}{0.42\textwidth}
    \centering
    \includegraphics[width=.9\linewidth, height=0.98\linewidth]{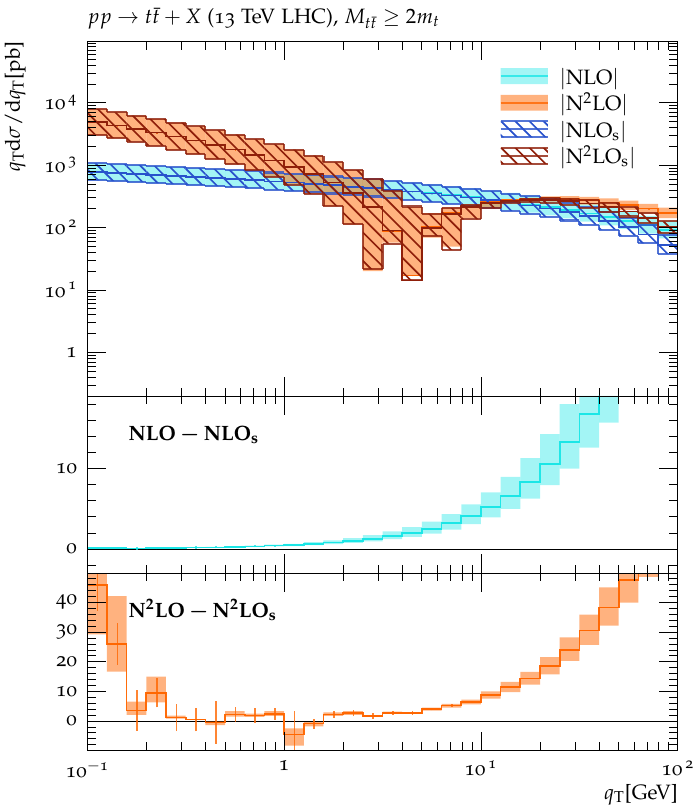}
    \caption{   }
    \label{fig:results:val:qTwt:FPS}
  \end{subfigure}
  \caption{
    The transverse momentum (left) and weighted transverse momentum (right)
    spectra of the \ttbar-pair
    at fixed-order QCD at \NLO\ and \NNLO\ accuracy
    in the process $pp\to\ttbar+X$ at $\sqrt{s}=13\,\text{TeV}$
    in the full phase space.
    The \NLOs\ and \NNLOs\ results
    encode the leading singular behaviour derived
    from SCET$+$HQET.
  }
  \label{fig:results:val:qT:FPS:all}
\end{figure}

Finally, we exhibit the \qT\ and weighted \qT\ distributions in
Fig.~\ref{fig:results:val:qT:FPS:all}, evaluated over the full
phase space.
Unsurprisingly, the behaviours observed in
Fig.~\ref{fig:results:val:qT:FPS} and
Fig.~\ref{fig:results:val:qTwt:FPS} closely resemble those
found in Fig.~\ref{fig:results:val:qT:c} and
Fig.~\ref{fig:results:val:qTwt:c}, respectively, as the slice
$\Mttbar\ge400\,\text{GeV}$ accounts for the bulk of contributions
in the phase space integrals.
With these findings, we are now in a position to determine the
coefficients $c_{\mathrm{m}}$ and $r_{\mathrm{m}}$ comprised
in the arguments of the transition function $f_{\mathrm{tran}}$ of
Eq.~\eqref{eq:def:mat} governing our matching procedure.
From the analysis above, we find that for all three invariant-mass
slices, the leading power approximation of
Eq.~\eqref{eq:methods:res:qT_dphi} is capable of reproducing
the asymptotic behaviour of the exact QCD calculation up to
$\qT\sim10~$GeV within percent level accuracy.
At a level of $\sim\!80\%$ of the full theory, this holds until
$\qT\sim50\,\text{GeV}$.
In light of this, we will make use of
\begin{align}\label{eq:scale:ftran:qT:central}
  \{c_{\rm{m}},r_{\mathrm{m}}\}&=\{50\,\text{GeV},35\,\text{GeV}\}
\end{align}
as our default choice during the implementation of the matching
procedure.
With these parameters, the resummation in Eq.~\eqref{eq:methods:res:qT_dphi}
is fully retained until $\qT=15\,\text{GeV}$, after which the
transition function $f_{\mathrm{tran}}$ phases out the resummation
gradually, reducing it to half its size at
$\qT=50\,\text{GeV}$ and completely terminating it at $\qT=85\,\text{GeV}$.
In order to estimate the theoretical uncertainties associated with
our matching procedure, we adopt the following alternative matching
parameters
\begin{align}
\label{eq:scale:ftran:qT:var}
  \{c_{\rm{m}},r_{\mathrm{m}}\}&=\{45\,\text{GeV},30\,\text{GeV}\},\;\{55\,\text{GeV},40\,\text{GeV}\} \,
\end{align}
and construct an envelope of the calculated spectra.

\begin{figure}[t!]
  \centering
  \begin{subfigure}{0.32\textwidth}
    \centering
    \includegraphics[width=.9\linewidth, height=0.98\linewidth]{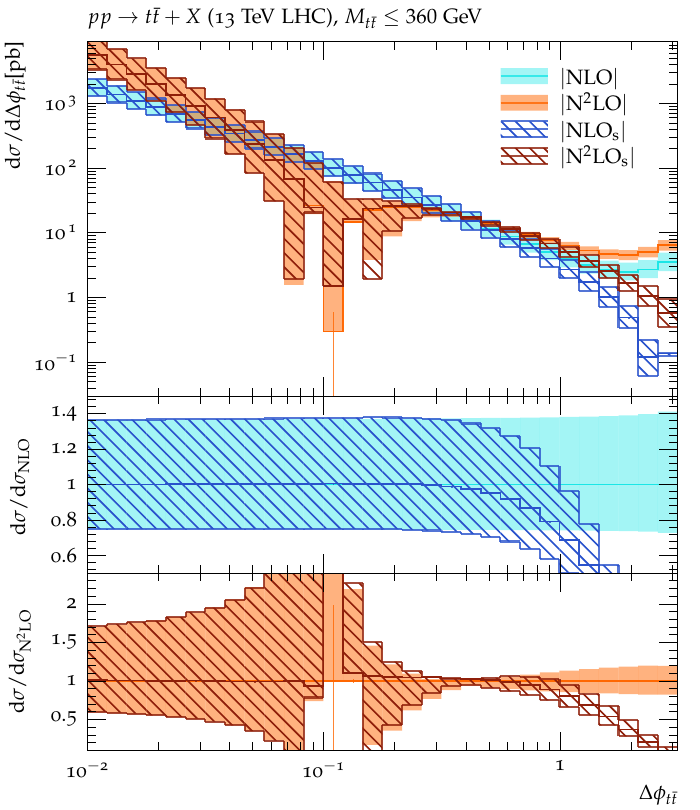}
    \caption{  }
    \label{fig:results:val:dphi:a}
  \end{subfigure}
  \begin{subfigure}{0.32\textwidth}
    \centering
    \includegraphics[width=.9\linewidth, height=0.98\linewidth]{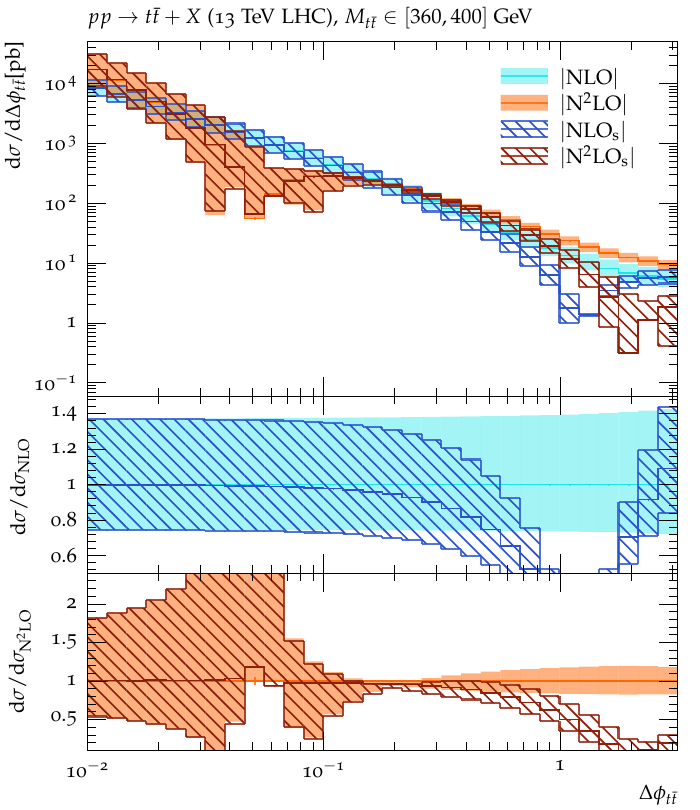}
    \caption{  }
    \label{fig:results:val:dphi:b}
  \end{subfigure} 
  \begin{subfigure}{0.32\textwidth}
    \centering
    \includegraphics[width=.9\linewidth, height=0.98\linewidth]{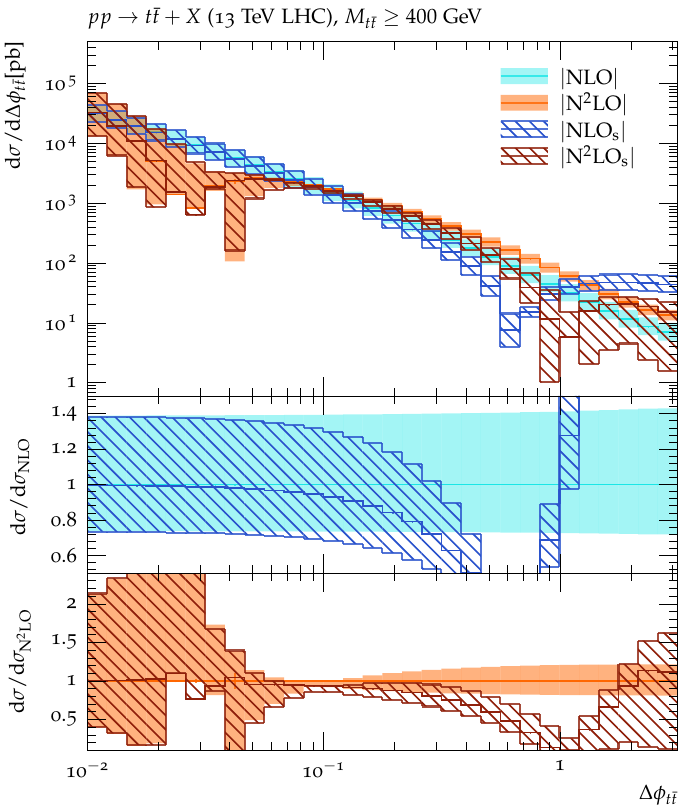}
    \caption{   }
    \label{fig:results:val:dphi:c}
  \end{subfigure}
  \caption{
    The spectrum of the azimuthal separation of the \ttbar-pair
    at fixed-order QCD at \NLO\ and \NNLO\ accuracy
    in the process $pp\to\ttbar+X$ at $\sqrt{s}=13\,\text{TeV}$ within the
    intervals $\Mttbar\le 360\,\text{GeV}$ (left),
    $\Mttbar\in[360,400]\,\text{GeV}$ (centre),
    and $\Mttbar\ge 400\,\text{GeV}$ (right).
    The \NLOs\ and \NNLOs\ results
    encode the leading singular behaviour derived
    from SCET$+$HQET.
  }
  \label{fig:results:val:dphi}
\end{figure}

\begin{figure}[t!]
  \centering
  \begin{subfigure}{0.32\textwidth}
    \centering
    \includegraphics[width=.9\linewidth, height=0.98\linewidth]{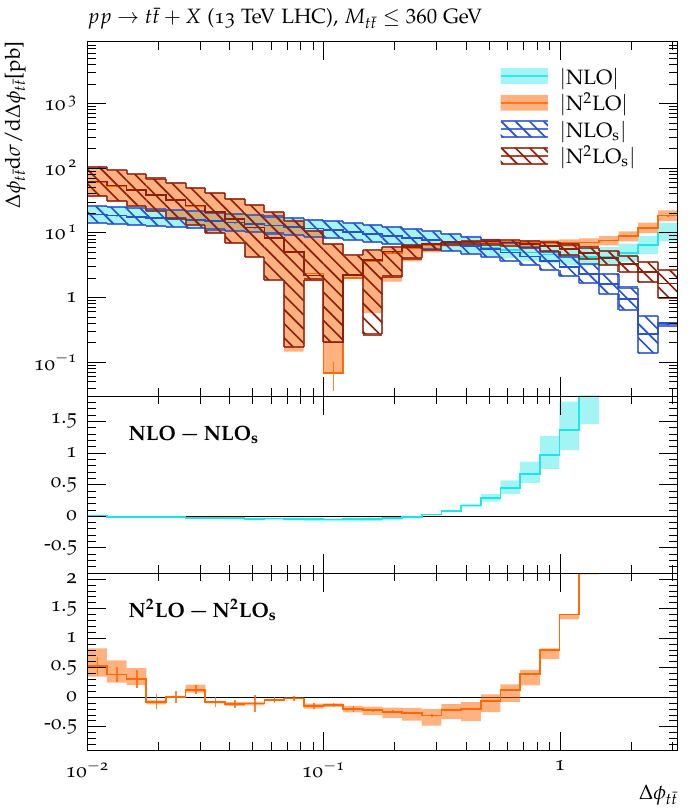}
    \caption{  }
    \label{fig:results:val:dphiwt:a}
  \end{subfigure}
  \begin{subfigure}{0.32\textwidth}
    \centering
    \includegraphics[width=.9\linewidth, height=0.98\linewidth]{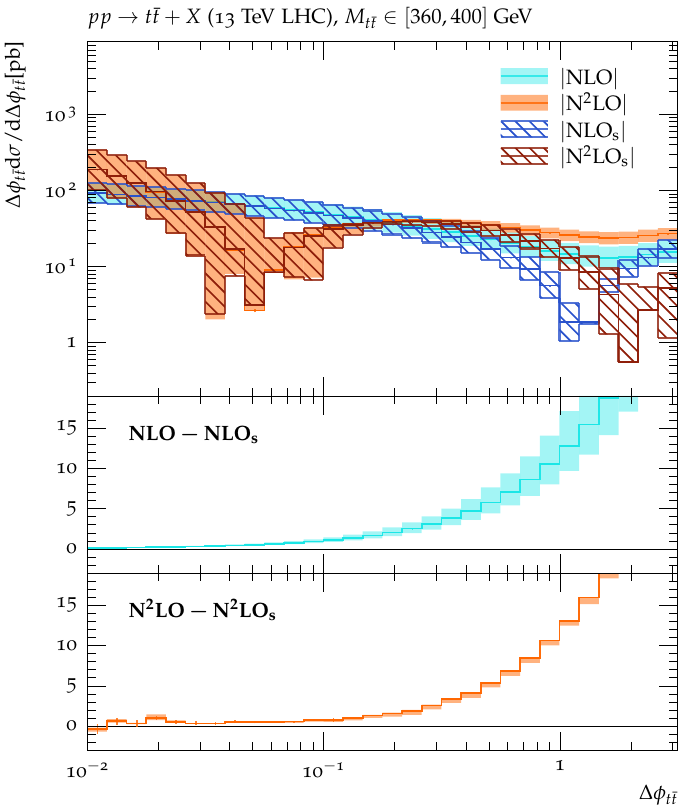}
    \caption{  }
    \label{fig:results:val:dphiwt:b}
  \end{subfigure} 
  \begin{subfigure}{0.32\textwidth}
    \centering
    \includegraphics[width=.9\linewidth, height=0.98\linewidth]{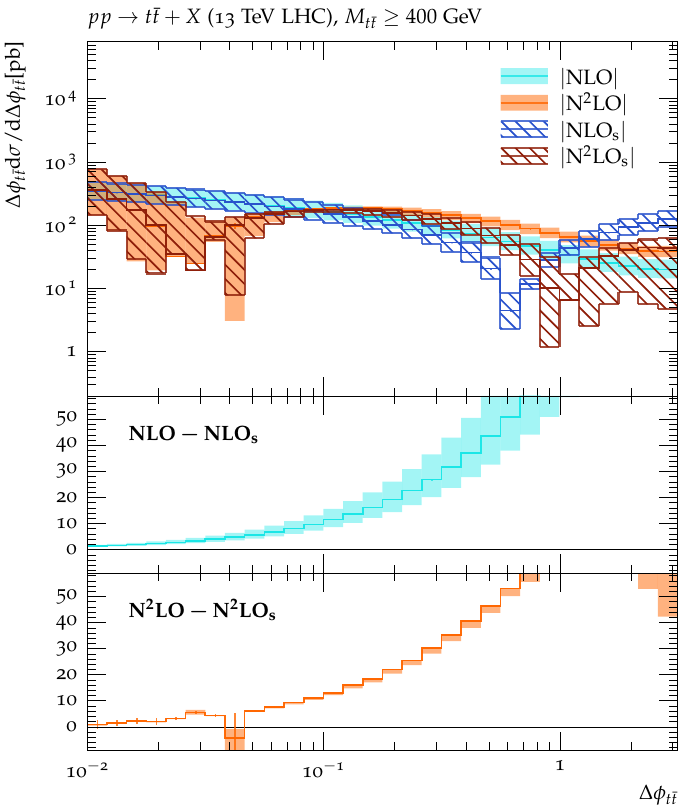}
    \caption{   }
    \label{fig:results:val:dphiwt:c}
  \end{subfigure}
  \caption{
    The weighted spectrum of the azimuthal separation of the \ttbar-pair
    at fixed-order QCD at \NLO\ and \NNLO\ accuracy
    in the process $pp\to\ttbar+X$ at $\sqrt{s}=13\,\text{TeV}$ within the
    intervals $\Mttbar\le 360\,\text{GeV}$ (left),
    $\Mttbar\in[360,400]\,\text{GeV}$ (centre),
    and $\Mttbar\ge 400\,\text{GeV}$ (right).
    The \NLOs\ and \NNLOs\ results
    encode the leading singular behaviour derived
    from SCET$+$HQET.
  }
  \label{fig:results:val:dphiwt}
\end{figure}

We now move on to the fixed-order results for the spectra of the
azimuthal separation of the top and anti-top, \dphittbar,
displayed in Fig.~\ref{fig:results:val:dphi}.\footnote{
  It should be noted that the results in
  Figs.~\ref{fig:results:val:dphi:c} and
  \ref{fig:results:val:dphiwt:c}, illustrating the region
  $\Mttbar\ge 400\,\text{GeV}$ have already been evaluated
  in \cite{Ju:2022wia}. To facilitate the comparison and later
  discussion, however, we exhibit them here once again.}

Akin to the \qT\ spectra in Fig.~\ref{fig:results:val:qT},
the $\done\sigma_{\ttbar}/\done\dphittbar$ in the main plots of
Fig.~\ref{fig:results:val:dphi} exhibit a similarly singular behaviour
in the low \dphittbar\ domain, at both \NLO\ and \NNLO.
However, comparing the exact and approximate results, even though
the SCET$+$HQET calculations are able to reproduce the correct
asymptotic behaviour in the region $\dphittbar\to 0$ in all
three \Mttbar\ slices, the size of the missing power corrections
are markedly larger than in the \qT\ case.
To be precise, at \NLO, while the SCET$+$HQET approximation agrees with the
full QCD result within percent level accuracy below
$\qT=10\,\text{GeV}$ in the \qT\ spectra in all three \Mttbar\ regions of
Fig.~\ref{fig:results:val:qT}, the \dphittbar\ distributions of
Fig.~\ref{fig:results:val:dphi} show a deviation of the EFT-based
approximation from the full theory of a few permille around
$\dphittbar\sim0.2$ in the $\Mttbar\le 360\,\text{GeV}$ slice,
increasing to approximately $5\%$ in $\Mttbar\in[360,400]\,\text{GeV}$,
and reaching more than $10\%$ in the interval $\Mttbar\ge 400\,\text{GeV}$.
An analogous behaviour can be found in the \NNLO\ results, although
the region where the approximate results deviate from the exact one
is shifted to slightly higher \dphittbar, around $\dphittbar\sim1$.
To interpret this phenomenon, it merits reminding that the derivation
of Eq.~\eqref{eq:methods:res:qT_dphi} is
subject to an asymptotic expansion of the differential cross section
in a given kinematic parameter, for instance,
$\lambda_{\mathrm{T}}=\qT/Q_h$ in the $\qT$ spectra and
$\lambda_{\tau}=\dphittbar\widetilde{P}_t^{\bot}/Q_h$ in the
\dphittbar\ ones, where $Q_h$ represents a hard scale of the
similar magnitude to \Mttbar\ and $m_t$.
In Fig.~\ref{fig:results:val:qT}, focusing on a constant value of
\qT, $\lambda_{\mathrm{T}}$ varies gently as $M_{\ttbar}$ changes when
progressing through our three slices since the PDFs
effectively suppress contributions from, individually, highly boosted top
and antitop quarks.
However, the situation for \dphittbar\ in Fig.~\ref{fig:results:val:dphi}
is different as $\lambda_{\tau}$ is sensitive to the variable
$\widetilde{P}_t^{\bot}$, the transverse momentum
of the top quark measured in the rest frame of the \ttbar\ system,
and is therefore proportional to \betattbar\ in the threshold
domain.
In turn, \betattbar\ scales with \Mttbar\ in the left panel, and thus
takes the typical values of around $0.2$ for $\Mttbar<360\,\text{GeV}$,
rising to values of about $0.5$ for $\Mttbar\in[360,400]\,\text{GeV}$,
and ultimately reaching $\betattbar\sim\mathcal{O}(1)$ for
$\Mttbar>400\,\text{GeV}$.
As a consequence of this additional kinematic suppression upon the
expansion parameter $\lambda_{\tau}$, weaker power corrections are
observed in Fig.~\ref{fig:results:val:dphi:a} than in
Fig.~\ref{fig:results:val:dphi:b}, with the largest power
corrections found in Fig.~\ref{fig:results:val:dphi:c},
for constant \dphittbar.

As before, we further assess the quality of the leading power
approximation by studying the weighted differential distributions
$\done\sigma_{\ttbar}/\done \ln \dphittbar$.
They are expected to observe the following power series in the
asymptotic domain,
\begin{align}\label{eq:AsyExp:dphiweighted}
  \begin{split}
  \lefteqn{
    \frac{\done{\sigma_{\ttbar}}}{\done \ln \dphittbar}
    =
    \dphittbar\,\frac{\done{\sigma_{\ttbar}}}{\done\dphittbar}
  }\\
  &
  \sim\;
    \sigma_\ttbar^\text{LO}\,
    \sum_{m,n}\,
    \left(\frac{\alpha_s }{4\pi}\right)^m\,
    \left[\,
      \underbrace{\tilde{c}^{(0)}_{m,n}\,{\ln^{n}(\dphittbar)}}_{\mathrm{LP}}
      +\,\underbrace{\tilde{c}^{(1)}_{m,n}\,\dphittbar\ln^{n}(\dphittbar)}_{\mathrm{NLP}}
      +\,\underbrace{\tilde{c}^{(2)}_{m,n}\,\dphittbar^2 \ln^{n}(\dphittbar)}_{\mathrm{N}^2\mathrm{LP}}
      +\dots
    \right]\,,
  \end{split}
\end{align}
wherein again $\sigma_\ttbar^\text{LO}$ is the LO cross section
and the $\tilde{c}_{m,n}^{(i)}$ are the coefficients at the
respective order of the expansion.
Numerical results for $\done\sigma_{\ttbar}/\done \ln \dphittbar$
in our three \Mttbar\ slices are displayed in
Fig.~\ref{fig:results:val:dphiwt}.
From the main plots therein, we find in all three regions that
the approximate SCET$+$HQET calculation can reproduce the desired
singular behaviour also of the weighted \dphittbar\ spectra of the
full QCD calculations at both \NLO\ and \NNLO.
Similarly, as illustrated in the middle and bottom subplots of
Fig.~\ref{fig:results:val:dphiwt}, their difference  progressively
decreases as \dphittbar\ decreases from $\dphittbar\sim\mathcal{O}(1)$,
until in the $\dphittbar\sim\mathcal{O}(10^{-2})$ non-negligible
integration uncertainties are encountered.
These observations demonstrate that at least up to \NNLO, the
fixed-order expansion of Eq.~\eqref{eq:methods:res:qT_dphi}
is able to describe the leading singular contributions of the
full theory.

\begin{figure}[t!]
  \centering
      \begin{subfigure}{0.42\textwidth}
    \centering
    \includegraphics[width=.9\linewidth, height=0.98\linewidth]{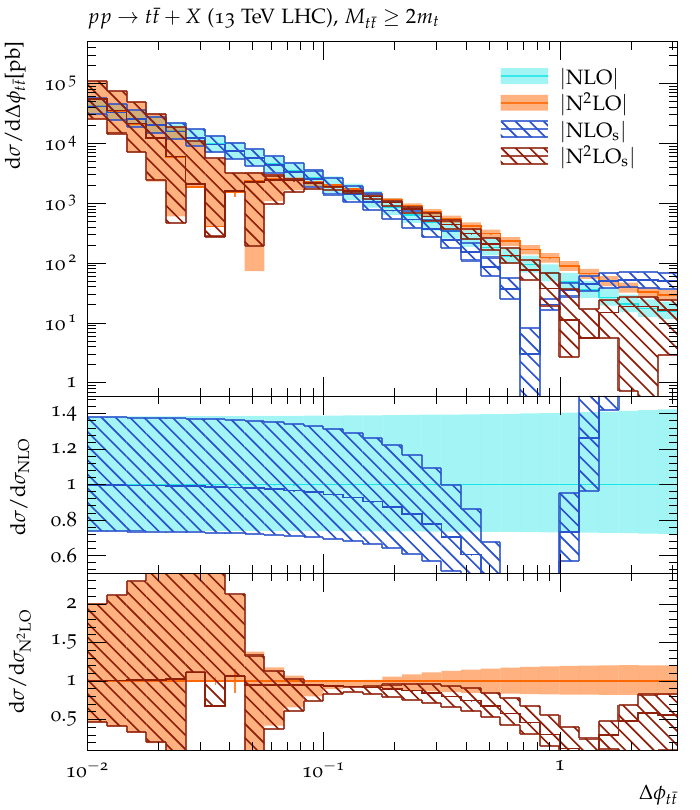}
    \caption{   }
    \label{fig:results:val:dphi:FPS}
  \end{subfigure}
      \begin{subfigure}{0.42\textwidth}
    \centering
    \includegraphics[width=.9\linewidth, height=0.98\linewidth]{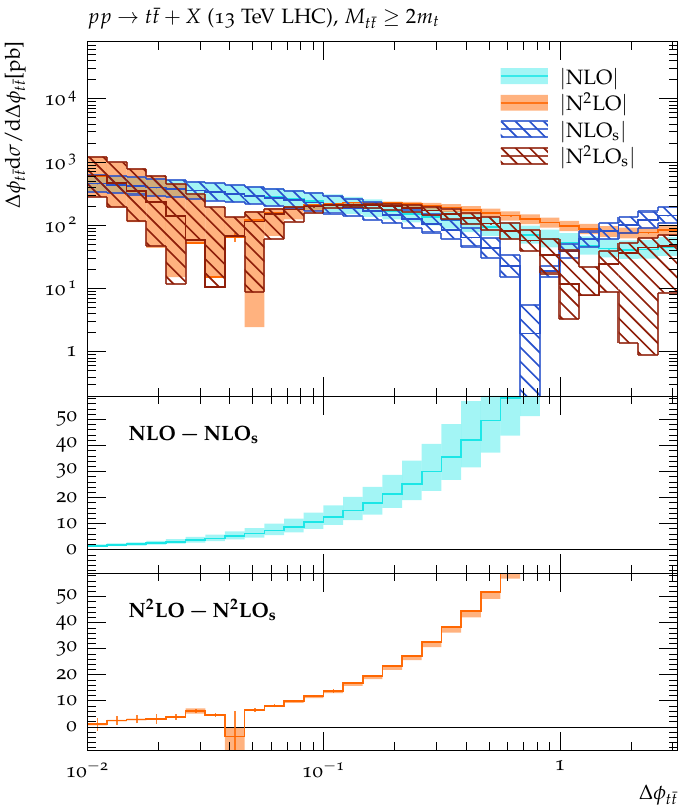}
    \caption{   }
    \label{fig:results:val:dphiwt:FPS}
  \end{subfigure}
  \caption{
    The spectrum (left) and weighted spectrum (right) of the
    azimuthal separation of the \ttbar-pair
    at fixed-order QCD at \NLO\ and \NNLO\ accuracy
    in the process $pp\to\ttbar+X$ at $\sqrt{s}=13\,\text{TeV}$ in the full phase space.
    The \NLOs\ and \NNLOs\ results
    encode the leading singular behaviour derived
    from SCET$+$HQET.
  }
  \label{fig:results:val:qT:dphi:FPS}
\end{figure}

Again, in a parallel to our appraisal of the approximate \qT\ spectra,
we examine the inclusive \dphittbar\ and weighted \dphittbar\ spectra
in Fig.~\ref{fig:results:val:qT:dphi:FPS}.
Once again, they effectively
reproduce the results for $\Mttbar>400\,\text{GeV}$ as this region
carries the bulk of the cross section.
With its help we can now determine the coefficients
$\tilde{c}_{\mathrm{m}}$ and $\tilde{r}_{\mathrm{m}}$ for the
transition function $f_{\mathrm{tran}}$ of Eq.~\eqref{eq:def:mat},
employed to match the resummed \dphittbar\ spectrum to its fixed-order
counter-part.
Considering that the size of the power corrections in the \dphittbar\
distribution is \Mttbar-dependent, the values of
$\tilde{c}_{\mathrm{m}}$ and $\tilde{r}_{\mathrm{m}}$ are chosen
differently for each region, i.e.,
\begin{equation}\label{eq:scale:ftran:dphi:central}
  \begin{array}{ll}
    \{\tilde{c}_{\rm{m}},\tilde{r}_{\mathrm{m}}\}&=\{0.5,0.3\}\,,
    \qquad\qquad
    \Mttbar\le 360~\mathrm{GeV}\,,\\
    \{\tilde{c}_{\rm{m}},\tilde{r}_{\mathrm{m}}\}&=\{0.3,0.2\}\,,\qquad\qquad
    \Mttbar\ge  360~\mathrm{GeV}\,.
  \end{array}
\end{equation}
Therein, in view of the excellent agreement between the approximate
and exact results in Fig.~\ref{fig:results:val:dphi:a}, we extend
the active range of the soft and collinear resummation in the region $M_{\ttbar}<360\,\text{GeV}$.
In consequence, here, the resummation is fully active for
$\dphittbar<0.2$ and then will be gradually turned off,
being reduced to half its strength at $\dphittbar=0.5$
and eliminated at $\dphittbar=0.8$.
Otherwise, a tightened choice of $\tilde{c}_{\mathrm{m}}$
and $\tilde{r}_{\mathrm{m}}$ is made for both other invariant
mass domains.
Here, the resummation is restricted to the region
$\dphittbar<0.1$, reduced to half-value at $\dphittbar=0.3$,
and terminated for $\dphittbar>0.5$.
To investigate the sensitivity of the final matched \dphittbar\ results
to the choice of $f_{\mathrm{tran}}$, we also embed the following
alternatives as matching parameters,
\begin{equation}\label{eq:scale:ftran:dphi:var}
  \begin{array}{ll}
    \{\tilde{c}_{\rm{m}},\tilde{r}_{\mathrm{m}}\}&=
    \{0.45,0.25\}\,,\{0.55,0.35\}\,,
    \qquad\qquad
    \Mttbar\le 360~\mathrm{GeV}\,,\\
    \{\tilde{c}_{\rm{m}},\tilde{r}_{\mathrm{m}}\}&=
    \{0.25,0.15\}\,,\{0.35,0.25\}\,,
    \qquad\qquad
    \Mttbar\ge  360~\mathrm{GeV}\,.
  \end{array}
\end{equation}

\subsection{Resummation-improved \texorpdfstring{\qT}{qT} and \texorpdfstring{\dphittbar}{dphi} distributions}
\label{sec:res:num}
  
In the following, we introduce the resummation-improved
\qT\ and \dphittbar\ spectra based on
Eq.~\eqref{eq:methods:res:qT_dphi}, including the
\changed{threshold region $\betattbar\to0$ by employing
two \textit{ad hoc} prescriptions, i.e.\ the D-\changed{prescription} of Eq.~\eqref{eq:LP:res:qqbar:gg:Dec} and the R-\changed{prescription} of Eq.~\eqref{eq:LP:res:qqbar:gg:Reexp}.
Therein,} as illustrated in Eqs.~\eqsref{eq:LP:res:qqbar}{eq:LP:res:gg},
our R(a)GE-based resummation is subject to two sets of auxiliary
scales, $\{\mu_h,\mu_b,\mu_s\}$ and $\{\nu_b,\nu_s\}$,
characterising the typical scales in the virtuality and rapidity
renormalisation, respectively.
In addition, the matching procedure of Eq.~\eqref{eq:def:mat},
introduces the fixed-order scale $\mu_\text{f.o.}$.
Their default choices are presented in
Eq.~\eqref{eq:scale:nat:qT_dphi} and Eq.~\eqref{eq:scale:mat}.
To estimate the corresponding theoretical uncertainties, we vary
all such scales within the intervals
$\mu\in[\tfrac{1}{2},2]\,\mu^{\mathrm{def}}\;(\mu=\mu_h,\mu_s,\mu_b,\mu_\text{f.o})$ and
$\nu\in[\tfrac{1}{2},2]\,\nu^{\mathrm{def}}\;(\nu=\nu_s,\nu_b)$.
We denote the resulting variation as $\delta_{\mathrm{scale}}$.
Furthermore, our matching procedure of Eq.~\eqref{eq:def:mat}
also introduces the coefficients
$\{c_{\rm{m}},r_{\mathrm{m}}\}$ (for \qT) and
$\{\tilde{c}_{\rm{m}},\tilde{r}_{\mathrm{m}}\}$ (for \dphittbar)
governing the active range of the soft and beam-collinear resummation.
Similarly, while the D-\changed{prescription} does not introduce further parameters,
the R-\changed{prescription} involves a second matching, see
Eq.~\eqref{eq:def:Vres:N2LL:Reexp}, as it embeds terms to mitigate
the threshold singularity in the resummation kernel.
Its associated parameters are
$\{c_{\rm{thr}},r_{\mathrm{thr}}\}$.
Their default choice has been presented in
Eqs.~\eqref{eq:scale:ftran:qT:central},
\eqref{eq:scale:ftran:dphi:central}, and
\eqref{eq:scale:ftran:beta_TT}, respectively.
We estimate the uncertainty of the corresponding matching procedure
using alternative matching parameter as defined in
Eqs.~\eqref{eq:scale:ftran:qT:var},
\eqref{eq:scale:ftran:dphi:var}, and
\eqref{eq:scale:ftran:beta_TT:var},
giving the combined matching uncertainty estimate
$\delta_{\mathrm{mat}}$.
Finally, both sources of uncertainties, $\delta_{\mathrm{scale}}$ and $\delta_{\mathrm{mat}}$, are combined in quadrature, giving the total
uncertainty, \footnote{\changed{%
  We note here that we have introduced a numerical cutoff into
  the impact-parameter space integrals of Eq.~\eqref{eq:methods:res:qT_dphi}
  to evade the QCD Landau divergence \cite{Neill:2015roa},
  $\bT^\text{cut}=2\,\text{GeV}^{-1}$.
  To estimate the uncertainties associated with this choice, we also
  evaluate \qT\ and \dphittbar\ spectra with the alternatives
  $\bT^\text{cut}=3\,\text{GeV}^{-1}$ and
  $\bT^\text{cut}=4\,\text{GeV}^{-1}$.
  We find the variation generally to be on the permille level,
  comparable to the statistical error from the Monte-Carlo integration.
  Therefore, we do not take them into account when calculating
  $\delta_{\mathrm{tot}}$ here.
}}
\begin{align}
\delta_{\mathrm{tot}}=\sqrt{\delta^2_{\mathrm{mat}}+\delta^2_{\mathrm{scale}}}\;.
\end{align}

\begin{figure}[t!]
  \centering
  \begin{subfigure}{0.42\textwidth}
    \centering
    \includegraphics[width=.9\linewidth, height=0.98\linewidth]{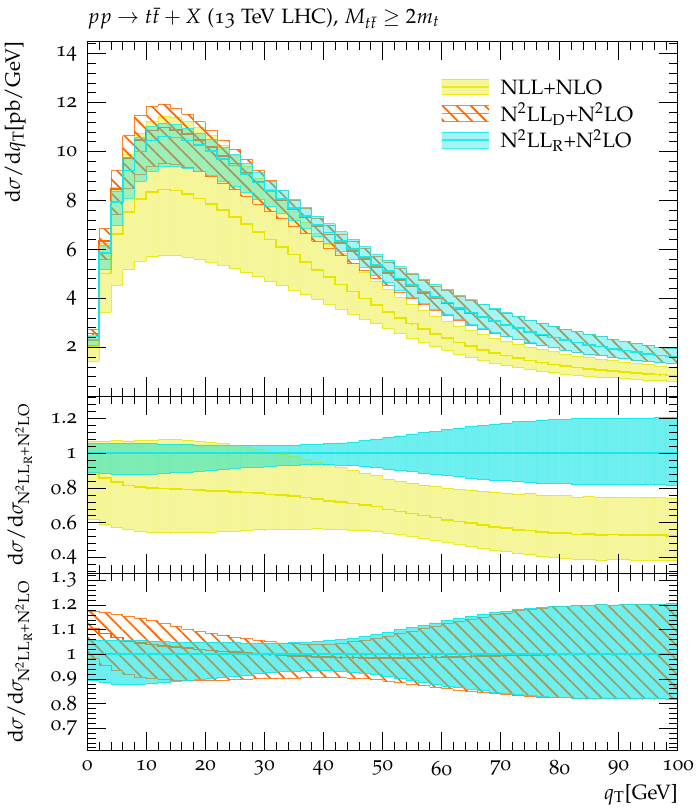}
    \caption{  }
  \label{fig:results:res:qT:dphi:FullPS:a}
  \end{subfigure}
  \begin{subfigure}{0.42\textwidth}
    \centering
    \includegraphics[width=.9\linewidth, height=0.98\linewidth]{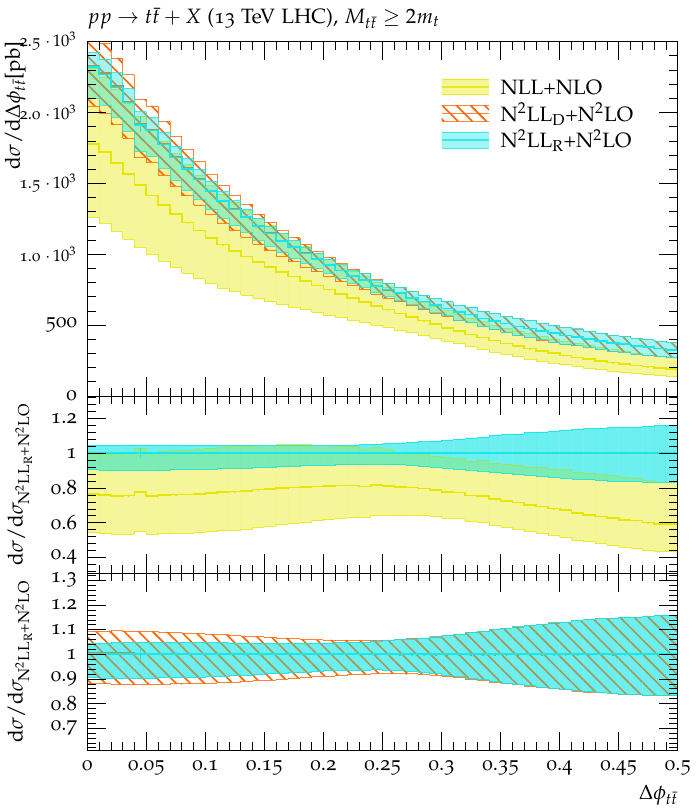}
    \caption{ }
  \label{fig:results:res:qT:dphi:FullPS:b}
  \end{subfigure} 
  \caption{
    The resummation-improved \qT\ (left) and \dphittbar\ (right) spectra of the
    process $pp\to\ttbar+X$ at $\sqrt{s}=13\,\text{TeV}$.
  }
  \label{fig:results:res:qT:dphi:FullPS}
\end{figure}

\changed{With this definition,}
we present in Fig.~\ref{fig:results:res:qT:dphi:FullPS}
the single differential distributions $\done\sigma_\ttbar/\done\qT$ and
$\done\sigma_\ttbar/\done\dphittbar$ over the whole phase space, integrated
over all \Mttbar\ values.
Therein, the results at \NLLNLO\
\changed{
are calculated from a literal implementation
of Eq.~\eqref{eq:methods:res:qT_dphi} since neither the \NLL\
non-cusp evolution kernel $\mathbf{V}_h^{[\kappa]}$ nor the tree-level
hard functions $\mathcal{C}^{[\kappa]}_{\alpha,\{h\}}$ induce any
singular behaviour in the limit $\betattbar\to0$, and thus no
modification of the resummation is required in the threshold regime.
These \NLLNLO\ results are depicted in the yellow bands} in Fig.~\ref{fig:results:res:qT:dphi:FullPS:a} (\qT) and
\ref{fig:results:res:qT:dphi:FullPS:b} (\dphittbar).
\changed{
Comparing the shapes of both distributions, however, we observe that
while a classical Sudakov peak is formed in the low \qT\ regime,
the \dphittbar\ spectrum grows monotonically as $\dphittbar\to0$.
To understand this structural difference, it is worth reminding that
although both the \qT\ and \dphittbar\ resummations comprise the same
partonic kernels in Eqs.~\eqsref{eq:LP:res:qqbar}{eq:LP:res:gg},
which approach constants as \qT\ and \dphittbar\ vanish,
the calculation of the \qT\ spectrum invokes an additional factor
$\propto\qT$ which is absent in the \dphittbar\ case, see
Eq.~\eqref{eq:methods:res:qT_dphi}.
In consequence, the resummed \qT\ spectrum experiences kinematical
suppression as $\qT\to0$ and in turn develops a Sudakov peak,
whereas the \dphittbar\ spectrum does not.
}

\changed{
At the core of this paper, however, are} the
\NNLLDNNLO\ and \NNLLRNNLO\ calculations \changed{shown in the red and
cyan in Fig.~\ref{fig:results:res:qT:dphi:FullPS}.
They are derived via the \textit{ad hoc} D- and R-prescriptions of
Eq.~\eqref{eq:LP:res:qqbar:gg:Dec} and
Eq.~\eqref{eq:LP:res:qqbar:gg:Reexp}, respectively.
It merits recalling that in these prescriptions two fundamentally
distinct methods are adopted to address the Coulomb-gluon-induced
threshold singularities of Eq.~\eqref{eq:asy:obs:qT_dphi}.
While the D-prescription simply shifts them to terms of higher logarithmic
order, \NNNLL\ and beyond, see Eq.~\eqref{eq:LP:res:qqbar:gg:Dec},
the R-prescription re-exponentiates the leading Coulomb singularities from
the non-cusp kernel $\mathbf{V}_h^{[\kappa]}$ via solving the respective
renormalisation group evolution equation in the $\betattbar\to0$ limit,
see Eq.~\eqref{eq:def:RGE:ReExp}, and leaves the singularities of the
hard scattering amplitudes $\mathcal{C}^{[\kappa]}_{\alpha;\{h\}}$ and
their complex conjugate to be cancelled by the respective phase space
suppression factors through Eq.~\eqref{eq:LP:res:qqbar:gg:Reexp}.
Therefore, comparing the results using the D- and R-prescriptions allows
us to assess the sensitivity of the \qT\ and \dphittbar\ spectra to the
details of how the Coulomb-singular contributions, that are not
systematically accommodated by the SCET$+$HQET based resummation in
Eq.~\eqref{eq:methods:res:qT_dphi} and can neither be full accessed
by scale variations, are addressed.
}

\changed{
Examining our results in detail, we find that the central value
at \NNLLRNNLO\ mostly coincides with that of \NNLLDNNLO\ for the
\dphittbar\ spectrum  of Fig.\ \ref{fig:results:res:qT:dphi:FullPS:b},
with the uncertainty band of the former being fully contained in
that of the latter.
Our findings are somewhat different for the \qT\ spectrum of
Fig.\ \ref{fig:results:res:qT:dphi:FullPS:a}, however.
We observe a difference of around  $10\%$ between the \NNLLDNNLO\
and \NNLLRNNLO\ as $\qT\to0$, while both calculations again agree
very well for the rest of the spectrum.
This difference can be understood from the kinematics in
Eqs.~\eqref{eq:methods:res:qT_dphi}, where
$\done\sigma_\ttbar/\done\dphittbar$ comprises an additional
factor of $|\widetilde{P}_{t}^{\perp}|$ compared to
$\done\sigma_\ttbar/\done\qT$.
As $\betattbar\to0$, this factor becomes of $\mathcal{O}(\betattbar)$
and effectively removes the threshold contribution that are the main
driver of the difference between the D- and R-prescriptions in the
phase space integration in the $\dphittbar$ spectrum.
Consequently, the well-separated domain, $\Delta E_\ttbar\sim\mathcal{O}(m_t)$
or $\betattbar\sim\order(1)$, dominates the single differential
\dphittbar\ spectrum, which leads to the converging uncertainty bands
of \NNLLRNNLO\ and \NNLLDNNLO\ and also the proximity of their central
values.
Nevertheless, the absence of such a dampening factor in
\qT\ spectrum emphasises the impact of threshold-enhanced terms in
the Coulomb limit, leading to a sizeable dependence on the details
of their treatment even after integrating over the entire $\Mttbar$
or $\betattbar$ range.%
}

\changed{
Given this observation, in order to further improve the
description of the \qT\ spectrum, an HQET$+$SCET-only-based analysis
will be insufficient to describe the entire phase space for lack
of an adequate description of higher-order Coulomb-enhanced terms.
To remove any ambiguity caused by the choice of \textit{ad hoc}
prescriptions, it will become necessary to develop a combined
resummation of the Coulomb, soft, and collinear corrections via
SCET$+$pNRQCD (or vNRQCD), at least in the threshold regime.
Conversely, the negligible difference between \NNLLRNNLO\ and
\NNLLDNNLO\ in the \dphittbar\ spectrum indicates a weaker
sensitivity to these Coulomb interactions.
Nonetheless, it is worth noting
that the inclusion of the threshold regime in
Eq.~\eqref{eq:methods:res:qT_dphi} is still subject to
\textit{ad hoc} prescriptions to regularise the threshold divergences.
We expect that, in this case as well, the introduction of the
aforementioned combined resummation will make the application of such
prescriptions unnecessary.
}

\section{Conclusions}
\label{sec:conclusions}

In this paper we presented the resummation-improved transverse
momentum and, for the first time, azimuthal separation spectra
of the \ttbar-pair at \NNLLNNLO\ accuracy, \changed{and studied their
predictions for top-antitop pair production at the 13\,TeV at the LHC}.
\changed{%
We based our calculation on the observation that as for such high
colliding energies the domain where the top and antitop quarks are
kinematically well separated, $\dEttbar\sim\mathcal{O}(m_t)$,
dominates the process $pp\to\ttbar+X$ and, therefore, the asymptotic
behaviour in the limits $\qT\to0$ and $\dphittbar\to0$ can be mostly
captured by the soft and beam-collinear resummation via HQET$+$SCET.
We then endeavored to extend this description over the entire phase space,
including the threshold regime, $\Delta E_\ttbar\to0$ or $\betattbar\to0$,
where the exchange of Coulomb gluons adds new potentially divergent
corrections to our process.
Implementing this method at \NLL, however, proved to be straightforward,
as the involved functions of the HQET$+$SCET resummation at this
accuracy were found to be regular in the threshold limit.
Nevertheless, starting from \NNLL, we demonstrated that Coulomb divergences
manifest themselves in both the hard function and its non-cusp
evolution kernels.
This, in turn prohibited a direct application of the soft and collinear
resummation result to the entire phase space.
}

\changed{
In order to address these Coulomb divergences emerging in our soft and
collinear resummation, we introduced two \textit{ad hoc} prescriptions,
referenced as D- and R-prescriptions, respectively.
While the threshold enhanced terms in D-prescription are simply shifted
to a higher logarithmic order, they are in part resummed in R-prescription.
Here, the leading singular terms induced by the two-loop non-cusp anomalous
are re-exponentiated via solving the hard renormalisation group equation
as $\betattbar\to0$. 
Comparing their respective predictions
allowed us to quantitatively assess the inherent  uncertainty of our
calculation and the sensitivity of both the \qT\ and \dphittbar\ spectra
to Coulomb interactions.
}

\changed{
Finally, we implemented both prescriptions to calculate
$\done\sigma_{t\bar{t}}/\done\dphittbar$ and
$\done\sigma_{t\bar{t}}/\done\qT$,  labelling their predictions
\NNLLDNNLO\ and \NNLLRNNLO.
We observed that the central values from D- and R-prescriptions mostly
coincide in $\done\sigma_{t\bar{t}}/\done\dphittbar$, with the
uncertainty band of the latter being fully contained by that of the
former.
Conversely, we found a deviation of around $10\%$ between both
prescriptions as $\qT\to0$ in $\done\sigma_{t\bar{t}}/\done\qT$.
This indicates that applying our soft-collinear resummation onto
the \qT\ spectrum introduces a non-negligible ambiguity at \NNLL\
in the absence of necessary higher-order contributions from the
threshold domain.
Therefore, to further improve the theoretical accuracy of the \qT\ spectrum,
a combined resummation of Coulomb, soft, and collinear corrections
via SCET and pNRQCD (or alternatively vNRQCD) in the threshold regime
seems warranted.
Similarly, while the negligible deviation between the D- and R-prescriptions
in the \dphittbar\ spectrum indicates a weaker sensitivity to Coulomb
interactions, it is worth noting that extending the soft and collinear
resummation coverage to the whole \Mttbar\ phase space still necessitates
\textit{ad hoc} prescriptions to regularise the arising Coulomb divergences
as $\dEttbar\to0$.
We expect that a combined resummation will be beneficial in this case
as well, removing the need for either prescription.
 }

\subsection*{Acknowledgements}

WJ would like to thank Li Lin Yang for sharing valuable details in
computing the hard evolution kernel in \cite{Zhu:2012ts,Li:2013mia,
  Ahrens:2010zv,Ahrens:2011mw}.
WJ is also grateful to Guoxing Wang for the helpful discussion on the
threshold soft function in \cite{Wang:2018vgu}.
WJ and MS would also like to express our gratitude to Ben Pecjak for
taking the time to respond to our numerous queries.
MS is funded by the Royal Society through a University Research Fellowship
(URF\textbackslash{}R1\textbackslash{}180549, URF\textbackslash{}R\textbackslash{}231031) and an Enhancement Award
(RGF\textbackslash{}EA\textbackslash{}181033,
 CEC19\textbackslash{}100349, and RF\textbackslash{}ERE\textbackslash{}210397)
as well as the STFC (ST/X003167/1 and ST/X000745/1).

\appendix

\section{\changed{Comparison of D- and R-prescriptions in the double differential distributions}}

\begin{figure}[t!]
  \centering
  \begin{subfigure}{0.32\textwidth}
    \centering
    \includegraphics[width=.9\linewidth, height=0.98\linewidth]{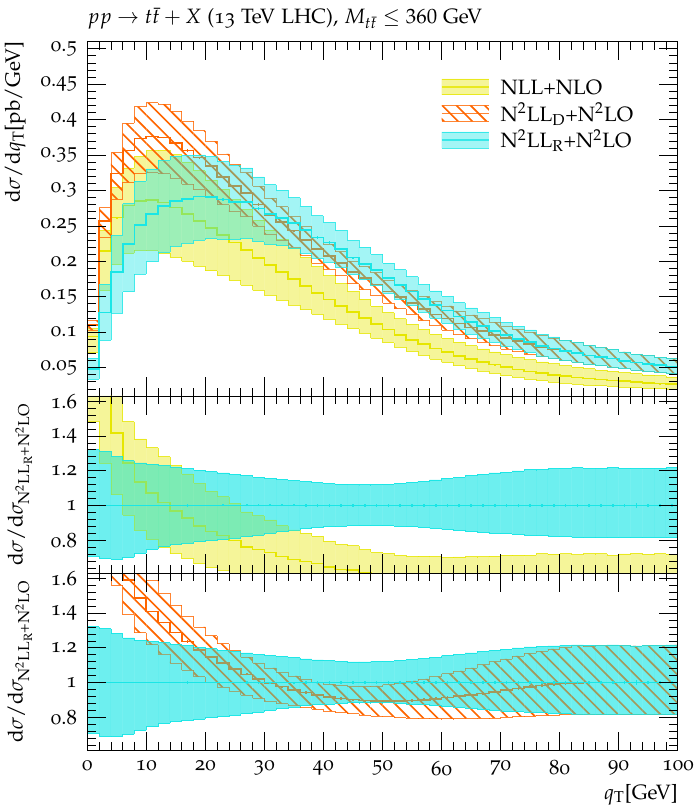}
    \caption{  }
\label{fig:results:res:qT:cs:a}
  \end{subfigure}
  \begin{subfigure}{0.32\textwidth}
    \centering
    \includegraphics[width=.9\linewidth, height=0.98\linewidth]{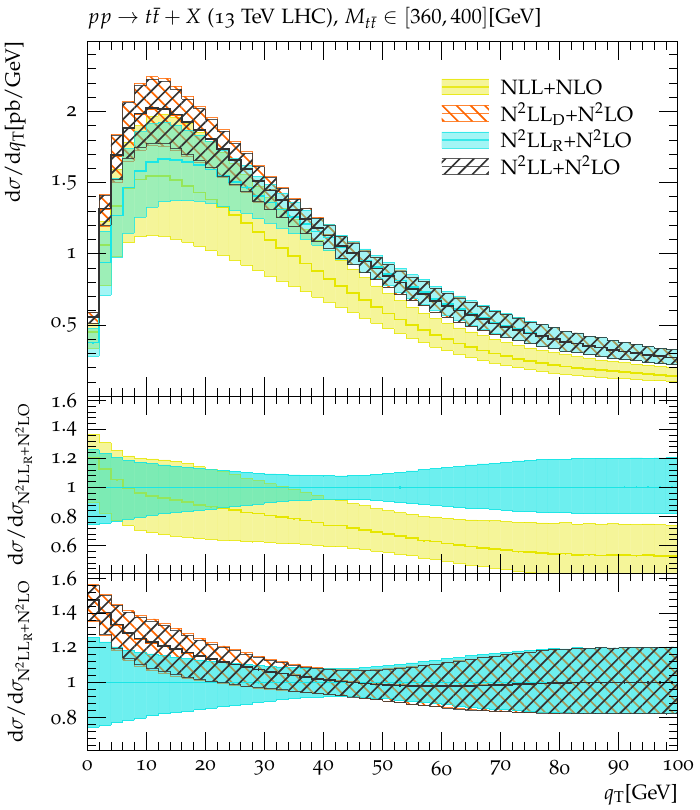}
    \caption{  }
\label{fig:results:res:qT:cs:b}
  \end{subfigure} 
  \begin{subfigure}{0.32\textwidth}
    \centering
    \includegraphics[width=.9\linewidth, height=0.98\linewidth]{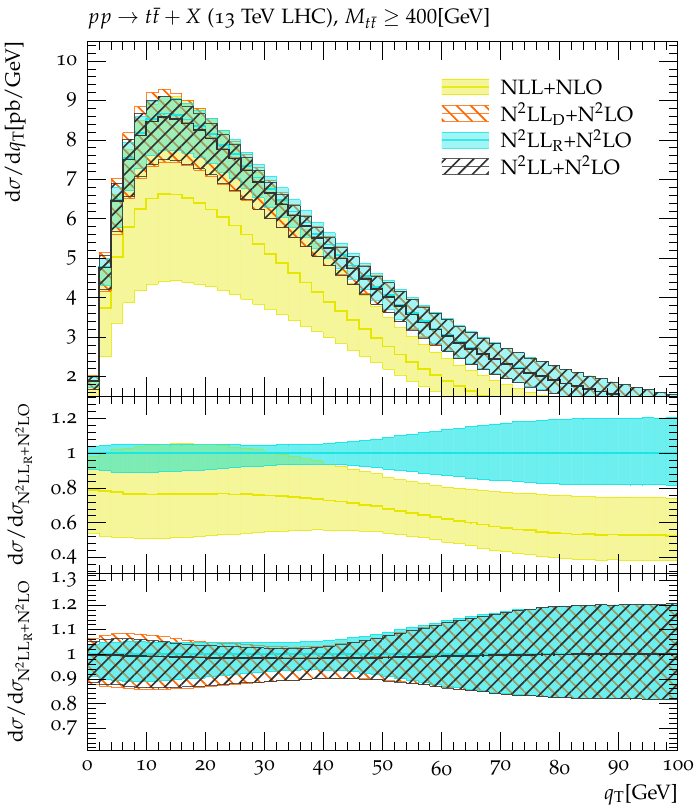}
    \caption{   }
\label{fig:results:res:qT:cs:c}
  \end{subfigure}
\\
  \centering
  \begin{subfigure}{0.32\textwidth}
    \centering
    \includegraphics[width=.9\linewidth, height=0.98\linewidth]{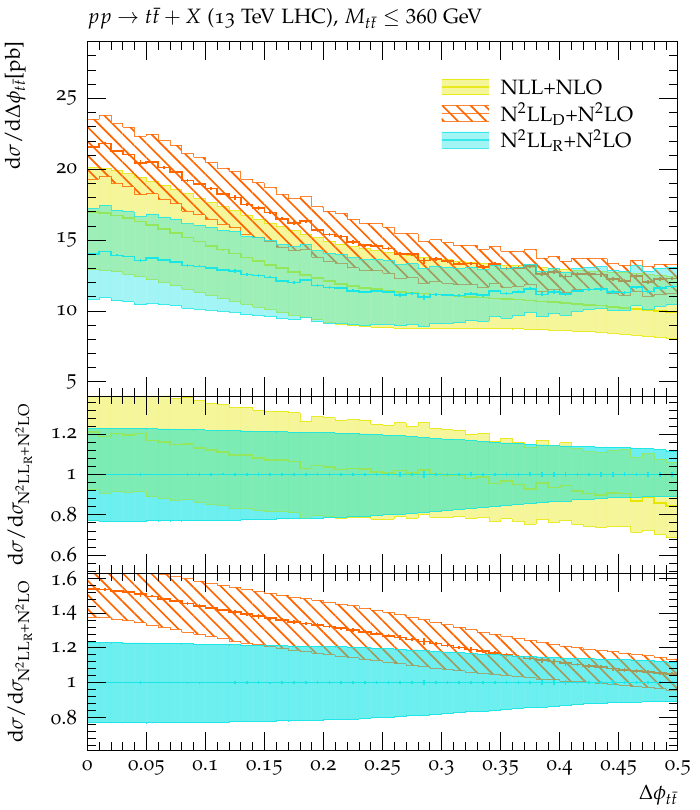}
    \caption{  }
    \label{fig:results:res:dphi:cs:a}
  \end{subfigure}
  \begin{subfigure}{0.32\textwidth}
    \centering
    \includegraphics[width=.9\linewidth, height=0.98\linewidth]{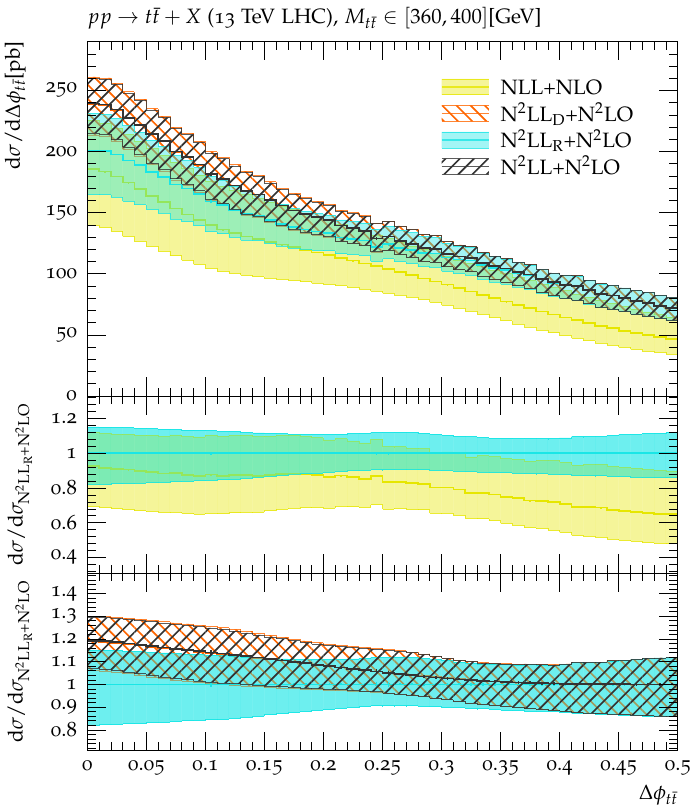}
    \caption{ }
    \label{fig:results:res:dphi:cs:b}
  \end{subfigure} 
  \begin{subfigure}{0.32\textwidth}
    \centering
    \includegraphics[width=.9\linewidth, height=0.98\linewidth]{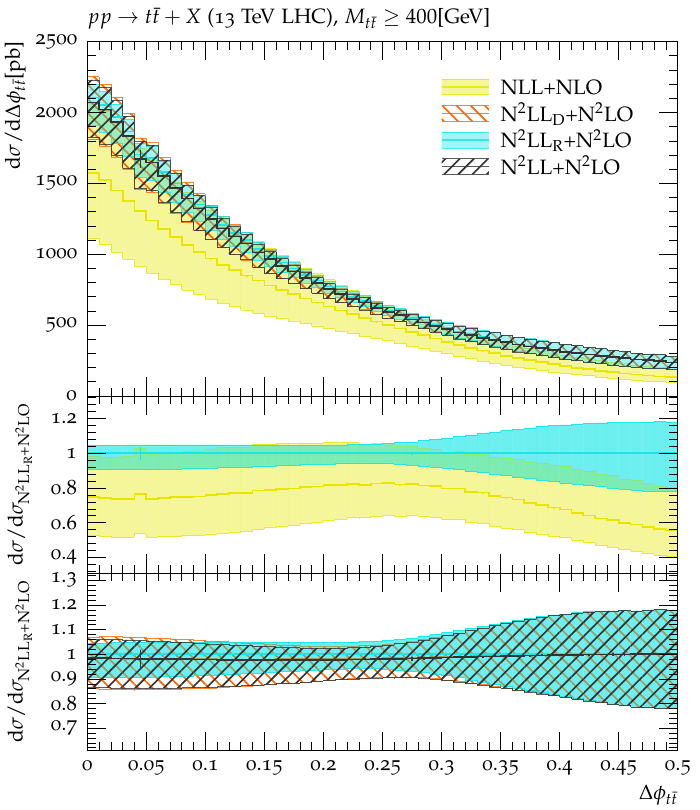}
    \caption{   }
    \label{fig:results:res:dphi:cs:c}
  \end{subfigure}
  \caption{
    Comparison of the D- and R-prescriptions within the intervals
    $\Mttbar\le 360\,\text{GeV}$ (left),  $\Mttbar\in[360,400]\,\text{GeV}$ (centre), and $\Mttbar\ge 400\,\text{GeV}$ (right).
  }
  \label{fig:results:res:qT:dphi:cs}
\end{figure}

\changed{
It is interesting to analyse our results double-differentially,
i.e.\ examining both the \qT\ and \dphittbar\ spectra in three
different regions of the \ttbar\ invariant mass \Mttbar,
$\Mttbar\le 360\,\text{GeV}$,
$\Mttbar\in[360,400]\,\text{GeV}$, and $\Mttbar\ge 400\,\text{GeV}$.
We can use these three regions, containing the \ttbar\ production
threshold and the Coulomb divergences, a transition region, and the
well-separated region where $\dEttbar\sim\order(m_t)$, respectively,
to investigate the development of the differences between the
D- and R-prescriptions as $\Mttbar$ varies.
The \qT\ and \dphittbar\ differential distributions within these
intervals are presented in Fig.~\ref{fig:results:res:qT:dphi:cs},
including the \NLLNLO, \NNLLDNNLO, and \NNLLRNNLO\ calculations
depicted in the yellow, red, and cyan bands, respectively.
As to the last two slices, due to the absence of Coulomb divergences,
we also add the results evaluated with the original evolution kernels
of Eq.~\eqref{eq:def:Vres:N2LL}, labelled \NNLLNNLO\ and shown
as a dark-grey hatched band.%
}\footnote{
  \changed{It should be noted that the \NLLNLO\ and \NNLLNNLO\
  results in Fig.~\ref{fig:results:res:dphi:cs:c} have
  been published in \cite{Ju:2022wia} and are shown
  here for comparison.}
}

\changed{
Fig.~\ref{fig:results:res:qT:cs:c} and
Fig.~\ref{fig:results:res:dphi:cs:c} illustrate
$\done\sigma_{t\bar{t}}/\done\qT$ and
$\done\sigma_{t\bar{t}}/\done\dphittbar$, respectively, for
$\Mttbar\ge 400\,\text{GeV}$, where the top and antitop quarks
are kinematically well separated and our soft-collinear resummation
possesses the best predictivity.
It is observed that the central values from \NNLL, \NNLLD, and
\NNLLR\ coincide  within a few percent of each other and their
uncertainty bands marginally overlap with that calculated at \NLL.
This phenomenon confirms our derivation in
Sec.~\ref{sec:thr:div:extrap} in that the difference amongst
D-prescription, R-prescription and the original evolution kernels
in Eq.~\eqref{eq:def:Vres:N2LL} is numerically of \NNLLp\ and
beyond for the separated domain and therefore the deviations between them can be well captured
by the scale variation in this domain.
}

\changed{
However, lowering $\Mttbar$ leads to distinct scenarios.
As exhibited in Fig.~\ref{fig:results:res:qT:cs:b} and Fig.~\ref{fig:results:res:dphi:cs:b}, although the unmodified
\NNLL\ result is still in good agreement with \NNLLD,
$50\%$ ($20\%$) deviations are observed in the \NNLLR\ calculation
in the low \qT\ (\dphittbar) domains.
To interpret this, it merits recalling that \NNLLD\ shifts
threshold enhanced terms in part from $\mathbf{V}_h^{[\kappa]}$
of \NNLL\ to higher logarithmic orders, while \NNLLR\ re-exponentiates
the leading Coulomb singularities from $\mathbf{V}_h^{[\kappa]}$ on
top of \NNLL\ and also embeds the NLO corrections from both
$\mathcal{C}^{[\kappa]}_{\alpha;\{h\}}$ and its complex conjugate.
Nevertheless,  the threshold enhancements generated by
$\mathbf{V}_h^{[\kappa]}$ are generally weaker than those from
the hard sector in the region $\Mttbar\in[360,400]\,\text{GeV}$.
For instance, according to Eq.~\eqref{eq:def:Chard:thr:qqb_qbq_gg:NLO}
and Eq.~\eqref{eq:def:Vres:N2LL:thr:qq_gg},
$\mathbf{V}_h^{[\kappa]}$ entails the leading Coulomb divergence
$\sim\frac{\alpha^2_s}{\beta^2_\ttbar}\left\{\left(\frac{55}{4761}\right)^2,\left(\frac{55}{38088}\right)^2 \right\}$
at \NNLO, whereas the product of $\mathcal{C}^{[\kappa]}_{\alpha;\{h\}}$
and its complex conjugate is
$\sim\frac{\alpha^2_s}{\beta^2_\ttbar}\left\{\left(\frac{\pi }{3}\right)^2,\left(\frac{\pi }{24}\right)^2\right\}$.
Here the first (second) entry refers to the color-singlet (color octet) contributions of the top-antitop quark pair.
Given this numerical hierarchy, the interval
$\Mttbar\in[360,400]\,\text{GeV}$ is dominated by enhancements in the
hard section and therefore gives rise to non-negligible deviation
between \NNLLR\ and \NNLLD\ as well as the unmodified \NNLL.
}

\changed{
Further decreasing $\Mttbar$ focusses on the threshold regime, as shown
in Fig.~\ref{fig:results:res:qT:cs:a} and
Fig.~\ref{fig:results:res:dphi:cs:a}, where the discrepancies between
\NNLLD\ and \NNLLR\ are observed to be up to $\sim 100\%$ in the low \qT\
realm and $\sim60\%$ in the low \dphittbar\ domain, owing to the growth of
threshold enhanced contributions.
This phenomenon, together with those in Fig.~\ref{fig:results:res:qT:cs:b}
and Fig.~\ref{fig:results:res:dphi:cs:b}, demonstrates the origin of
the difference between D- and R-prescriptions observed in
Fig.~\ref{fig:results:res:qT:dphi:FullPS}.
}

\section{Numerical results of hard-scale evolution kernel}
\label{app:vali:Vthr}

In this appendix we deliver a numerical comparison amongst
the original non-cusp evolution kernel in Eqs.~\eqsref{eq:def:Vres:NLL}{eq:def:Vres:N2LL} as proposed in~\cite{Buras:1991jm,Buchalla:1995vs},
the leading singular approximation from
Eqs.~\eqsref{eq:def:Vres:NLL:thr:qq_gg}{eq:def:Vres:N2LL:thr:qq_gg},
and the expansion of the re-exponentiated kernel in
Eq.~\eqref{eq:def:Vres:N2LL:Reexp:exp}.
During our computation, we fix $\mu_h=M_{\ttbar}$ and
$\mu_s=1\,\text{GeV}$ as well as the scattering angle of the top
quark $\theta_t={\pi}/{3}$.

\begin{figure}[t!]
  \centering
  \begin{subfigure}{0.49\textwidth}
    \centering
    \includegraphics[width=.9\linewidth, height=0.6\linewidth]{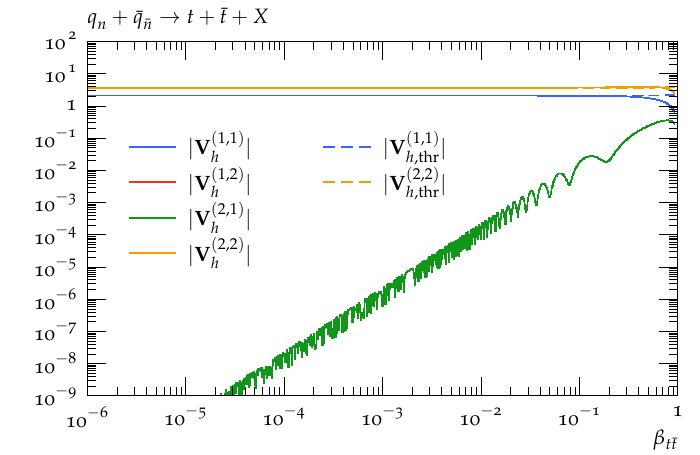}
    \caption{  }
    \label{fig:Suk:Exp:qqbar1:lp:nll}
  \end{subfigure}
  \begin{subfigure}{0.49\textwidth}
    \centering
    \includegraphics[width=.9\linewidth, height=0.6\linewidth]{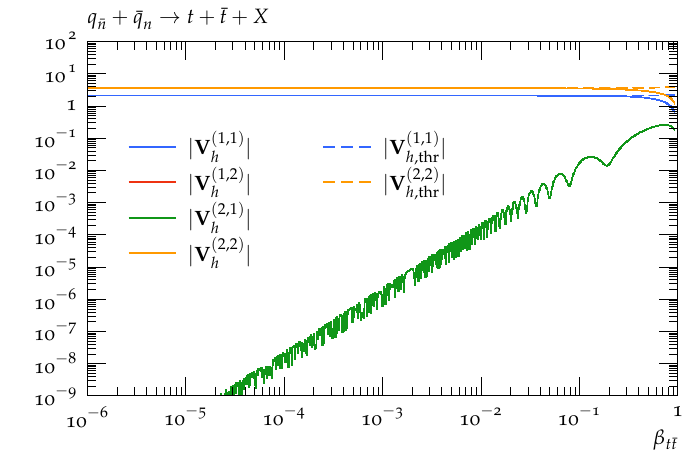}
    \caption{   }
    \label{fig:Suk:Exp:qbarq1:lp:nll}
  \end{subfigure}
  \\
   \centering
  \begin{subfigure}{0.49\textwidth}
    \centering
    \includegraphics[width=.9\linewidth, height=0.6\linewidth]{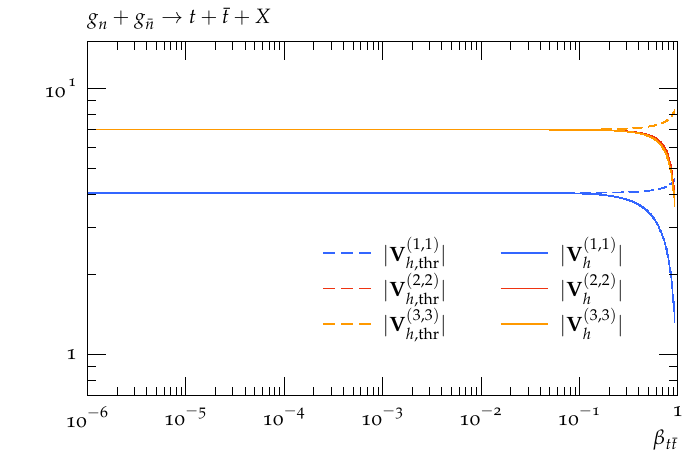}
    \caption{  }
    \label{fig:Suk:Exp:gg1a:lp:nll}
  \end{subfigure}
  \begin{subfigure}{0.49\textwidth}
    \centering
    \includegraphics[width=.9\linewidth, height=0.6\linewidth]{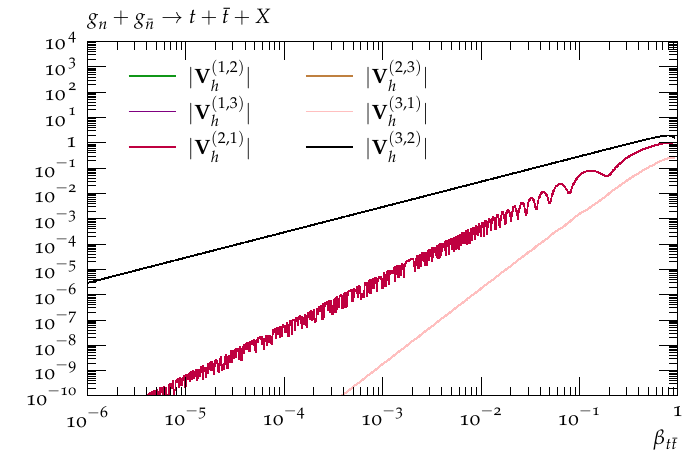}
    \caption{   }
    \label{fig:Suk:Exp:gg1b:lp:nll}
  \end{subfigure}
  \caption{
    Comparison of the non-cusp evolution kernel $\mathbf{V}_h$ in Eq.~\eqref{eq:def:Vres:NLL} and its leading terms $\mathbf{V}_{h,\mathrm{thr}}$ in Eq.~\eqref{eq:def:Vres:NLL:thr:qq_gg} in the threshold regime at \NLL. Therein,  $|\mathbf{V}_{h,(\mathrm{thr})}^{(m,n)}|$ denotes the absolute value of the entry of $\mathbf{V}_{h,(\mathrm{thr})}$ in the $m$-th row and  $n$-th column from respective partonic process.
  }
  \label{fig:Suk:Exp:lp:nll}
\end{figure}

Fig.~\ref{fig:Suk:Exp:lp:nll} exhibits the \NLL\ results of
$\mathbf{V}_h$ in Eq.~\eqref{eq:def:Vres:NLL} and
$\mathbf{V}_{h,\mathrm{thr}}$ in Eq.~\eqref{eq:def:Vres:NLL:thr:qq_gg}
in solid and dashed lines, respectively.
Therein, the dependence of all the entries of $\mathbf{V}_h$ is
displayed with respect to $\betattbar$. For $\mathbf{V}_{h,\mathrm{thr}}$,
we show the non-zero components only.
Due to the facts that the leading threshold enhanced terms in
the \NLO\ anomalous dimension only manifest themselves in the
imaginary parts and that the complete Coulomb singular behaviour
has been exponentiated at \NLL\ in both Eq.~\eqref{eq:def:Vres:NLL}
and Eq.~\eqref{eq:def:Vres:NLL:thr:qq_gg}, the \NLL\ kernel
illustrated in Fig.~\ref{fig:Suk:Exp:lp:nll} invokes no
divergence as $\betattbar\to0$.
Comparing $\mathbf{V}_h$ with $\mathbf{V}_{h,\mathrm{thr}}$, we
find that the leading approximation is capable of replicating
the correct asymptotic behaviour of all the diagonal entries of
$\mathbf{V}_h$, while the non-diagonal elements of $\mathbf{V}_h$
become progressive smaller in the low $\betattbar$ region.
This indicates the non-diagonal entries of $\mathbf{V}_h$
are all power suppressed in magnitude and it is thus in
agreement with the absence of the off-diagonal contributions
in $\mathbf{V}_{h,\mathrm{thr}}$ in
Eq.~\eqref{eq:def:Vres:NLL:thr:qq_gg}.

\begin{figure}[t!]
  \centering
    \begin{subfigure}{0.49\textwidth}
    \centering
    \includegraphics[width=.9\linewidth, height=0.6\linewidth]{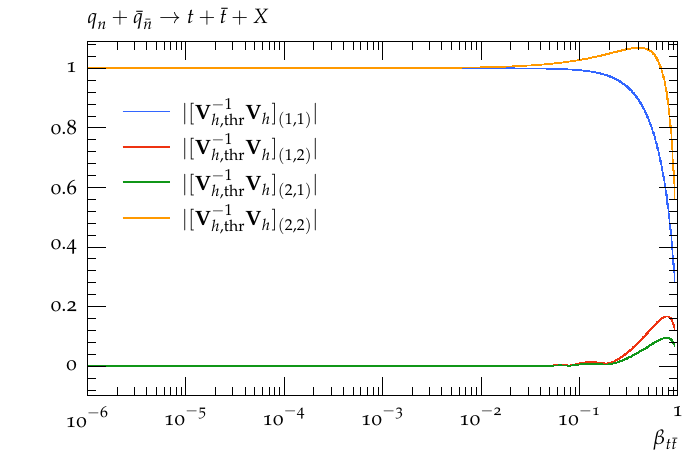}
    \caption{  }
    \label{fig:Suk:Exp:qqbar2:lp:nll:ratio}
  \end{subfigure}
  \begin{subfigure}{0.49\textwidth}
    \centering
    \includegraphics[width=.9\linewidth, height=0.6\linewidth]{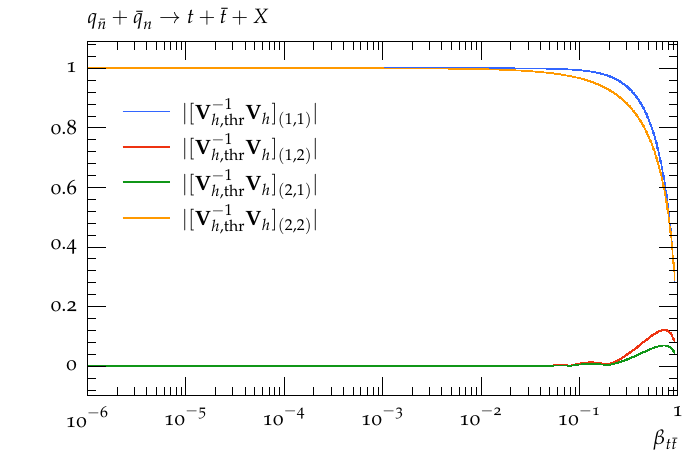}
    \caption{  }
    \label{fig:Suk:Exp:qbarq2:lp:nll:ratio}
  \end{subfigure}  
  \\
   \begin{subfigure}{0.49\textwidth}
    \centering
    \includegraphics[width=.9\linewidth, height=0.6\linewidth]{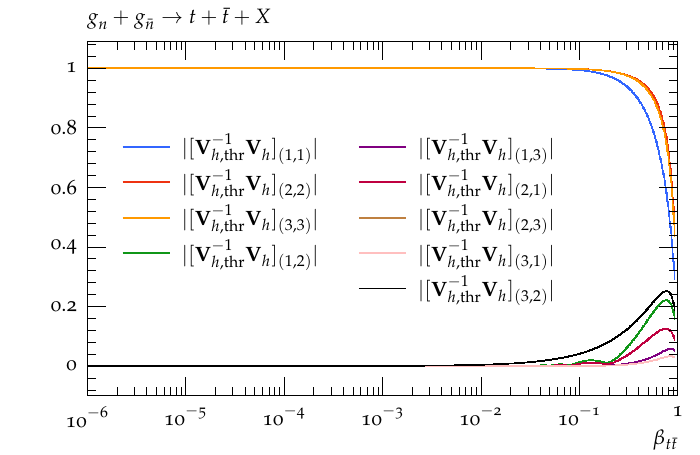}
    \caption{   }
    \label{fig:Suk:Exp:gg2:lp:nll:ratio}
  \end{subfigure}
    \caption{
Product of the non-cusp evolution $\mathbf{V}_h$ in Eq.~\eqref{eq:def:Vres:NLL} and the inverse matrix  $\mathbf{V}^{-1}_{h,\mathrm{thr}}$ in Eq.~\eqref{eq:def:Vres:NLL:thr:qq_gg} at \NLL. Therein,  $|[\mathbf{V}^{-1}_{h,\mathrm{thr}}\mathbf{V}_h]_{m,n}|$ denotes the absolute value of the entry of $(\mathbf{V}^{-1}_{h,\mathrm{thr}}\mathbf{V}_h)$ in the $m$-th row and  $n$-th column from respective partonic process.
  }
  \label{fig:Suk:Exp:lp:nll:ratio}
\end{figure}
 
In a bid to scrutinise our leading approximation in
Eq.~\eqref{eq:def:Vres:NLL:thr:qq_gg} further, we plot
the result of the product
$(\mathbf{V}^{-1}_{h,\mathrm{thr}}\mathbf{V}_h)$ in
Fig.~\ref{fig:Suk:Exp:lp:nll:ratio}.
We observe that $(\mathbf{V}^{-1}_{h,\mathrm{thr}}\mathbf{V}_h)$
approaches the unity matrix for all three partonic processes of
interest.
This phenomenon shows that $\mathbf{V}^{-1}_{h,\mathrm{thr}}$
is able to serve as a qualified inverse matrix of $\mathbf{V}_h$
in the vicinity of $\betattbar=0$.
Further, it details that the approximation in
Eq.~\eqref{eq:def:Vres:NLL:thr:qq_gg} indeed manages to
reproduce the leading asymptotic behaviour of
Eq.~\eqref{eq:def:Vres:NLL}.

\begin{figure}[t!]
  \centering
  \begin{subfigure}{0.49\textwidth}
    \centering
    \includegraphics[width=.9\linewidth, height=0.6\linewidth]{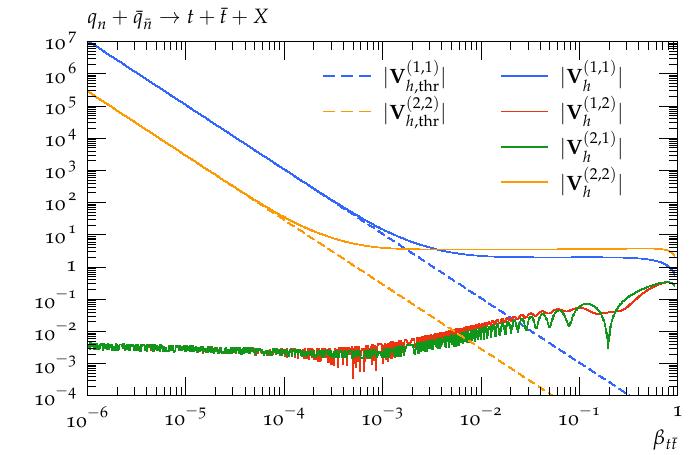}
    \caption{  }
    \label{fig:Suk:Exp:qqbar1:lp:n2ll}
  \end{subfigure}
  \begin{subfigure}{0.49\textwidth}
    \centering
    \includegraphics[width=.9\linewidth, height=0.6\linewidth]{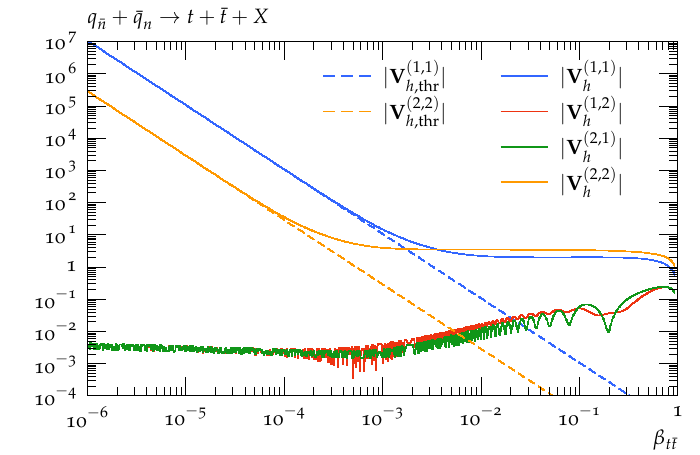}
    \caption{   }
    \label{fig:Suk:Exp:qbarq1:lp:n2ll}
  \end{subfigure}
  \\
   \centering
  \begin{subfigure}{0.49\textwidth}
    \centering
    \includegraphics[width=.9\linewidth, height=0.6\linewidth]{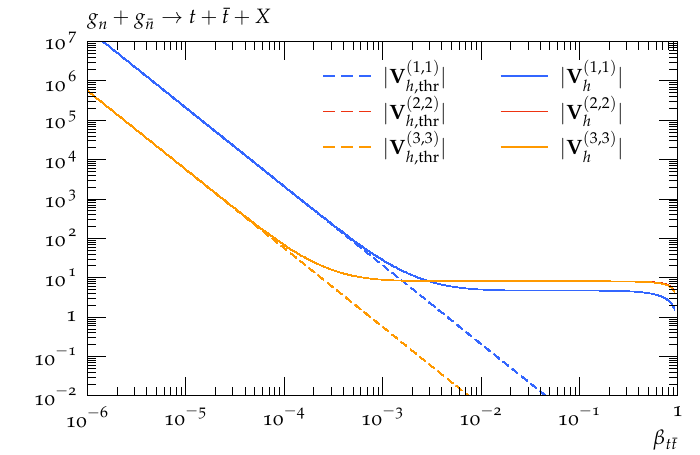}
    \caption{ }
    \label{fig:Suk:Exp:gg1a:lp:n2ll}
  \end{subfigure}
  \begin{subfigure}{0.49\textwidth}
    \centering
    \includegraphics[width=.9\linewidth, height=0.6\linewidth]{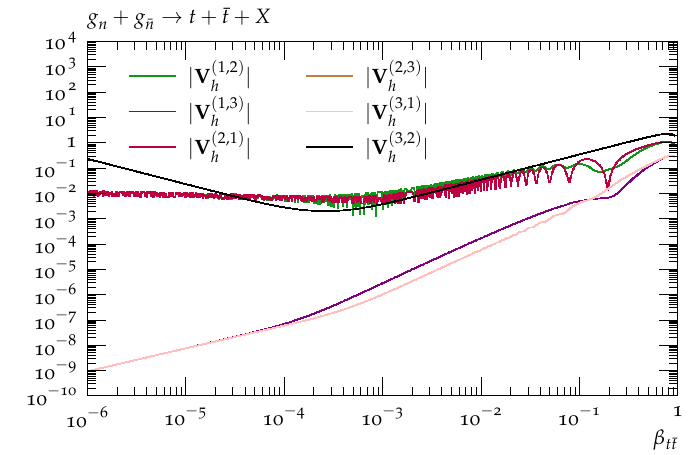}
    \caption{ }
    \label{fig:Suk:Exp:gg1b:lp:n2ll}
  \end{subfigure}
   \caption{
Comparison of the non-cusp evolution $\mathbf{V}_h$ in Eq.~\eqref{eq:def:Vres:N2LL} and its leading terms $\mathbf{V}_{h,\mathrm{thr}}$ in Eq.~\eqref{eq:def:Vres:N2LL:thr:qq_gg} in the threshold regime at \NNLL. Therein,  $|\mathbf{V}_{h,(\mathrm{thr})}^{(m,n)}|$ denotes the absolute value of the entry of $\mathbf{V}_{h,(\mathrm{thr})}$ in the $m$-th row and  $n$-th column from respective partonic process.
  }
  \label{fig:Suk:Exp:lp:n2ll}
\end{figure}

\begin{figure}[t!]
  \centering
    \begin{subfigure}{0.49\textwidth}
    \centering
    \includegraphics[width=.9\linewidth, height=0.6\linewidth]{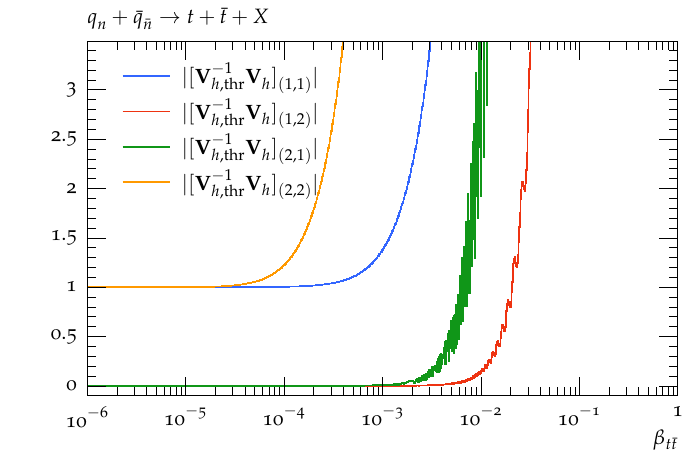}
    \caption{  }
    \label{fig:Suk:Exp:qqbar2:lp:n2ll:ratio}
  \end{subfigure}
  \begin{subfigure}{0.49\textwidth}
    \centering
    \includegraphics[width=.9\linewidth, height=0.6\linewidth]{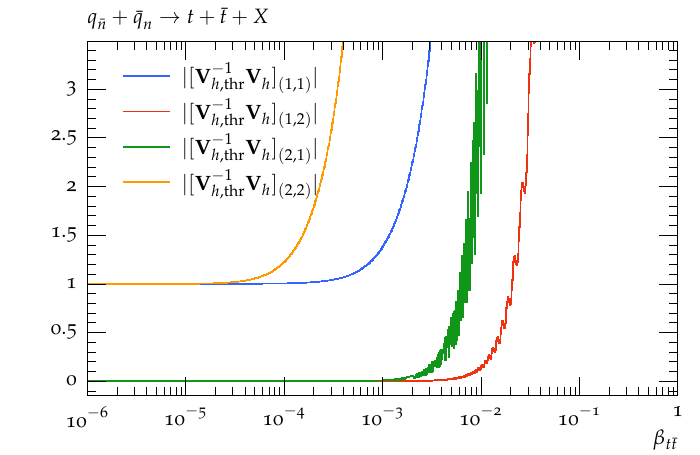}
    \caption{  }
    \label{fig:Suk:Exp:qbarq2:lp:n2ll:ratio}
  \end{subfigure}  
  \\
   \begin{subfigure}{0.49\textwidth}
    \centering
    \includegraphics[width=.9\linewidth, height=0.6\linewidth]{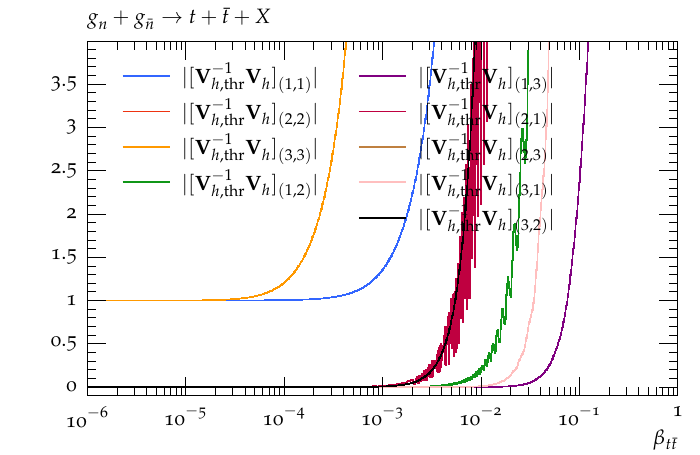}
    \caption{ . }
    \label{fig:Suk:Exp:gg2:lp:n2ll:ratio}
  \end{subfigure}
    \caption{
Product of the non-cusp evolution $\mathbf{V}_h$ in Eq.~\eqref{eq:def:Vres:N2LL} and the inverse matrix  $\mathbf{V}^{-1}_{h,\mathrm{thr}}$ in Eq.~\eqref{eq:def:Vres:N2LL:thr:qq_gg} at \NNLL. Therein,  $|[\mathbf{V}^{-1}_{h,\mathrm{thr}}\mathbf{V}_h]_{m,n}|$ denotes the absolute value of the entry of $(\mathbf{V}^{-1}_{h,\mathrm{thr}}\mathbf{V}_h)$ in the $m$-th row and  $n$-th column from respective partonic process.
  }
  \label{fig:Suk:Exp:lp:n2ll:ratio}
\end{figure}

In Figs.~\ref{fig:Suk:Exp:lp:n2ll} we confront $\mathbf{V}_h$
of Eq.~\eqref{eq:def:Vres:N2LL} with $\mathbf{V}_{h,\mathrm{thr}}$
of Eq.~\eqref{eq:def:Vres:N2LL:thr:qq_gg} at \NNLL.
At variance with the findings of Figs.~\ref{fig:Suk:Exp:lp:nll}
and \ref{fig:Suk:Exp:lp:nll:ratio}, the diagonal entries of
$\mathbf{V}_h$ develop divergent behaviour in the threshold domain,
as a result of the Coulomb singularity residing in
Eq.~\eqref{eq:Jh:thr:exp:qq_gg}, which $\mathbf{V}_{h,\mathrm{thr}}$
is able to replicate at \NNLL.
Further, differing from $\mathbf{V}_h$ at \NLL, where all the
non-diagonal entries in Fig.~\ref{fig:Suk:Exp:lp:nll} generally
decline in magnitude as \betattbar\ reduces, the non-diagonal
constituents in Fig.~\ref{fig:Suk:Exp:lp:n2ll} can experience
enhancements in the threshold domain, such as $|\mathbf{V}_h^{(3,2)}|$
in Fig.~\ref{fig:Suk:Exp:gg1b:lp:n2ll}.
The reason for this phenomenon is that in the expression of
Eq.~\eqref{eq:def:Vres:N2LL:thr:qq_gg}, only the leading singular
terms of Eq.~\eqref{eq:def:Vres:N2LL:thr:qq_gg}, which are of
$\mathcal{O}(\betattbar^{-2})$, have been taken into account.
Divergence of $\mathcal{O}(\betattbar^{-1})$ are, however, still possible.
To verify that there is no stronger divergent behaviour in non-diagonal
elements other than that of Eq.~\eqref{eq:def:Vres:N2LL:thr:qq_gg},
we present the results of $(\mathbf{V}^{-1}_{h,\mathrm{thr}}\mathbf{V}_h)$
in Fig.~\ref{fig:Suk:Exp:lp:n2ll:ratio}.
It is found that all but the diagonal elements are reduced substantially
as $\betattbar\to0$, while all the diagonal elements of the product
$(\mathbf{V}^{-1}_{h,\mathrm{thr}}\mathbf{V}_h)$  approach unity.
This unambiguously shows that our leading approximation in
Eq.~\eqref{eq:def:Vres:N2LL:thr:qq_gg} can describe the asymptotic
behaviour of Eq.~\eqref{eq:def:Vres:N2LL} at \NNLL\ as well.
  
\begin{figure}[t!]
  \centering
  \begin{subfigure}{0.49\textwidth}
    \centering
    \includegraphics[width=.9\linewidth, height=0.6\linewidth]{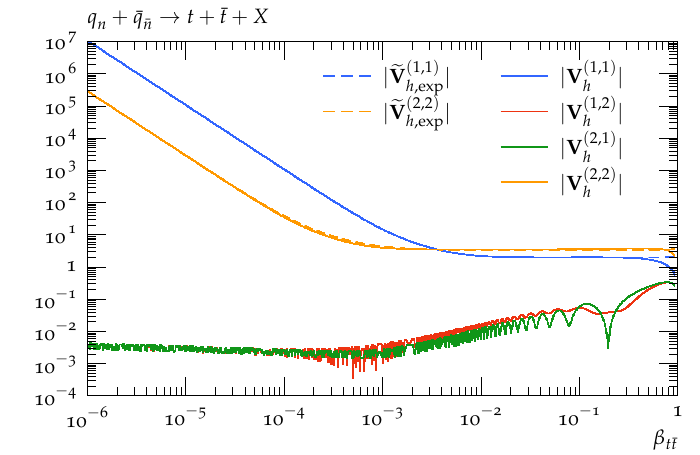}
    \caption{   }
    \label{fig:Suk:Exp:qqbar1:n2ll}
  \end{subfigure}
  \begin{subfigure}{0.49\textwidth}
    \centering
    \includegraphics[width=.9\linewidth, height=0.6\linewidth]{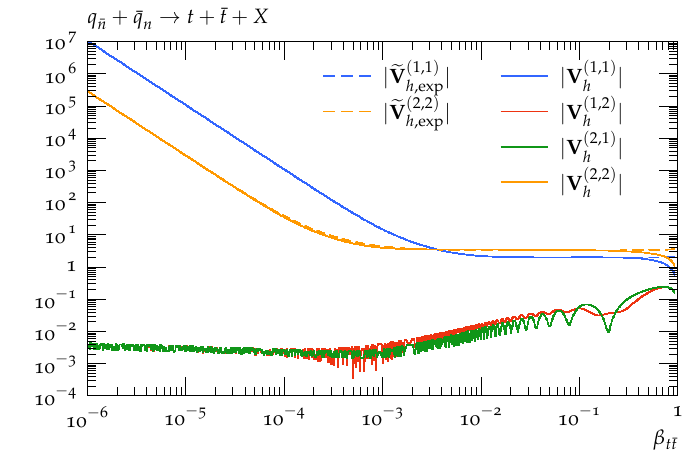}
    \caption{   }
    \label{fig:Suk:Exp:qbarq1:n2ll}
  \end{subfigure}
  \\
   \centering
  \begin{subfigure}{0.49\textwidth}
    \centering
    \includegraphics[width=.9\linewidth, height=0.6\linewidth]{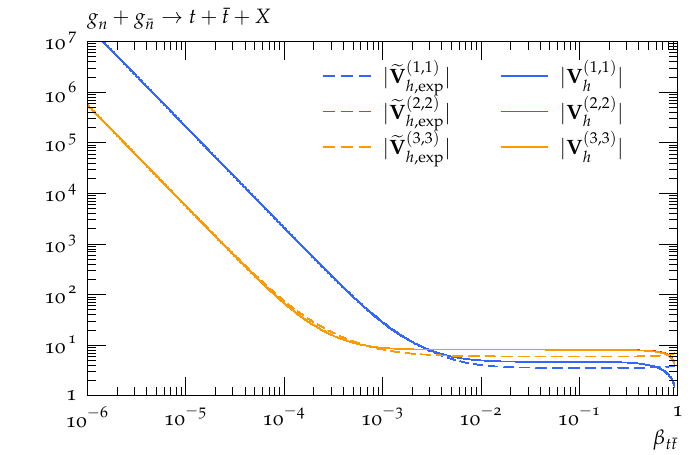}
    \caption{  }
    \label{fig:Suk:Exp:gg1a:n2ll}
  \end{subfigure}
  \begin{subfigure}{0.49\textwidth}
    \centering
    \includegraphics[width=.9\linewidth, height=0.6\linewidth]{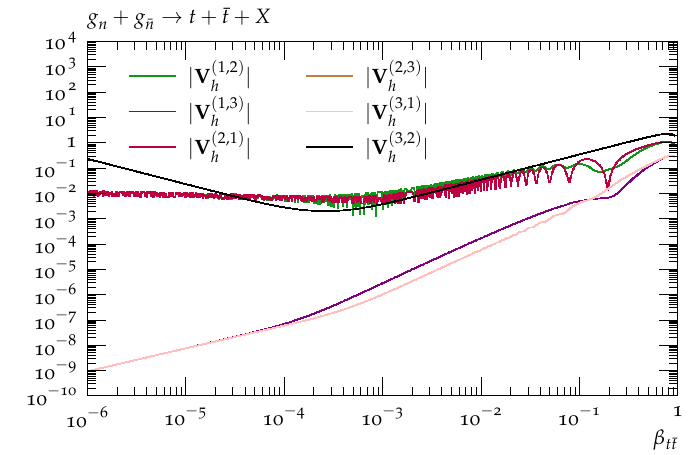}
    \caption{   }
    \label{fig:Suk:Exp:gg1b:n2ll}
  \end{subfigure}
   \caption{
Comparison of the non-cusp evolution $\mathbf{V}_h$ in Eq.~\eqref{eq:def:Vres:N2LL} and  $\widetilde{\mathbf{V}}_{h,\mathrm{exp}}$ in Eq.~\eqref{eq:def:Vres:N2LL:Reexp:exp} at \NNLL. Therein,  $|\mathbf{V}_{h}^{(m,n)}|$  and  $|\widetilde{\mathbf{V}}_{h,\mathrm{exp}}^{(m,n)}|$  denote  the absolute values of the entries of $\mathbf{V}_h$ and $\widetilde{\mathbf{V}}_{h,\mathrm{exp}}$ in the $m$-th row and  $n$-th column, respectively. 
  } 
  \label{fig:Suk:Exp:n2ll}
\end{figure}

\begin{figure}[t!]
   \centering
  \begin{subfigure}{0.49\textwidth}
    \centering
    \includegraphics[width=.9\linewidth, height=0.6\linewidth]{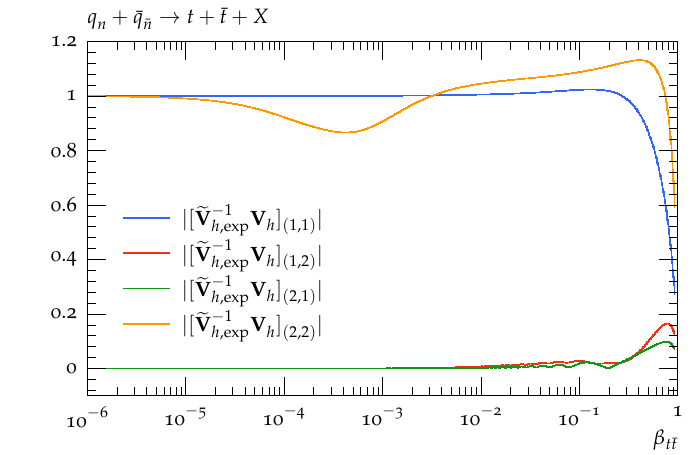}
    \caption{ .}
    \label{fig:Suk:Exp:qqbar2:n2ll:ratio}
  \end{subfigure}
  \begin{subfigure}{0.49\textwidth}
    \centering
    \includegraphics[width=.9\linewidth, height=0.6\linewidth]{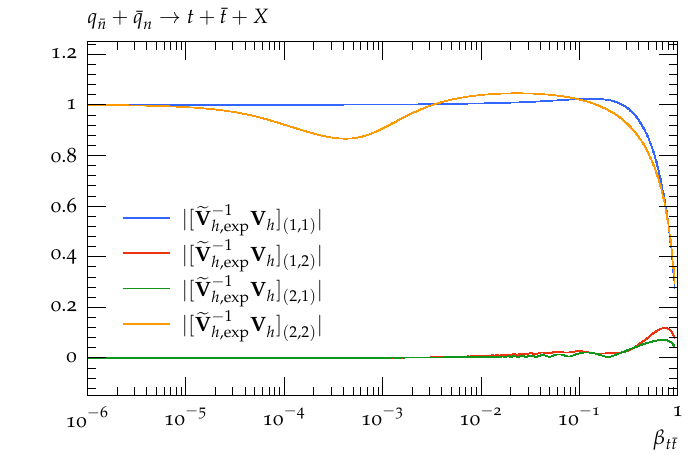}
    \caption{ . }
    \label{fig:Suk:Exp:qbarq2:n2ll:ratio}
  \end{subfigure}
\\
  \centering
  \begin{subfigure}{0.49\textwidth}
    \centering
    \includegraphics[width=.9\linewidth, height=0.6\linewidth]{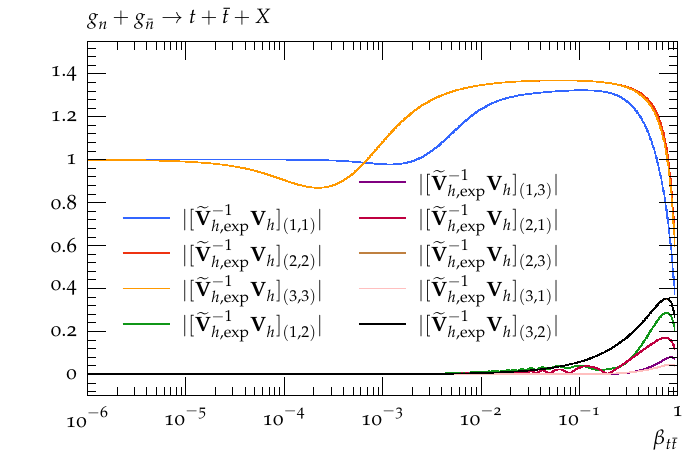}
    \caption{ .}
    \label{fig:Suk:Exp:gg2:n2ll:ratio}
  \end{subfigure}
    \caption{
Product of the non-cusp evolution $\mathbf{V}_h$ in Eq.~\eqref{eq:def:Vres:N2LL} and the inverse matrix  $\widetilde{\mathbf{V}}^{-1}_{h,\mathrm{exp}}$ in Eq.~\eqref{eq:def:Vres:N2LL:Reexp:exp} at \NNLL. Therein,  $|[\widetilde{\mathbf{V}}^{-1}_{h,\mathrm{exp}}\mathbf{V}_h]_{m,n}|$ denotes the absolute value of the entry of $(\widetilde{\mathbf{V}}^{-1}_{h,\mathrm{exp}}\mathbf{V}_h)$ in the $m$-th row and  $n$-th column from respective partonic process.
  }
  \label{fig:Suk:Exp:n2ll:ratio}
\end{figure}

Eventually, Figs.~\ref{fig:Suk:Exp:n2ll} and
\ref{fig:Suk:Exp:n2ll:ratio} depict the results
of $\mathbf{V}_h$ of Eq.~\eqref{eq:def:Vres:N2LL}
and $\widetilde{\mathbf{V}}_{h,\mathrm{exp}}$ in
Eq.~\eqref{eq:def:Vres:N2LL:Reexp:exp} at \NNLL,
the expansion of the re-exponentiated Sudakov factor.
Owing to the fact that in comparison with the leading
singular result of $\mathbf{V}_{h,\mathrm{thr}}$ in
Eq.~\eqref{eq:def:Vres:N2LL:thr:qq_gg},
$\widetilde{\mathbf{V}}_{h,\mathrm{exp}}$ is embedded
with more power corrections, the agreement between
$\mathbf{V}_h$ and $\widetilde{\mathbf{V}}_{h,\mathrm{exp}}$
is considerably improved compared to
Figs.~\ref{fig:Suk:Exp:lp:n2ll} and
\ref{fig:Suk:Exp:lp:n2ll:ratio}.
For instance, focussing on the quark-induced process,
while in Fig.~\ref{fig:Suk:Exp:qqbar2:lp:n2ll:ratio},
the deviation between the leading approximation
$\mathbf{V}_{h,\mathrm{thr}}$ in
Eq.~\eqref{eq:def:Vres:N2LL:thr:qq_gg} and
$\mathbf{V}_h$ in Eq.~\eqref{eq:def:Vres:N2LL} rapidly
surges above $\betattbar\sim 10^{-3}$, in
Fig.~\ref{fig:Suk:Exp:qqbar2:n2ll:ratio}, numerical
agreement of $\mathbf{V}_h$ and
$\widetilde{\mathbf{V}}_{h,\mathrm{exp}}$ holds up to
$\betattbar\sim 10^{-1}$ within around $10\%$.
In light of this excellent agreement, in matching the
re-exponentiated Sudakov factor onto $\mathbf{V}_h$ in
Eq.~\eqref{eq:def:Vres:N2LL:Reexp}, we choose the
matching parameters $c^{\mathrm{def}}_{\rm{thr}}=0.4$
and $ r^{\mathrm{def}}_{\mathrm{thr}}=0.1$ as defined
in Eq.~\eqref{eq:scale:ftran:beta_TT}.
In this way, the re-exponentiation impact is fully switched
on in the domain $\betattbar\le0.3$ but then gets gradually
faded out until the total shutdown at $\betattbar=0.5$.
This choice of this active range brings the difference
between $\mathbf{V}_h$ and $\widetilde{\mathbf{V}}_{h,\mathrm{exp}}$
under control, i.e.\ generally below $40\%$ for all three
partonic channels, and also steers clear of the tail region
in Fig.~\ref{fig:Suk:Exp:n2ll:ratio} where the power
correction to the small \betattbar\ expansion escalates
dramatically.

\section{\changed{Definitions of the color and helicity bases}}
\label{app:def:col:spin}
\changed{
Here we present the colour bases \cite{Beneke:2009rj} used in calculating $\mathcal{C}_{\alpha;\{h\}}^{[\kappa]}$ and $\mathcal{S}^{\alpha\beta}_{[\kappa]}$ in Eqs.\ \eqsref{eq:LP:res:qqbar}{eq:LP:res:gg},
\begin{equation}\label{eq:def:color:basis:qq_gg}
  \begin{split}
  c^{qq,(1)}_{a_1a_2a_3a_4}
  &=
    \frac{1}{3}\,\delta_{a_1a_2}\delta_{a_3a_4}\,,\;\;
  c^{qq,(2)}_{a_1a_2a_3a_4}
  =
    \frac{1}{\sqrt{2}}\sum_c T^{c}_{a_1a_2}T^{c}_{a_3a_4}\,,\\
  c^{gg,(1)}_{a_1a_2a_3a_4}
  &=
    \frac{1}{2\sqrt{6}}\,\delta_{a_1a_2}\delta_{a_3a_4}\,,\;\;
  c^{gg,(2)}_{a_1a_2a_3a_4}
  =
    \frac{i}{2\sqrt{3}}\,\sum_c f^{a_1 c a_2}T^{c}_{a_3a_4}\,,\;\;
  c^{gg,(3)}_{a_1a_2a_3a_4}
  =
    \frac{1}{2}\sqrt{\frac{3}{5}}\sum_cd^{a_1 c a_2}T^{c}_{a_3a_4}\,,
  \end{split}
\end{equation}
where $T^{c}_{ab}$ stands for the  generator in the fundamental
representation of the SU$(3)$ group. $f_{abc}$ and $d_{abc}$ mark
the antisymmetric and symmetric structure constants for the SU$(3)$
group, respectively.
}

\changed{We also specify the helicity bases concerned by the gluon-initialised channel \cite{Actis:2012qn,Actis:2016mpe}, 
\begin{align} \label{eq:Bgg:def:hel:space}
  \epsilon^{\mu}_{n,\pm}\equiv \left\{0, \frac{\mp1}{\sqrt{2}},\frac{-\mathrm{i}}{\sqrt{2}},0\right\}\,,
  \qquad \epsilon^{\mu}_{\nbar,\pm}\equiv \left\{0, \frac{\pm1}{\sqrt{2}},\frac{-\mathrm{i}}{\sqrt{2}},0\right\}\,.
\end{align}
}

\clearpage

\bibliographystyle{amsunsrt_mod}
\bibliography{ref_generator}

  \end{document}